\documentclass[]{aastex631}

\usepackage{hyperref}
\usepackage{graphicx}
\usepackage{subfigure}
\usepackage[graphicx]{realboxes}
\usepackage{appendix}
\usepackage{booktabs}
\usepackage{tabularx}
\usepackage{multirow}
\usepackage{amsmath}
\usepackage{hhline}
\usepackage[clockwise]{rotating}
\usepackage{stackengine}
\usepackage{orcidlink}
\usepackage{bm}
\usepackage[marginal]{footmisc}

\begin{document}
	\title{Calibrating the Color–Magnitude Relation of M Giants by Using Open Clusters}
	\author{\textup{X. Y. Tang}}
	\affiliation{School of Physics and Astronomy, China West Normal University, 1 ShiDa Road, Nanchong 637002, People’s Republic of China; \href{mailto:zzyan@bao.ac.cn}{zzyan@bao.ac.cn}, \href{mailto:lijing@shao.ac.cn}{lijing@shao.ac.cn}}
	\affiliation{Purple Mountain Observatory, Chinese Academy of Sciences, Nanjing 210023, People’s Republic of China; \href{mailto:xuye@pmo.ac.cn}{xuye@pmo.ac.cn}}
	
	\author{\textup{C. J. Hao}}
	\affiliation{Purple Mountain Observatory, Chinese Academy of Sciences, Nanjing 210023, People’s Republic of China; \href{mailto:xuye@pmo.ac.cn}{xuye@pmo.ac.cn}}
	
	\author{\textup{J. Li}~\orcidlink{0000-0002-4953-1545}}
	\affiliation{School of Physics and Astronomy, China West Normal University, 1 ShiDa Road, Nanchong 637002, People’s Republic of China; \href{mailto:zzyan@bao.ac.cn}{zzyan@bao.ac.cn}, \href{mailto:lijing@shao.ac.cn}{lijing@shao.ac.cn}}
	
	\author{\textup{Z. Z. Yan}~\orcidlink{0000-0003-3571-6060}}
	\affiliation{School of Physics and Astronomy, China West Normal University, 1 ShiDa Road, Nanchong 637002, People’s Republic of China; \href{mailto:zzyan@bao.ac.cn}{zzyan@bao.ac.cn}, \href{mailto:lijing@shao.ac.cn}{lijing@shao.ac.cn}}
	
	\author{\textup{Y. Xu}~\orcidlink{0000-0001-5602-3306}}
	\affiliation{Purple Mountain Observatory, Chinese Academy of Sciences, Nanjing 210023, People’s Republic of China; \href{mailto:xuye@pmo.ac.cn}{xuye@pmo.ac.cn}}
	\affiliation{School of Astronomy and Space Science, University of Science and Technology of China, Hefei 230026, People’s Republic of China}
	
	\author{\textup{J. Zhong}~\orcidlink{0000-0001-5245-0335}}
	\affiliation{Astrophysics division, Shanghai Astronomical Observatory, Chinese Academy of Sciences, 80 Nandan Road, Shanghai 200030, People’s Republic of China}
	
	\author{\textup{Z. H. Lin}}
	\affiliation{Purple Mountain Observatory, Chinese Academy of Sciences, Nanjing 210023, People’s Republic of China; \href{mailto:xuye@pmo.ac.cn}{xuye@pmo.ac.cn}}
	
	\author{\textup{Y. J. Li}~\orcidlink{0000-0001-7526-0120}}
	\affiliation{Purple Mountain Observatory, Chinese Academy of Sciences, Nanjing 210023, People’s Republic of China; \href{mailto:xuye@pmo.ac.cn}{xuye@pmo.ac.cn}}
	
	\author{\textup{D. J. Liu}}
	\affiliation{College of Science, China Three Gorges University, Yichang 443002, People's Republic of China}
	
	\author{\textup{L. F. Ding}}
	\affiliation{School of Physics and Astronomy, China West Normal University, 1 ShiDa Road, Nanchong 637002, People’s Republic of China; \href{mailto:zzyan@bao.ac.cn}{zzyan@bao.ac.cn}, \href{mailto:lijing@shao.ac.cn}{lijing@shao.ac.cn}}
	
	\author{\textup{X. F. Long}}
	\affiliation{School of Physics and Astronomy, China West Normal University, 1 ShiDa Road, Nanchong 637002, People’s Republic of China; \href{mailto:zzyan@bao.ac.cn}{zzyan@bao.ac.cn}, \href{mailto:lijing@shao.ac.cn}{lijing@shao.ac.cn}}	
	\begin{abstract}
		M giants, with their distinctive properties such as high luminosity, serve as excellent indicators for mapping the structure of the Milky Way. The distance to distant M giants can be determined by using the color-magnitude relation (CMR), which is derived from color-magnitude diagrams of specific systems in previous studies. In this work, we aimed to achieve more accurate distance determination for M giants by focusing on open clusters (OCs) with a large number of member stars and thus improve the CMR. For the first time, we compiled a census of OCs harboring M giants using Gaia Data Release 3 (DR3) and Large Sky Area Multi-Object Fiber Spectroscopic Telescope Data Release 9. We identified 58 M giants associated with 43 OCs and obtained their astrometric and photometric parameters from Gaia DR3. Using the distances of these OCs, we derived the CMR for M giants as a linear correlation, expressed as $M_{K_s}=3.85-8.26(J-K_s)$. This linear relation proved superior to the empirical distance relation in characterizing the CMR of M giants. The photometric distances of M giants derived from the CMR are consistent with the parallax distances from Gaia and known spectroscopic distances, with median deviations of 1.5\% and 2.3\%, respectively. Using the distances of M giants derived from the CMR, we computed their radial velocity ($V_R$), azimuthal velocity ($V_\phi$), and vertical velocity ($V_Z$), respectively. The distributions of these velocities revealed key features of the Galactic disk, including oscillation, north–south rotational asymmetry, and warp. These ﬁndings are consistent with previous studies and further validate the reliability of the derived CMR.
	\end{abstract}
 
	\keywords{open clusters and associations: general - stars: M giants - stars: distances - methods: data analysis}
	
	\section{Introduction} 
	M giants, located at the tip of the red giant branch (RGB) with high luminosity~\citep[$\log~L/L_\odot \sim$~3-4;][]{Gray_2009}, can be detected at large distances, making them excellent tracers for revealing accretion and merger events in the Milky Way~\citep[e.g.,][]{Ibata_1994, Martin_2004, Rocha_2003}. Determining the distances to M giants is crucial, as it not only helps to understand their properties but also unveils the structure of the Milky Way through these stars.	
	
	The color-magnitude relation (CMR) of M giants is an important method for determining their distances and is mainly studied through two types of relations.~One is the linear relation, where the absolute magnitude $M_{K_s}$ of M giants is linearly related to their intrinsic $(J-K_s)$ color, i.e.,~$M_{K_s}=A+B(J-K_s)$.~\cite{Sharma_2010} fitted this relation based on stellar models.~The other is an empirical distance relation provided by \cite{Li_2016_2}, where the absolute magnitude $M_{J}$ and intrinsic $(J-K)_0$ color of M giants are power-law related, i.e., $M_J=A_1\left[(J-K)_0^{A_2}-1\right]+A_3$.~\cite{Li_2016_2} fitted this relation using distances from the LMC, SMC, and Sgr core region as standards to estimate $M_J$.~They also found that different metallicities and star formation histories affect the fit. Therefore, for Milky Way stars, \cite{Li_2023} refitted the relation using M giants, mostly Milky Way disk stars, with distances provided by~\cite{Bailer_2021} from Gaia Data Release 3 (DR3) parallaxes.~\cite{Bailer_2021} noted that individual star distances are less accurate than those derived from star clusters.~Therefore, this work aims at calibrating the CMR of M giants using cluster distances.
	
	The known ages of open clusters (OCs) range from a few million years to several billion years. As shown by~\cite{Bellazzini_2006}, the typical ages of M giants are over 1 Gyr, overlapping with those of OCs. Therefore, M giants are expected to be found in OCs. To calibrate the CMR of M giants using their host OCs, large observational data sets from both are required.
	
	Over the last 20 years, several large-scale survey projects have gathered photometric data for thousands of M giants, including the Two Micron All Sky Survey \citep[2MASS;][]{Skrutskie_2006}, UKIRT Infrared Deep Sky Survey \citep[UKIDSS;][]{Lawrence_2007}, Wide-field Infrared Survey Explorer \citep[WISE;][]{Wright_2010}, and Panoramic Survey Telescope and Rapid Response System \citep[Pan-STARRS;][]{Kaiser_2002}. Based on these photometric data, samples of M giants were established~\citep[e.g.,][]{Majewski_2003, Bochanski_2014}. M-giant samples need to be enlarged, and this has been done by using the Large sky Area Multi Object
	ﬁber Spectroscopic Telescope (LAMOST, also known as the Guo
	Shou Jing Telescope). In the last 10 yr, thousands
	of spectra of M giants have been obtained by LAMOST~\citep{Wang_1996,Cui_2012,Su_2004,Luo_2012,Zhao_2012,Yan_2022}. To recognize M-type stars from LAMOST, \cite{Zhong_20151} developed an automatic template-fitting algorithm and \cite{Zhong_2015} identified 8639 M giants from LAMOST DR1. Recently, \cite{Li_2023} selected 44,036 M giants from the low-resolution spectral data of LAMOST DR9, making it the largest M-giant sample from LAMOST.
		
	With the continuous release of high-precision astrometric and photometric data from Gaia \citep{Gaia_Collaboration_2016}, the size of the OC catalog is continuously expanding, and the quality of OC parameters has also improved. For example, based on Gaia Data Release 2 (DR2) \citep{Gaia_2018}, \cite{Cantat-Gaudin_2018} recalculated the astrometric parameters and membership probabilities of 1229 OCs, and discovered 60 new OCs. \cite{Castro_Ginard_2020, Castro-Ginard_2022} discovered 1210 new OCs in the Galactic disk based on Gaia DR2 and Gaia Early Data Release 3 (EDR3). \cite{Hao_2022_1} discovered 704 new OCs based on Gaia EDR3. \cite{Hunt_2023} conducted a comprehensive all-sky search for Milky Way star clusters using Gaia DR3, identifying a total of 7167 clusters. More recently, they reported that 5647 of these clusters (79\%) are consistent with being gravitationally bound OCs \citep{Hunt_2024}. 
	
	The large sample of M giants and OCs enables us to search for OC and M-giant pairs in the Milky Way. Therefore, we can fit the CMR for M giants using the distances of OCs from the identified pairs. This paper is organized as follows. In Sect. \ref{sec2}, we describe the data used in this work. In Sect. \ref{sec3}, we present the methods and results of searching for OCs harboring M giants, including the relevant parameters and the selection of these OC-M giant pairs. The calibration of the CMR of M giants is presented in Sect. \ref{sec4}. Sect. \ref{sec6} demonstrates the application of the derived CMR, and we summarize this work in Sect. \ref{sec5}.
	
	\section{Data} \label{sec2}
	The latest Gaia data release, Gaia DR3~\citep{Gaia_2023}, was adopted to search for OC-M giant pairs. Gaia DR3 contains 1.47 billion sources with five- or six-parameter astrometry, and the accuracies of parallax and proper motions can reach 0.02–0.03 mas and 0.02–0.03 $\text{ mas yr}^{-1}$, respectively, offering a 30\% improvement in parallax precision and roughly doubled accuracy in proper motions compared to Gaia DR2~\citep{Gaia_2021}.
	
\begin{figure}[h]
		\centering 
		\includegraphics[width=\textwidth]{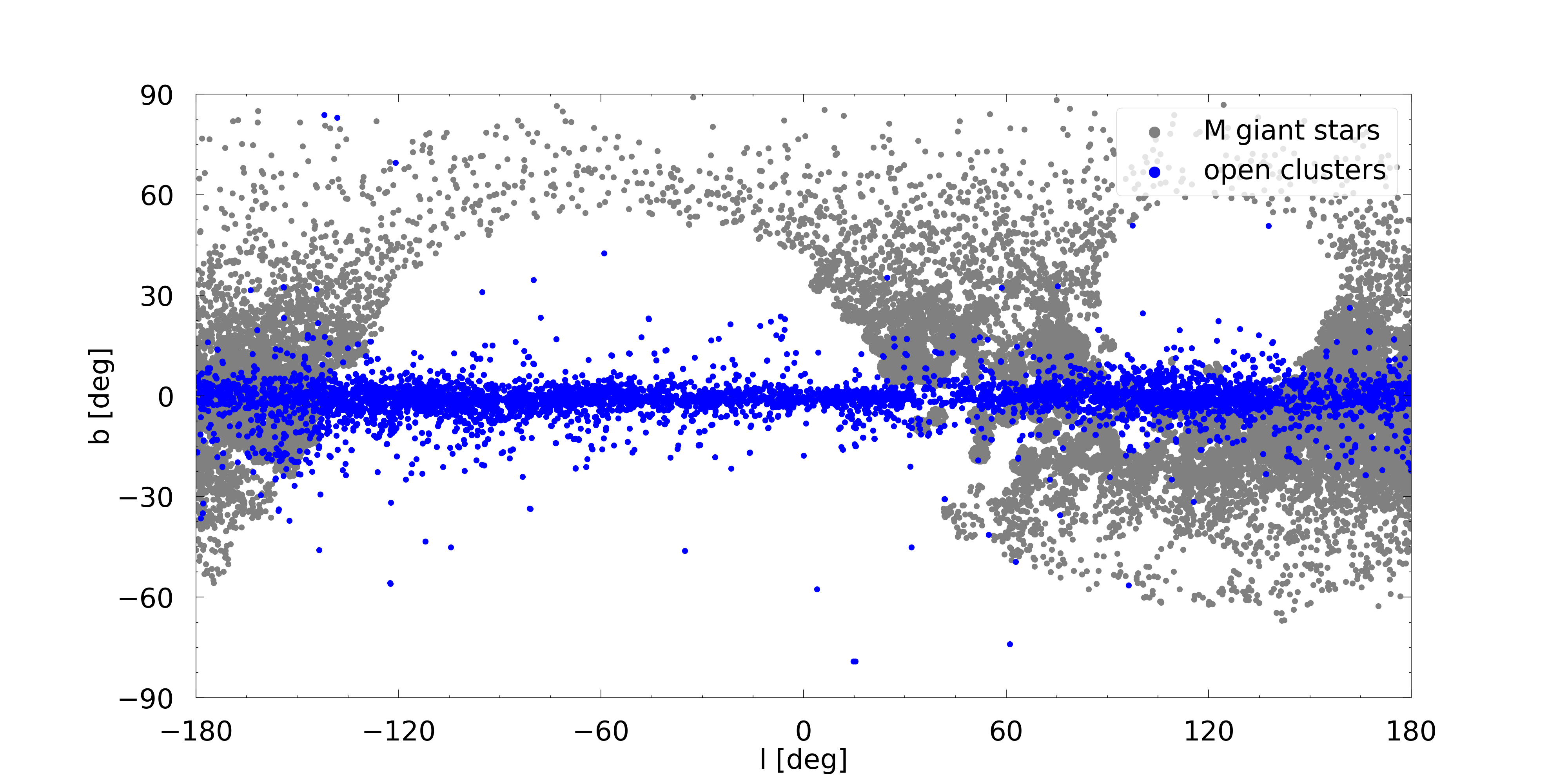}
		\caption{Distributions of M giants (shown as gray dots) from LAMOST DR9 and OCs (shown as blue dots) from Gaia DR3 in Galactic coordinates.}
		\label{fig1}
	\end{figure}

	The M giants used in this work came from the catalog compiled by \cite{Li_2023} from LAMOST DR9 low-resolution spectra. We crossmatched these 44,036 M giants with the Gaia DR3 catalog, using a radius of $\text{1''}$, and obtained 34,657 M giants. These M giants are mainly located within the Galactic disk, and their distributions in Galactic longitude ($l$) and latitude ($b$) coordinates are shown as gray dots in Figure~\ref{fig1}.
	
	The OCs utilized in this study were sourced from \cite{Cantat-Gaudin_2020}, \cite{Castro-Ginard_2022}, \cite{Hao_2022_1}, and \cite{Hunt_2024}, resulting in a total of 6879 OCs. The member stars of these clusters were crossmatched with Gaia DR3, and all $source\_ids$ for the OC member stars included in this analysis were derived from Gaia DR3. As illustrated by the blue dots in Figure ~\ref{fig1}, these OCs are mainly distributed within the Galactic disk, concentrated at latitudes of $|b|\leqslant20^{\circ}$. %\vspace{4\baselineskip}

	\section{Open Clusters harboring M giants} \label{sec3}
	
	\subsection{Methods and Results}
	
	We identified OCs harboring M giants using two criteria similar to \cite{Hao_2022_2}: (1) where M giants are confirmed as member stars of OCs; or (2) where the 5D astrometric parameters of M giants and OCs, which include stellar celestial positions ($l$, $b$), parallaxes ($\varpi$), and proper motions ($\mu_{\alpha^*}$ and~$\mu_\delta$), are consistent within 3 times the corresponding standard deviations. For the first criterion, we crossmatched the list of M giants with the member stars of OCs in this study using the Gaia DR3 $source\_id$, resulting in 19 OCs containing 27 M giants. Under the second criterion, we compared the 5D astrometric parameters of the remaining M giants with all OCs, identifying 31 M giants associated with 24 OCs.
	
	Overall, we identified 43 OCs harboring 58 M giants (see Appendix \ref{appendixA}), with the names of these OC-M giant pairs listed in Table \ref{table1}. Figure~\ref{fig2} illustrates selected examples of OC-M giant pairs, depicting their distributions in astrometric spaces and color–magnitude diagrams (CMDs). Additional cases are detailed in Appendix \ref{appendixB}, as shown in Figures \ref{figB1}–\ref{figB5}.\vspace{2\baselineskip}
	
	\subsection{Parameters}
	Out of the 43 OCs containing M giants, seven were reported by \cite{Cantat-Gaudin_2020}, three by \cite{Castro-Ginard_2022}, two by \cite{Hao_2022_1}, and 31 by \cite{Hunt_2024}. Based on the catalogs from these four studies, we listed the astrometric parameters ($\alpha, \delta, \varpi, \mu_{\alpha^*}, \mu_\delta$) and their corresponding standard deviations for the OCs in Table \ref{table1}, all derived from Gaia DR3.
	
	The distances to OCs hosting M giants were primarily obtained from the \cite{Hunt_2024} catalog, which employed the maximum likelihood estimation method. For clusters not included in \cite{Hunt_2024}, we independently calculated their distances using the same approach.
	
	The ages of the OCs were primarily sourced from the most recent determinations based on Gaia data. When OCs from the catalogs of \cite{Cantat-Gaudin_2020}, \cite{Hao_2022_1}, or \cite{Castro-Ginard_2022} overlapped with \cite{Hunt_2024}, we relied exclusively on the results from \cite{Hunt_2024}. For clusters not included in \cite{Hunt_2024}, we utilized the ages provided in the aforementioned three catalogs. During our review, we noted that the age uncertainty for Berkeley 29, as reported in \cite{Hunt_2024}, is substantial (7.88--9.12 Gyr). For this cluster, we referred to previous dedicated studies, which consistently indicate an age exceeding 1 Gyr \citep[e.g.,][]{Tosi_2004, Salaris_2004, Perren_2022}.
	
	 The astrometric parameters and uncertainties for M giants are also listed in Table \ref{table1} from the data set of Gaia DR3. The apparent angular radius ($\theta$) for each OC was estimated as $\theta$ = $\sqrt{{\sigma_l}^{2}+{\sigma_b}^{2}}$ (${\sigma_l}$ and ${\sigma_b}$ are the standard deviations of $l$ and $b$, respectively) following~\cite{Castro-Ginard_2022}, and the apparent diameter was calculated as twice $\theta$. Angular separations between M giants and their host OCs were also estimated, with cluster centers derived from mean astrometric positions of member stars. Additionally, the parameters of the OCs listed in Table \ref{table1} are accompanied by their corresponding reference sources.
	 
	\subsection{Selection\label{subsec:Selection}}
	
\begin{figure}[h]
		\centering  
		\subfigure{
			\includegraphics[width=0.32\textwidth]{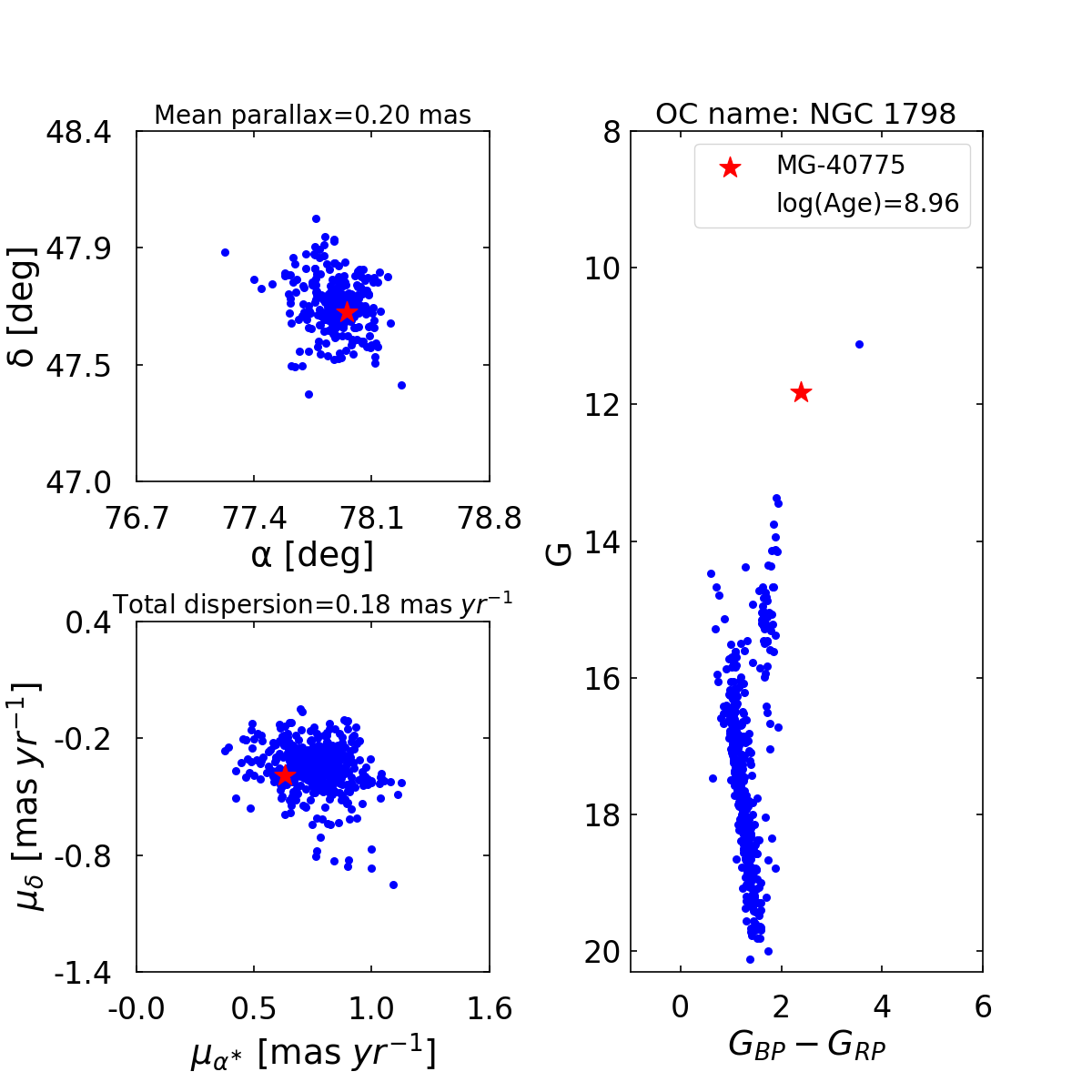}}
		\subfigure{
			\includegraphics[width=0.32\textwidth]{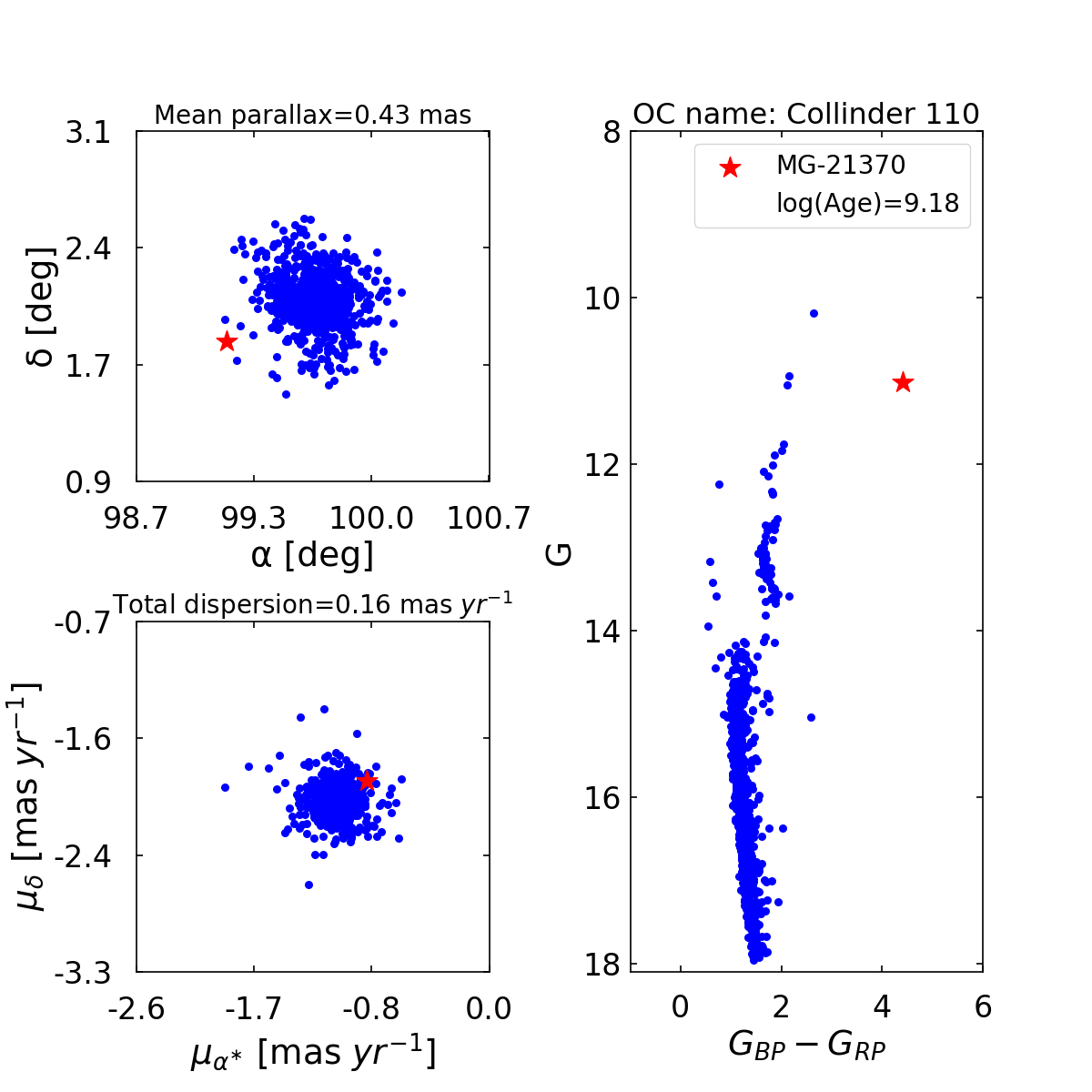}}
		\subfigure{
			\includegraphics[width=0.32\textwidth]{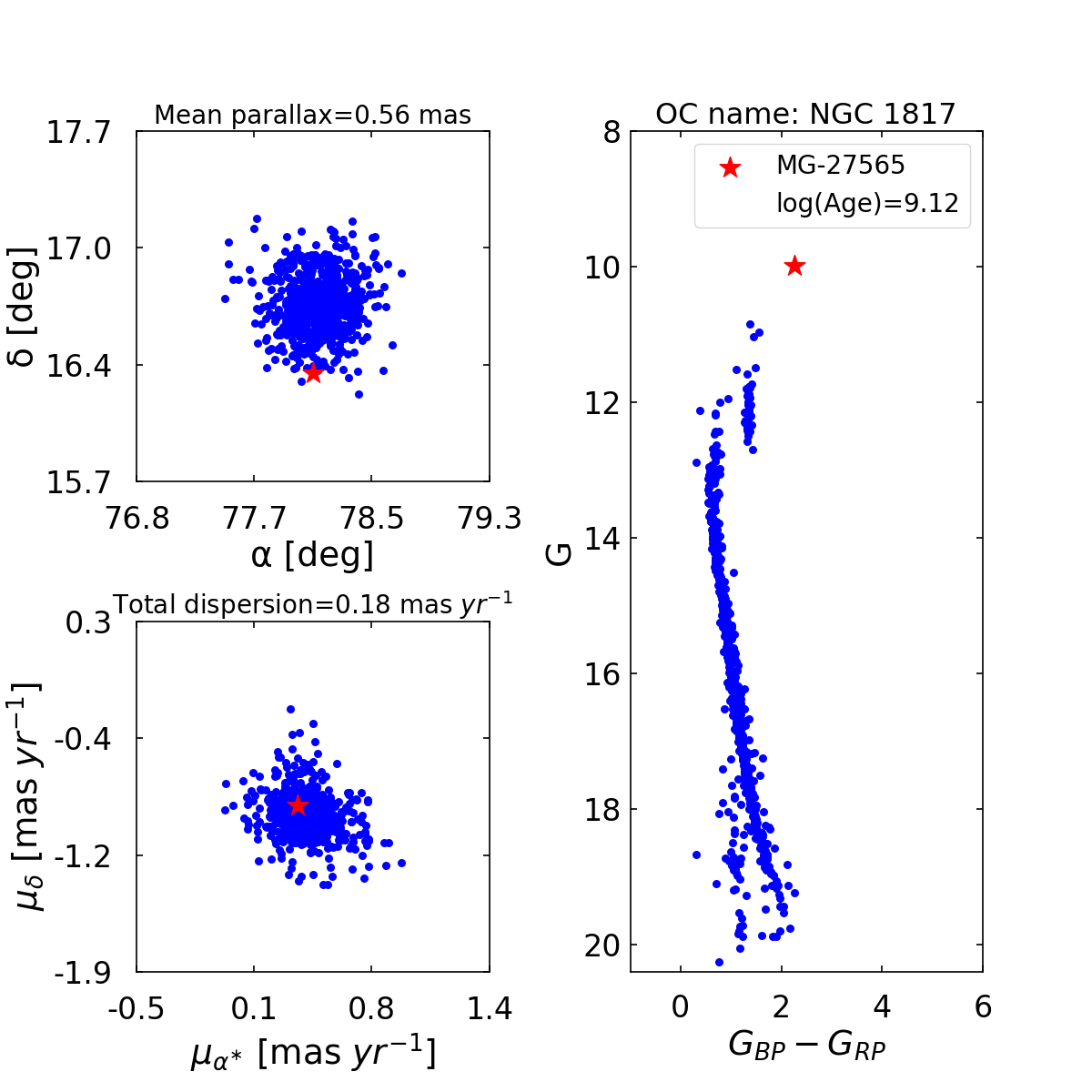}}\\
		\centering  
		\subfigure{
			\includegraphics[width=0.32\textwidth]{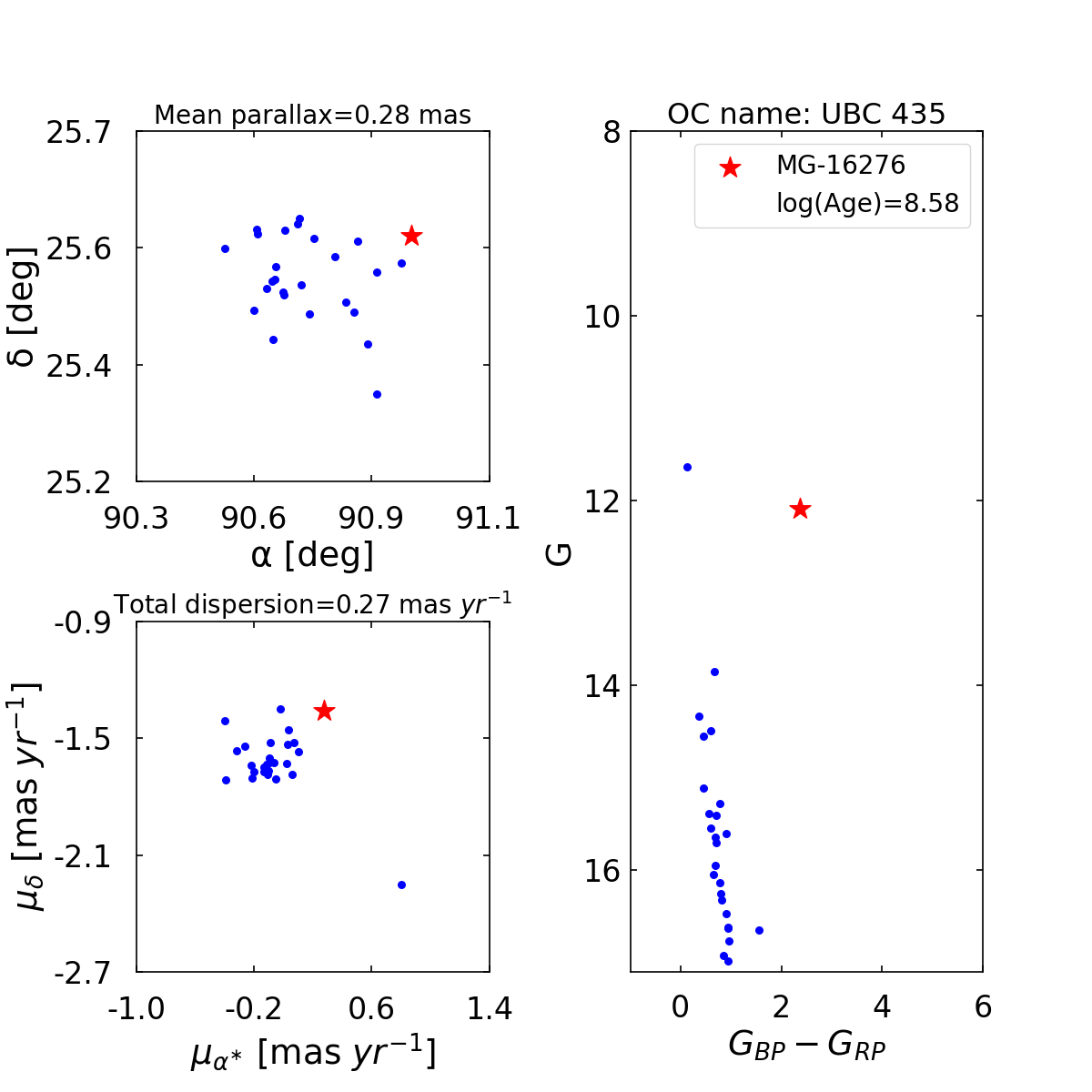}}
		\subfigure{
			\includegraphics[width=0.32\textwidth]{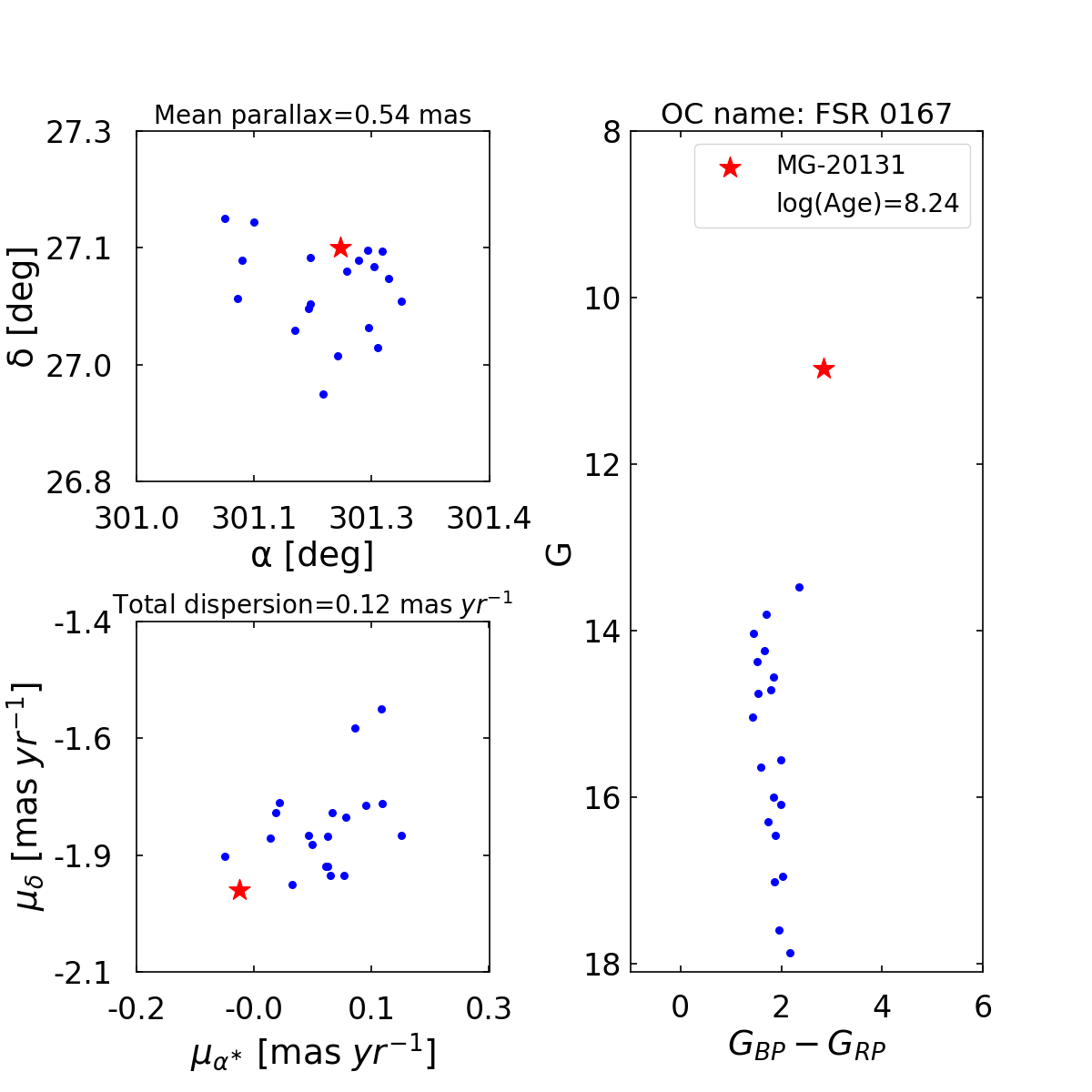}}
		\subfigure{
			\includegraphics[width=0.32\textwidth]{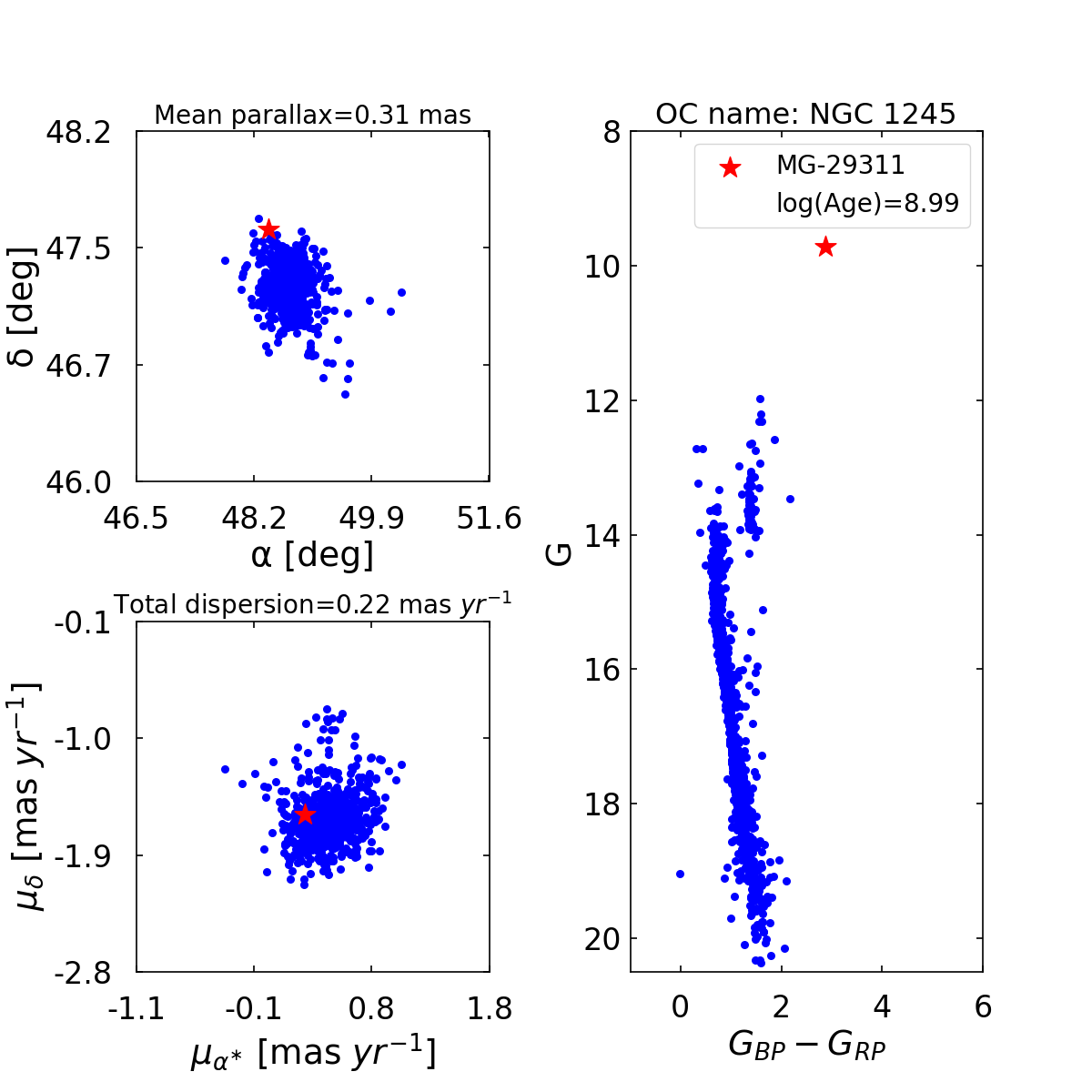}}
		\caption{Examples of OCs (blue dots) harboring M giants (red stars). The columns of each panel represent the distributions of the member stars of OCs and M giants for positions ($\alpha$ and $\delta$), proper motions (${\mu}_{\alpha^*}$ and ${\mu}_\delta$), and the CMD, as well as the mean parallax and total proper-motion dispersion of OCs.~Here, the listed OCs are NGC~1798, Collinder~110, NGC~1817, UBC~435, FSR~0167,~and~NGC~1245. The identification numbers of M giants and the ages of OCs are displayed in the upper right of the CMD.}
		\label{fig2}
	\end{figure}

	As M giants are RGB stars, they are generally located at the tip of the RGB in the CMDs of OCs that harbor them. We visually inspected the CMDs of the 58 OC-M giant pairs identified in this work (as depicted in Figure ~\ref{fig2} and Appendix \ref{appendixB}), focusing on the 57 pairs that exhibit one of the following characteristics: (1) those with a distinct OC RGB and the M giants are clearly positioned at its tip; or (2) those with an indistinct OC RGB due to a limited number of OC member stars, yet the M giants are positioned to the right of the OC member-star distribution in the CMD. Considering the ages of M giants are typically over 1 Gyr, as noted by~\cite{Bellazzini_2006}, we further refined our selection to 17 pairs where the ages of the OCs are also over 1 Gyr. These OCs are marked with ($\star$) in Table~\ref{table1}.
		
	Next, we will calibrate the CMR for M giants using the distances derived from all 58 OC-M giant pairs, as well as the subset of 17 pairs where the OC ages exceed 1 Gyr.
	
\begin{longrotatetable}
	\addtolength{\tabcolsep}{-5pt}
	\begin{deluxetable*}{ccccccccccccccccccc}
		%\centering
		\tablecaption{Parameters of the 67 OC-M giant pairs\label{table1}}
		\tablewidth{2000pt}
		\tabletypesize{\scriptsize}
		\tablehead{
			\colhead{MG} & \colhead{MG: $\alpha$}&\colhead{MG: $\delta$}&\colhead{OC name} & 
			\colhead{~~OC: $\alpha$} & \colhead{~OC: $\delta$} & 
			\colhead{Sep} & \colhead{OC: AD} & 
			\colhead{MG: $\mu_{\alpha^*}$} & \colhead{MG: $\mu_{\delta}$} & \colhead{OC: $\mu_{\alpha^*}$} & \colhead{OC: $\mu_{\delta}$} & \colhead{MG: $\varpi$} & \colhead{OC: $\varpi$} & \colhead{OC: log(age)} &\colhead{OC: N} &\colhead{~Num} \\ 
			\colhead{} & \colhead{[deg]} & \colhead{[deg]} & \colhead{} & 
			\colhead{[deg]} & \colhead{[deg]} & \colhead{[arcmin]} &
			\colhead{[arcmin]} & \colhead{[mas $yr^{-1}$]} & \colhead{[mas $yr^{-1}$]} & \colhead{[mas $yr^{-1}$]} & \colhead{[mas $yr^{-1}$]} & \colhead{[mas]} & \colhead{[mas]} & \colhead{[yr]}
		}  
		\startdata
		31646 & 359.26(0.01) & 56.77(0.01) & $\text{NGC 7789}^a$($\star$) & 359.33(0.34) & 56.72(0.16) & 3.30 & 29.75 & -0.85(0.01) & -2.04(0.01) & -0.91(0.12) & -1.95(0.12) & 0.53(0.01) & 0.48(0.03) & $9.19^{+0.15}_{-0.17}$ & 4125 & 1 &\\
		31740 & 359.38(0.01) & 56.85(0.01) & $\text{NGC 7789}^a$($\star$) & 359.33(0.34) & 56.72(0.16) & 7.59 & 29.75 & -0.95(0.01) & -2.33(0.01) & -0.91(0.12) & -1.95(0.12) & 0.53(0.01) & 0.48(0.03) & $9.19^{+0.15}_{-0.17}$ & 4125 & 2 &\\
		41506 & 290.32(0.02) & 37.78(0.02) & $\text{NGC 6791}^a$($\star$) & 290.22(0.11) & 37.77(0.10) & 4.86 & 15.51 & -0.33(0.02) & -2.36(0.03) & -0.42(0.10) & -2.28(0.10) & 0.25(0.02) & 0.22(0.07) & $9.58^{+0.18}_{-0.14}$ & 4193 & 3 &\\
		41460 & 290.27(0.01) & 37.79(0.01) & $\text{NGC 6791}^a$($\star$) & 290.22(0.11) & 37.77(0.10) & 2.63 & 15.51 & -0.42(0.01) & -2.34(0.01) & -0.42(0.10) & -2.28(0.10) & 0.21(0.01) & 0.22(0.07) & $9.58^{+0.18}_{-0.14}$ & 4193 & 4 &\\
		41308 & 290.39(0.01) & 37.84(0.01) & $\text{NGC 6791}^a$($\star$) & 290.22(0.11) & 37.77(0.10) & 9.27 & 15.51 & -0.36(0.01) & -2.41(0.02) & -0.42(0.10) & -2.28(0.10) & 0.22(0.01) & 0.22(0.07) & $9.58^{+0.18}_{-0.14}$ & 4193 & 5 &\\
		21954 & 290.21(0.02) & 37.73(0.02) & $\text{NGC 6791}^a$($\star$) & 290.22(0.11) & 37.77(0.10) & 2.93 & 15.51 & -0.46(0.02) & -2.35(0.02) & -0.42(0.10) & -2.28(0.10) & 0.23(0.02) & 0.22(0.07) & $9.58^{+0.18}_{-0.14}$ & 4193 & 6 &\\
		41267 & 290.22(0.01) & 37.74(0.01) & $\text{NGC 6791}^a$($\star$) & 290.22(0.11) & 37.77(0.10) & 2.14 & 15.51 & -0.45(0.01) & -2.18(0.01) & -0.42(0.10) & -2.28(0.10) & 0.22(0.01) & 0.22(0.07) & $9.58^{+0.18}_{-0.14}$ & 4193 & 7 &\\
		32601 & 290.23(0.01) & 37.77(0.01) & $\text{NGC 6791}^a$($\star$) & 290.22(0.11) & 37.77(0.10) & 0.52 & 15.51 & -0.38(0.01) & -2.50(0.02) & -0.42(0.10) & -2.28(0.10) & 0.20(0.01) & 0.22(0.07) & $9.58^{+0.18}_{-0.14}$ & 4193 & 8 &\\
		22819 & 295.38(0.01) & 40.15(0.01) & $\text{NGC 6819}^a$($\star$) & 295.32(0.12) & 40.19(0.10) & 3.48 & 17.06 & -2.53(0.01) & -3.66(0.01) & -2.90(0.09) & -3.87(0.09) & 0.37(0.01) & 0.38(0.03) & $9.23^{+0.20}_{-0.13}$ & 2311 & 9 &\\
		32009 & 295.28(0.01) & 40.33(0.01) & $\text{NGC 6819}^a$($\star$) & 295.32(0.12) & 40.19(0.10) & 8.35 & 17.06 & -2.81(0.02) & -3.86(0.02) & -2.90(0.09) & -3.87(0.09) & 0.36(0.01) & 0.38(0.03) & $9.23^{+0.20}_{-0.13}$ & 2311 & 10 &\\
		24452 & 99.31(0.02) & -1.00(0.02) & $\text{Berkeley 24}^a$($\star$) & 99.45(0.06) & -0.88(0.06) & 10.73 & 9.78 & 0.30(0.03) & 1.28(0.02) & 0.33(0.12) & 1.27(0.12) & 0.18(0.02) & 0.19(0.07) & $9.15^{+0.19}_{-0.16}$ & 306 & 11 &\\
		24559 & 99.45(0.02) & -0.84(0.02) & $\text{Berkeley 24}^a$($\star$) & 99.45(0.06) & -0.88(0.06) & 2.15 & 9.78 & 0.24(0.03) & 1.32(0.03) & 0.33(0.12) & 1.27(0.12) & 0.22(0.02) & 0.19(0.07) & $9.15^{+0.19}_{-0.16}$ & 306 & 12 &\\
		27565 & 78.09(0.02) & 16.30(0.01) & $\text{NGC 1817}^a$($\star$) & 78.14(0.19) & 16.69(0.17) & 24.30 & 29.78 & 0.38(0.03) & -0.86(0.02) & 0.43(0.10) & -0.93(0.09) & 0.55(0.02) & 0.57(0.05) & $9.12^{+0.15}_{-0.19}$ & 718 & 13 &\\
		21370 & 99.17(0.05) & 1.81(0.06) & $\text{Collinder 110}^b$($\star$) & 99.68(0.14) & 2.07(0.16) & 33.91 & 26.00 & -0.88(0.07) & -1.89(0.06) & -1.10(0.12) & -2.04(0.11) & 0.53(0.07) & 0.43(0.06) & $9.18^{+0.15}_{-0.15}$ & 881 & 14 &\\
		11061 &~103.27(0.02) &~16.93(0.01) & $\text{Berkeley 29}^b$($\star$) & 103.27(0.02) & 16.93(0.01) & 0.21 & 2.27 & 0.12(0.02) & -1.07(0.02) & 0.09(0.06) & -1.03(0.04) & 0.04(0.02) & 0.03(0.05) & $9.54^e, 9.57^f, 9.64^g$ & 11 & 15 &\\
		14581 &~298.07(0.01) &~ 44.91(0.01) &~$\text{UBC 1108}^c$($\star$) &~298.09(0.06) &~44.81(0.06) & 6.05 & 8.88 & -2.01(0.02) & -3.80(0.02) & -1.91(0.10) & -3.73(0.12) & 0.21(0.01) & 0.21(0.02) & $9.26^{+0.22}_{-0.21}$ & 56 & 16 &\\
		40175 & 73.93(0.02) & 36.71(0.01) & $\text{OC 0285}^d$($\star$) & 73.94(0.06) & 36.78(0.04) & 4.50 & 7.99 & 0.06(0.03) & -1.35(0.03) & 0.07(0.07) & -1.57(0.10) & 0.48(0.03) & 0.45(0.02) & $9.00^{+0.22}_{-0.18}$ & 28 & 17 &\\
		24827 & 81.53(0.02) & 41.88(0.01) & $\text{Berkeley 70}^a$ & 81.45(0.07) & 41.95(0.05) & 5.57 & 9.08 & 0.67(0.02) & -1.68(0.02) & 0.83(0.12) & -1.87(0.11) & 0.20(0.02) & 0.21(0.08) & $8.96^{+0.21}_{-0.19}$ & 291 & 18 &\\
		40775 & 77.95(0.02) & 47.67(0.01) & $\text{NGC 1798}^a$ & 77.91(0.12) & 47.69(0.08) & 1.92 & 13.88 & 0.65(0.02) & -0.40(0.02) & 0.79(0.12) & -0.37(0.12) & 0.23(0.02) & 0.20(0.07) & $8.96^{+0.17}_{-0.17}$ & 453 & 19 &\\
		29311 & 48.40(0.02) & 47.57(0.02) & $\text{NGC 1245}^a$ & 48.71(0.20) & 47.24(0.11) & 23.97 & 21.21 & 0.29(0.03) & -1.59(0.02) & 0.47(0.06) & -1.66(0.06) & 0.28(0.02) & 0.30(0.03) & $8.99^{+0.17}_{-0.18}$ & 1165 & 20 &\\
		35948 &~45.61(0.02) &~48.02(0.02) &~$\text{UBC 1246}^a$ &~45.52(0.18) &~47.97(0.14) &~4.57 &~22.40 & ~0.44(0.02) &~ -1.21(0.02) & ~0.50(0.11) &~ -1.23(0.14) & ~0.35(0.02) & ~0.34(0.05) & $8.94^{+0.21}_{-0.18}$ & ~105 & 21 &\\
		20083 & 97.68(0.03) & 15.17(0.02) & $\text{UBC 203}^a$ & 97.81(0.08) & 15.06(0.12) & 9.86 & 17.53 & -0.24(0.03) & -1.07(0.02) & -0.15(0.10) & -1.07(0.11) & 0.26(0.03) & 0.32(0.06) & $8.76^{+0.17}_{-0.18}$ & 145 & 22 &\\
		5631 & 92.11(0.03) & 28.11(0.03) & $\text{CWNU 1295}^a$ & 92.08(0.09) & 28.07(0.26) & 2.71 & 33.26 & -0.002(0.03) & -1.61(0.03) & -0.22(0.12) & -1.67(0.10) & 0.30(0.04) & 0.26(0.04) & $8.75^{+0.20}_{-0.21}$ & 101 & 23 &\\
		39206 & 105.08(0.02) & -0.24(0.02) & $\text{Berkeley 34}^b$ & 105.10(0.02) & -0.24(0.02) & 0.92 & 3.29 & -1.36(0.02) & 0.18(0.02) & -1.36(0.16) & 0.21(0.08) & 0.11(0.02) & 0.15(0.08) & $8.29^{+0.29}_{-0.25}$ & 30 & 24 &\\
	    19191 & 61.24(0.03) & 51.36(0.02) & $\text{UBC 1256}^c$ & 61.35(0.13) & 51.51(0.09) & 10.07 & 14.86 & -0.31(0.05) & -1.14(0.04) & -0.38(0.11) & -1.27(0.10) & 0.31(0.04) & 0.34(0.04) & $7.74^{+0.22}_{-0.21}$ & 57 & 25 &\\
	    49820 & 96.99(0.02) & 16.97(0.01) & $\text{UBC 1314}^c$ & 97.03(0.11) & 16.80(0.14) & 10.54 & 21.23 & -0.09(0.02) & -0.29(0.02) & 0.003(0.07) & -0.24(0.10) & 0.18(0.02) & 0.21(0.02) & $7.95^{+0.15}_{-0.15}$ & 66 & 26 &\\
		38564 & 300.83(0.01) & 30.80(0.02) & $\text{OC 0094}^d$ & 300.84(0.09) & 30.78(0.05) & 0.85 & 11.41 & -2.83(0.02) & -5.44(0.02) & -2.94(0.20) & -5.31(0.15) & 0.24(0.02) & 0.24(0.003) & $7.10^{+0.16}_{-0.16}$ & 15 & 27 &\\
		23669 & 12.47(0.02) & 55.72(0.02) & $\text{CWNU 1697}^a$ & 12.01(0.37) & 55.91(0.14) & 19.26 & 30.06 & -2.07(0.02) & -0.52(0.03) & -1.92(0.10) & -0.75(0.15) & 0.38(0.03) & 0.26(0.05) & $8.77^{+0.20}_{-0.19}$ & 136 & 28 &\\
		11413 & 72.67(0.01) & 45.84(0.01) & $\text{FSR 0702}^a$ & 72.92(0.07) & 45.78(0.06) & 11.06 & 9.54 & 0.18(0.02) & -0.50(0.02) & 0.35(0.09) & -0.39(0.11) & 0.16(0.02) & 0.20(0.04) & $7.31^{+0.17}_{-0.17}$ & 55 & 29 &\\
		19038 & 81.89(0.04) & 40.11(0.03) & $\text{HSC 1322}^a$ & 82.04(0.20) & 39.52(0.20) & 36.55 & 30.67 & -0.09(0.04) & -0.79(0.03) & 0.10(0.08) & -0.70(0.09) & 0.11(0.04) & 0.18(0.04) & $7.89^{+0.25}_{-0.23}$ & 73 & 30 &\\
		23538 & 82.17(0.03) & 38.86(0.02) & $\text{HSC 1322}^a$ & 81.98(0.20) & 39.52(0.20) & 40.07 & 30.67 & -0.11(0.04) & -0.45(0.03) & 0.10(0.08) & -0.70(0.09) & 0.23(0.03) & 0.18(0.04) & $7.89^{+0.25}_{-0.23}$ & 73 & 31 &\\
		16054 & 89.15(0.03) & 26.16(0.02) & $\text{HSC 1450}^a$ & 89.00(0.17) & 25.72(0.12) & 27.76 & 23.63 & 0.15(0.04) & -0.88(0.02) & 0.30(0.10) & -0.65(0.12) & 0.11(0.03) & 0.19(0.04) & $8.05^{+0.25}_{-0.21}$ & 82 & 32 &\\
		30443 & 99.30(0.01) & 16.83(0.01) & $\text{HSC 1552}^a$ & 99.14(0.16) & 17.16(0.23) & 21.86 & 33.01 & -0.13(0.02) & -0.55(0.01) & 0.001(0.18) & -0.51(0.21) & 0.11(0.02) & 0.22(0.04) & $8.37^{+0.20}_{-0.21}$ & 126 & 33 &\\
		30411 & 99.17(0.03) & 16.55(0.03) & $\text{HSC 1552}^a$ & 99.14(0.16) & 17.16(0.23) & 36.77 & 33.01 & -0.46(0.04) & -0.19(0.03) & 0.001(0.18) & -0.51(0.21) & 0.25(0.04) & 0.22(0.04) & $8.37^{+0.20}_{-0.21}$ & 126 & 34 &\\
		30610 & 98.72(0.04) & 17.44(0.03) & $\text{HSC 1552}^a$ & 99.14(0.16) & 17.16(0.23) & 29.58 & 33.01 & -0.36(0.05) & -0.95(0.04) & 0.001(0.18) & -0.51(0.21) & 0.32(0.04) & 0.22(0.04) & $8.37^{+0.20}_{-0.21}$ & 126 & 35 &\\
		39037 & 61.88(0.02) & 46.90(0.01) &  $\text{Juchert 20}^a$ & 62.70(0.24) & 46.87(0.13) & 33.67 & 24.79 & 0.92(0.03) & -1.53(0.02) & 1.09(0.11) & -1.66(0.12) & 0.38(0.02) & 0.31(0.07) & $8.33^{+0.18}_{-0.20}$ & 167 & 36 &\\
		27611 & 92.90(0.01) & 26.12(0.01) & $\text{Koposov~53}^a$ & 92.23(0.21) & 26.26(0.22) & 37.19 & 34.84 & 0.29(0.02) & -0.66(0.01) & 0.24(0.09) & -0.63(0.09) & 0.12(0.02) & 0.20(0.04) & $8.32^{+0.21}_{-0.20}$ & 131 & 37 &\\
		42133 & 91.42(0.01) & 26.37(0.01) & $\text{Koposov~53}^a$ & 92.23(0.21) & 26.26(0.22) & 43.86 & 34.84 & 0.23(0.02) & -0.44(0.01) & 0.24(0.09) & -0.63(0.09) & 0.26(0.02) & 0.20(0.04) & $8.32^{+0.21}_{-0.20}$ & 131 & 38 &\\
		867 & 323.99(0.02) & 53.86(0.01) & $\text{Kronberger 84}^a$ & 323.89(0.15) & 53.51(0.10) & 21.11 & 16.31 & -3.12(0.02) & -3.00(0.02) & -2.92(0.09) & -3.03(0.07) & 0.15(0.02) & 0.19(0.03) & $8.04^{+0.21}_{-0.23}$ & 79 & 39 &\\
		8900 & 90.84(0.01) & 28.11(0.01) & $\text{LP 1375}^a$ & 91.04(0.19) & 28.05(0.18) & 10.96 & 29.66 & 0.57(0.02) & -1.11(0.01) & 0.40(0.10) & -1.01(0.11) & 0.09(0.02) & 0.21(0.05) & $8.71^{+0.21}_{-0.19}$ & 105 & 40 &\\
		5966 & 91.57(0.02) & 27.61(0.02) & $\text{LP 1375}^a$ & 91.04(0.19) & 28.05(0.18) & 38.55 & 29.66 & 0.60(0.02) & -0.84(0.02) & 0.40(0.10) & -1.01(0.11) & 0.19(0.02) & 0.21(0.05) & $8.71^{+0.21}_{-0.19}$ & 105 & 41 &\\
		21826 & 95.74(0.01) & 22.23(0.01) & $\text{LP 1973}^a$ & 95.45(0.12) & 22.41(0.09) & 19.40 & 17.73 & 1.29(0.02) & -2.65(0.01) & 1.00(0.12) & -2.58(0.11) & 0.23(0.02) & 0.36(0.06) & $8.71^{+0.19}_{-0.18}$ & 178 & 42 &\\
		14156 & 80.09(0.02) & 42.10(0.01) & $\text{PHOC 9}^a$ & 79.59(0.23) & 41.94(0.13) & 24.39 & 26.33 & 0.25(0.03) & -0.71(0.02) & 0.06(0.14) & -0.48(0.09) & 0.11(0.02) & 0.19(0.04) & $7.58^{+0.21}_{-0.20}$ & 107 & 43 &\\
		24066 & 79.87(0.01) & 41.55(0.01) & $\text{PHOC 9}^a$ & 79.59(0.23) & 41.94(0.13) & 26.48 & 26.33 & 0.49(0.02) & -0.55(0.01) & 0.06(0.14) & -0.48(0.09) & 0.17(0.01) & 0.19(0.04) & $7.58^{+0.21}_{-0.20}$ & 107 & 44 &\\
		42382 & 79.80(0.02) & 41.83(0.01) & $\text{PHOC 9}^a$ & 79.59(0.23) & 41.94(0.13) & 11.36 & 26.33 & 0.40(0.02) & -0.56(0.02) & 0.06(0.14) & -0.48(0.09) & 0.09(0.02) & 0.19(0.04) & $7.58^{+0.21}_{-0.20}$ & 107 & 45 &\\
		32887 & 87.21(0.02) & 25.47(0.01) & $\text{Teutsch 57}^a$ & 87.00(0.08) & 25.50(0.20) & 11.80 & 25.92 & 0.13(0.02) & -0.52(0.01) & 0.33(0.11) & -0.43(0.09) & 0.16(0.02) & 0.17(0.03) & $8.49^{+0.19}_{-0.23}$ & 43 & 46 &\\
		23698 & 65.12(0.03) & 52.27(0.02) & $\text{UBC 52}^a$ & 64.79(0.14) & 52.33(0.09) & 12.52 & 15.16 & -0.62(0.04) & 0.47(0.03) & -0.84(0.09) & 0.57(0.08) & 0.33(0.03) & 0.43(0.06) & $8.88^{+0.19}_{-0.14}$ & 50 & 47 &\\
		12281 & 62.55(0.01) & 53.09(0.01) & $\text{UBC 1255}^a$ & 62.97(0.14) & 53.01(0.09) & 16.28 & 15.34 & 0.19(0.02) & -0.49(0.01) & 0.19(0.09) & -0.61(0.11) & 0.28(0.02) & 0.23(0.05) & $8.19^{+0.25}_{-0.21}$ & 86 & 48 &\\
		41783 & 96.56(0.02) & 17.40(0.02) & $\text{UBC 1313}^a$ & 96.89(0.16) & 17.19(0.12) & 22.80 & 23.58 & 0.48(0.02) & 0.05(0.02) & 0.71(0.08) & -0.002(0.08) & 0.15(0.02) & 0.18(0.04) & $8.58^{+0.21}_{-0.17}$ & 81 & 49 &\\
		29451 & 107.50(0.02) & 0.62(0.02) & $\text{UBC 1344}^a$ & 107.35(0.10) & 0.53(0.08) & 10.76 & 15.40 & -0.82(0.03) & -0.80(0.02) & -1.01(0.08) & -0.67(0.06) & 0.24(0.02) & 0.29(0.03) & $8.46^{+0.22}_{-0.23}$ & 57 & 50 &\\ 
		10261 & 64.74(0.02) & 53.15(0.02) & $\text{Waterloo 1}^a$ & 64.65(0.27) & 52.86(0.11) & 17.67 & 23.43 & 0.31(0.03) & -0.92(0.02) & 0.49(0.12) & -0.78(0.06) & 0.19(0.03) & 0.23(0.03) & $7.29^{+0.25}_{-0.21}$ & 122 & 51 &\\
		42720 & 87.69(0.03) & 22.25(0.02) & $\text{Berkeley 72}^b$ & 87.56(0.07) & 22.25(0.07) & 7.15 & 11.65 & 0.83(0.03) & -0.54(0.02) & 0.75(0.19) & -0.24(0.17) & 0.12(0.03) & 0.16(0.10) & $8.47^{+0.17}_{-0.24}$ & 55 & 52 &\\
		20131 & 301.22(0.01) & 27.11(0.01) & $\text{FSR 0167}^b$ & 301.21(0.06) & 27.06(0.06) & 3.21 & 9.15 & -0.07(0.02) & -1.97(0.02) & 0.06(0.07) & -1.83(0.10) & 0.52(0.02) & 0.54(0.03) & $8.24^{+0.21}_{-0.20}$ & 19 & 53 &\\
		10715 & 45.39(0.02) & 57.28(0.03) & $\text{SAI 25}^b$ & 45.09(0.13) & 57.35(0.09) & 10.69 & 13.35 & 0.92(0.03) & -0.73(0.04) & 0.38(0.29) & -1.00(0.32) & 0.28(0.03) & 0.33(0.09) & $7.27^{+0.19}_{-0.22}$ & 23 & 54 &\\
		16276 & 90.96(0.01) & 25.59(0.01) & $\text{UBC 435}^b$ & 90.70(0.11) & 25.52(0.06) & 14.41 & 14.20 & 0.26(0.02) & -1.40(0.01) & -0.08(0.22) & -1.67(0.16) & 0.20(0.02) & 0.29(0.04) & $8.58^{+0.10}_{-0.10}$ & 25 & 55 &\\
		27415 & 60.95(0.02) & 54.74(0.01) & $\text{HSC 1181}^a$ & 59.88(0.57) & 55.41(0.27) & 54.62 & 50.15 & 0.02(0.03) & -0.26(0.02) & -0.01(0.10) & -0.04(0.09) & 0.27(0.02) & 0.23(0.04) & $7.71^{+0.20}_{-0.23}$ & 491 & 56 &\\
		35768 & 97.82(0.05) & 12.37(0.05) & $\text{HSC 1593}^a$ & 98.98(0.27) & 11.82(0.48) & 76.05 & 66.34 & -0.22(0.07) & -0.48(0.05) & -0.04(0.12) & -0.31(0.14) & 0.16(0.06) & 0.20(0.03) & $8.50^{+0.19}_{-0.19}$ & 394 & 57 &\\
		32025 & 66.42(0.02) & 48.43(0.01) & $\text{Theia 3318}^a$ & 65.61(0.50) & 49.59(0.38) & 76.49 & 59.51 & -0.43(0.02) & -0.79(0.02) & -0.65(0.19) & -0.79(0.09) & 0.29(0.02) & 0.31(0.04) & $7.72^{+0.22}_{-0.22}$ & 215 & 58 &\\
		\enddata
	\end{deluxetable*} 
	\vspace{-30pt}
	\begin{flushleft}
	\begin{nolinenumbers}
		\tablecomments{The columns show the parameters of 58 OC–Mgiant pairs identiﬁed in this work, including the identification number of each M-giant star (MG), astrometric parameters (i.e., positions ($\alpha$ and $\delta$), proper motions (${\mu}_{\alpha^*}$ and ${\mu}_\delta$) and parallaxes ($\varpi$)) of each M-giant star and OC, angular distance between the M-giant star and its host OC (Sep), apparent angular diameter of the OC (AD), age of the OC (log(age)), and the number of member stars in the OC (N). An OC name marked with ($\star$) indicates that the OC is older than 1 Gyr. Values in parentheses for M giants are uncertainties, while those for OCs are standard deviations.\\
		Reference:\\
		(a) Parameters of OCs sourced from \cite{Hunt_2024}.\\
		(b) Parameters of OCs sourced from \cite{Cantat-Gaudin_2020} but recalculated based on Gaia DR3.\\
		(c) Parameters of OCs sourced from \cite{Castro-Ginard_2022} but recalculated based on Gaia DR3.\\
		(d) Parameters of OCs sourced from \cite{Hao_2022_1} but recalculated based on Gaia DR3.\\		
		(e) Log(age) of Berkeley 29 determined as 9.54 yr by \cite{Tosi_2004}.\\
		(f) Log(age) of Berkeley 29 determined as 9.57 yr by \cite{Perren_2022}.\\
		(g) Log(age) of Berkeley 29 determined as 9.64 yr by \cite{Salaris_2004}.\\}
	\end{nolinenumbers}
	\end{flushleft}
\end{longrotatetable}
	
	\vspace{30\baselineskip}
	
	\section{Color-Magnitude Relation of M giants} \label{sec4}
	
	\subsection{Previous Results of Color–Magnitude Relations}
	\label{1}
	There are two main types of CMRs for M giants.~One is a linear relation, i.e., $M_{K_s}=A+B(J-K_s)$.~\cite{Sharma_2010} fitted this relation using a range of stellar models with ages and metallicities consistent with simulated stellar halos \citep{Bullock_2005}.~The relevant parameters can be found in rows (3)--(5) of Table~\ref{table2}, and these relations are shown as gray lines in Figure~\ref{fig4}(a).~The other type is a power-law relation that does not rely on stellar models, i.e., $M_J=A_1\left[(J-K)_0^{A_2}-1\right]+A_3$.~\cite{Li_2016_2} fitted this relation using the distances of the LMC, SMC and Sgr core region as standards to estimate $M_J$.~The relevant parameters are presented in rows (7)--(9) of Table~\ref{table2}.~However, these relations are not applicable to the Milky Way stars, so \cite{Li_2023} refitted the relation for Milky Way stars using the distances of some well-calibrated M giants as the standard to estimate $M_J$. The relevant parameters can be found in row (10) of Table~\ref{table2}, and this relation is shown in Figure~\ref{fig4}(b).
	
	\begin{deluxetable*}{ccccc}[h]
	\tablenum{2}
	\tablecaption{The CMR Fitting of M giants}
	\label{table2}
	\tablewidth{\textwidth} 
	\tablehead{
		\colhead{Color-Magnitude Relation} & \colhead{Different Work} & \colhead{Fitting Sample} &
		\colhead{$A$,~~~$B$} & \colhead{Number} 
	}
	
	\startdata
	$M_{K_s}=A+B(J-K_s)$ & This work & Full sample \textsuperscript{[1]} & -1.21,~~~-3.40 & (1) \\
	& & Subsample \textsuperscript{[2]} & 3.77,~~~-8.14 & (2) \\
	& \cite{Sharma_2010} & Based on stellar models & 2.16,~~~-9.42 & (3) \\
	& & & 3.26,~~~-9.42 & (4) \\
	& & & 4.36,~~~-9.42 & (5) \\
	%\hline
	… & … & … & $A_1$,~~~$A_2$,~~~$A_3$ & … \\
	%\hline
	$M_J=A_1\left[(J-K)_0^{A_2}-1\right]+A_3$ & This work & Full sample \textsuperscript{[1]} & cannot be fitted & … \\
	& & Subsample \textsuperscript{[2]} & ~~~3.09,~~~-2.39,~~~-3.58~~~ & (6) \\
	& ~~~\cite{Li_2016_2}~~~ & M giants in Sgr core & 3.12,~~~-2.6,~~~~-4.61 & (7)\\
	&  & M giants in LMC & 3.12,~~~-2.6,~~~~-4.78 & (8)\\
	&  & M giants in SMC & 3.12,~~~-2.6,~~~~-5.08 & (9)\\
	& ~~~\cite{Li_2023}~~~ & Conditioned M giants \textsuperscript{[3]}  & ~~6.07,~~~-1.13,~~~-3.34~ & (10) \\
	\enddata
	\hspace{1.5cm}
	\begin{minipage}{\linewidth}
		\tablecomments{\textsuperscript{1}All OC-M giant pairs (58 pairs) identified in this work.\\
			\textsuperscript{2}Seventeen OC-M giant pairs selected from the full sample, with the ages of OCs all over 1 Gyr.\\
			\textsuperscript{3}M giants with Gaia EDR3 distances~\citep{Bailer_2021} closer than 4 kpc and parallax/parallax errors $>$ 5.}
	\end{minipage}
	\end{deluxetable*}
	
	\subsection{The Fitted Color-Magnitude Relations}
\begin{figure}[h]
		\centering  
		\subfigure[]{
			\includegraphics[width=0.49\textwidth]{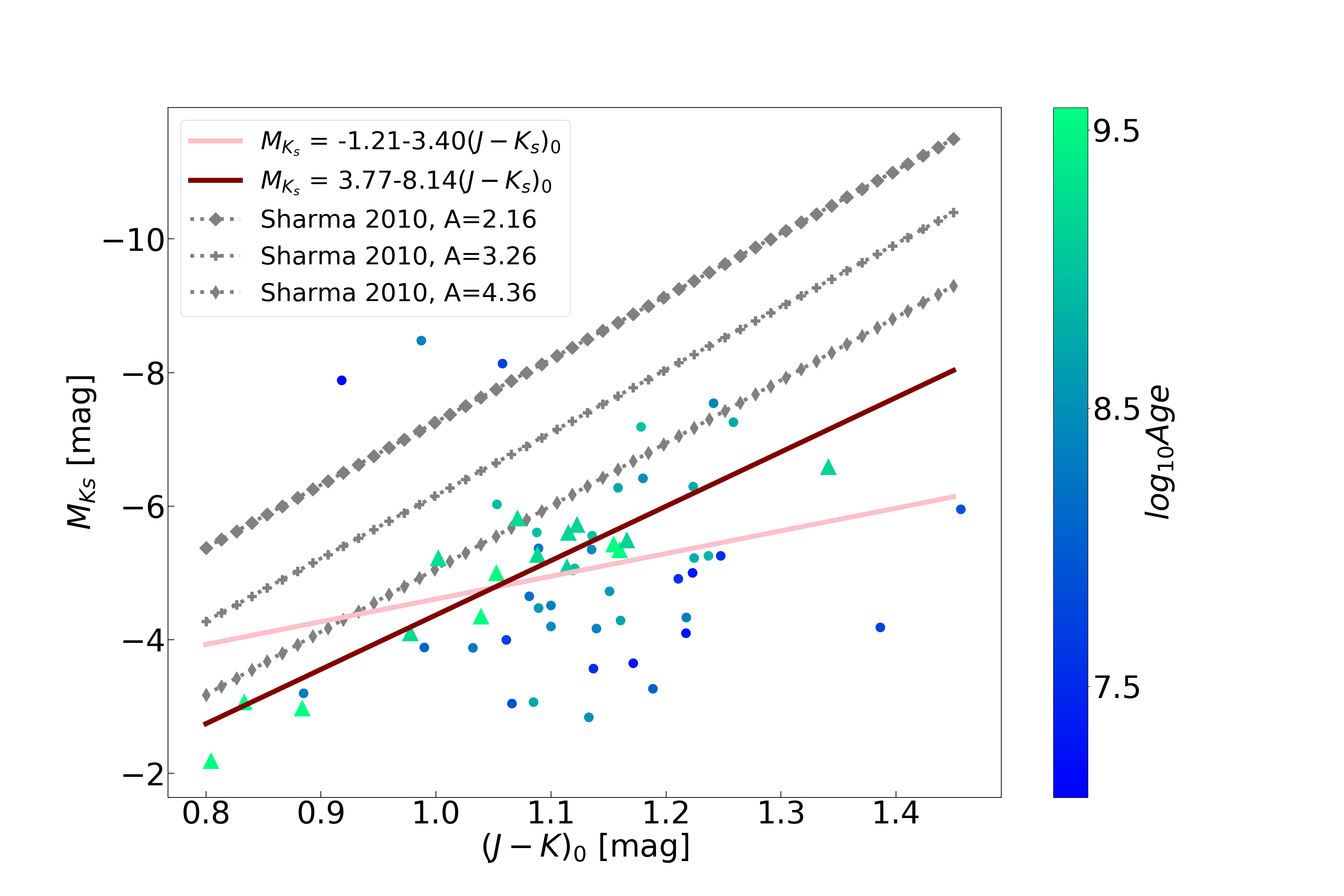}}
		\subfigure[]{
			\includegraphics[width=0.49\textwidth]{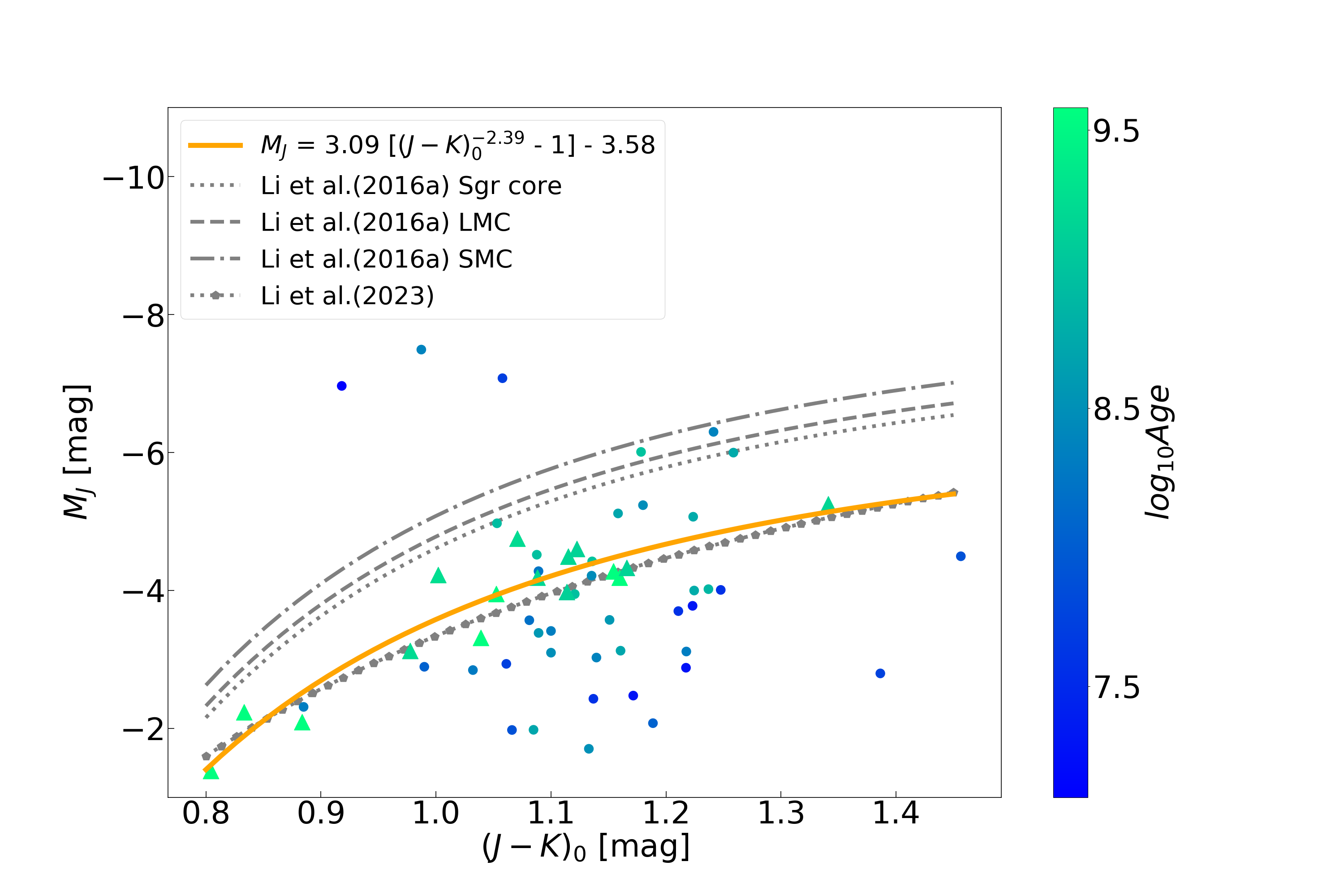}}
		\caption{Comparison of our color-absolute magnitude distribution to various literature sources, as presented in rows (1)--(5) (panel (a)) and (6)--(10) (panel (b)) of Table~\ref{table2}. The dots and triangles represent the distributions of all identified M giants. The ages of host OCs are color-coded, with triangles indicating those over 1 Gyr.}
		\label{fig4}
	\end{figure}

	We fitted the two types of CMRs of M giants using the full sample and a subsample, respectively. The full sample consists of all 58 OC-M giant pairs we identified, while the subsample consists of 17 OC-M giant pairs selected from the full sample, with the ages of these OCs all over 1 Gyr.~The parameters of these fittings are presented in rows (1), (2), and (6) of Table~\ref{table2}, with the corresponding relation distributions shown in Figure~\ref{fig4}. As depicted in Figure~\ref{fig4}, the relation in row (2) of Table~\ref{table2}, fitted using the subsample, exhibits stronger consistency with the relations in rows (3)--(5). Similarly, the relation in row (6), also ﬁtted using the subsample, is strongly correlated with relations in rows (7)--(10). These suggest that the subsample is better suited for calibrating the CMR of M giants.
	
\begin{figure}[h]
		\centering  
		\subfigure{
			\includegraphics[width=\textwidth]{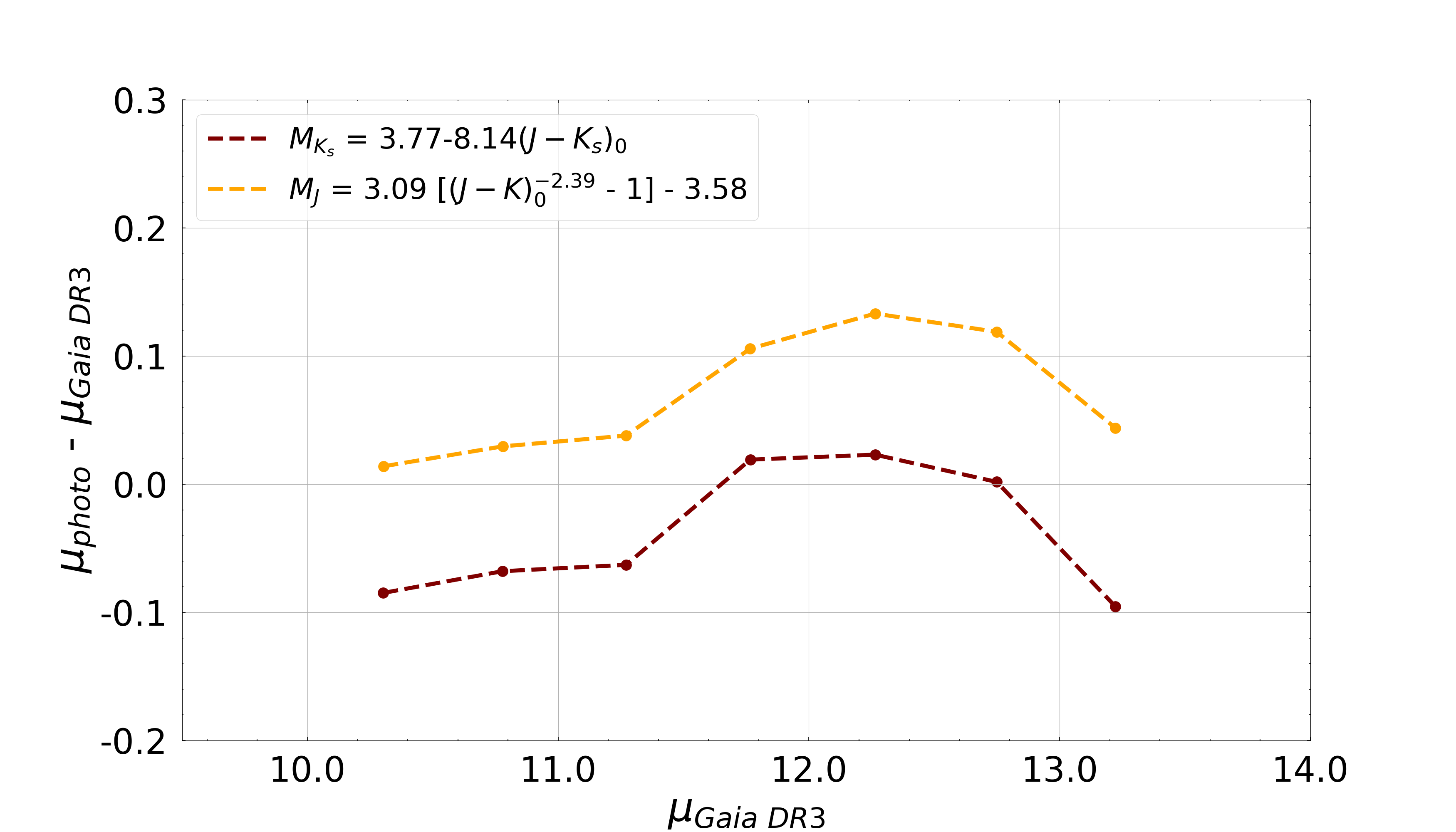}}
		\caption{Distance modulus comparison between the relations presented in rows (2) (brown) and (6) (yellow) of Table~\ref{table2} vs. the parallax distances from Gaia DR3, segmented based on the latter distance modulus.~The dots represent the median of each bin.}
		\label{fig5}
	\end{figure}

	To verify the feasibility of the relations presented in rows (2) and (6) of Table~\ref{table2}, we compared the distance moduli of M giants derived from these relations with those obtained from Gaia DR3 parallaxes.~Considering that the distances of OCs we used to fit the CMRs are almost all within 5 kpc, we focused on the M giants with parallax distances from Gaia closer than 5 kpc and parallax/parallax errors $>$ 10, totaling 19,340 M giants.~As shown in Figure~\ref{fig5}, there are systematic offsets between these two types of distance moduli.~We calculated the overall offset by summing the absolute differences of these two distance moduli, segmented based on the distance modulus derived from Gaia parallaxes. The result is presented in column (8) of Table~\ref{table3a}. It can also be seen from Figure~\ref{fig5} that the overall offset of the brown line is smaller than that of the yellow line, indicating the overall offset derived from the linear relation is smaller than that obtained from the empirical distance relation.
	
	Considering that the $J_0$ and $K_0$ values we used to fit the CMRs in Table~\ref{table2} were dereddened using the 3D dust map from~\cite{Green_2019}, which has a median uncertainty of around 30\%, the systematic offsest in Figure~\ref{fig5} may be due to the reddening uncertainty.~Therefore, for all M giants, we added errors ranging from -30\% to 30\% in steps of 5\% to the reddening values provided by \cite{Green_2019}, and refitted the relations presented in rows (2) and (6) of Table~\ref{table2}.~The overall offsets of the two types of distance moduli are presented in Table~\ref{table3a}.~As shown in Table~\ref{table3a}, when we fit these two types of CMRs using the subsample, the overall offset derived from the linear relation is almost less than that obtained from the empirical distance relation.~This suggests that the linear relation is superior to the empirical distance relation in characterizing the CMR of M giants.
	
\begin{figure}[h]
	\centering  
	\subfigure{
	\includegraphics[width=\textwidth]{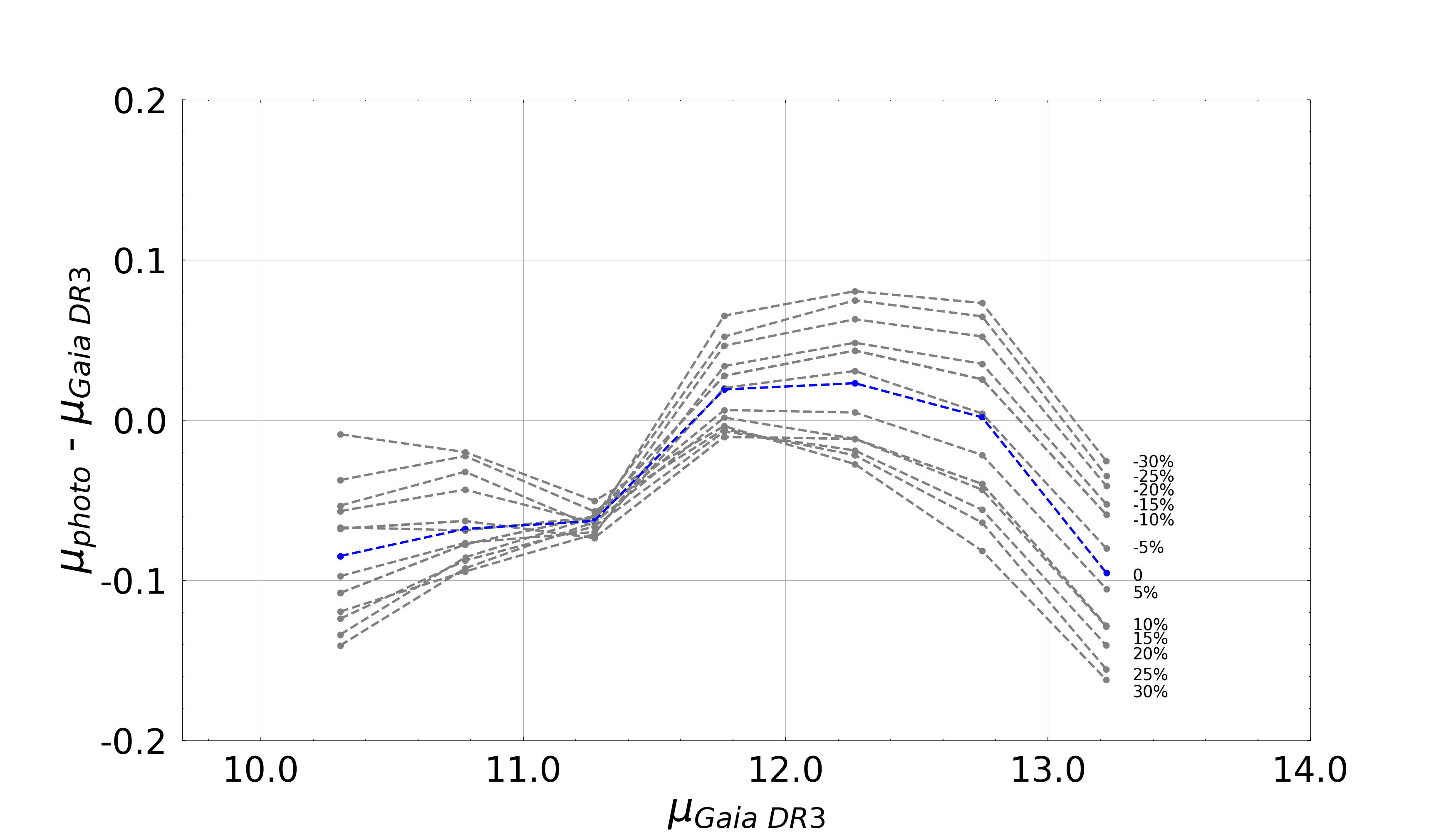}}
	\caption{Distance modulus comparison between the linear relation fitted using the subsample vs. the parallax distances from Gaia DR3, with the reddening errors added ranging from -30\% to 30\% in steps of 5\%. The blue line depicts the comparison between the two types of distance moduli without any reddening correction. The dots represent the median of each bin. }
	\label{fig6}
\end{figure}

	In addition, as shown in Table~\ref{table3a}, the overall offset of these two distance moduli is minimized when we fit the linear relation using a subsample of 17 OC-M giant pairs selected from the full sample, with a 5\% error added to the reddening values of M giants.~Figure~\ref{fig6} shows the comparison of these two types of distance moduli when the subsample is used to fit the linear relation, with different errors added to the reddening values of M giants.~The parameters $A$ = 3.85~$\pm$~0.02
		and $B$ = $-$8.26~$\pm$~0.03 are adopted in this work, and the errors of these two parameters are from the OC distance errors.
		
   	We assessed the reliability of the 17 OC-M giant pairs in the subsample.~In the all sample, M giants in 27 pairs are identified as OC member stars, while the remaining 31 pairs are identified using 5D astrometric parameters.~In the subsample, all 17 OC-M giant pairs consist of M giants that were identified as OC member stars.~Therefore, the pairs in the subsample demonstrate a high degree of reliability.~Further investigation is warranted for other identified OC-M giant pairs, particularly those identified via 5D astrometric parameters, e.g., the inaccuracies in individual M-giant distance measurements and the uncertainties in OC age estimates.
   	
\begin{deluxetable*}{cccccccccccccc}
	\tablenum{3}
	\tablecaption{Overall Offset of Distance Moduli Derived from CMRs Based on Subsample and Parallax Distances}
	\label{table3a}
	\tablewidth{\textwidth} 
	\tablehead{
		\colhead{Color-magnitude relation} &
		\multicolumn{12}{c}{~~Overall Offset of the Two Distance Moduli when Adding Different Reddening Errors (mag)} \\
		\colhead{ } &
		\colhead{-30\%} & \colhead{-25\%} & \colhead{-20\%} & \colhead{-15\%} & \colhead{-10\%} & \colhead{-5\%} &
		\colhead{0\%} & \colhead{5\%} & \colhead{10\%} & \colhead{15\%} & \colhead{20\%} & \colhead{25\%} & \colhead{30\%}
	}
	\decimalcolnumbers
	\startdata
	$M_{K_s}=A+B(J-K_s)$ & 0.54 & 0.51 & 0.48 & 0.46 & 0.40 & 0.38 & 0.36 & 0.33 & 0.39 & 0.35 & 0.37 & 0.36 & 0.35 \\
	$M_J=A_1\left[(J-K)_0^{A_2}-1\right]+A_3$ &  0.93 & 0.85 & 0.74 & 0.67 & 0.64 & 0.51 & 0.48 & 0.43 & 0.38 & 0.38 & 0.43 & … & … \\
	\enddata
	\tablecomments{The subsample consists of 17 OC-M giant pairs selected from the full sample, with the ages of OCs all over 1 Gyr. The empirical distance relation cannot be fitted with a 25\% and 30\% error added to the reddening values of M giants.}
\end{deluxetable*}

	\clearpage
	\subsection{Veriﬁcation of the Color–Magnitude Relations in This Work}
\begin{figure}[h]
		\centering 
		\subfigure{
			\includegraphics[width=0.49\textwidth]{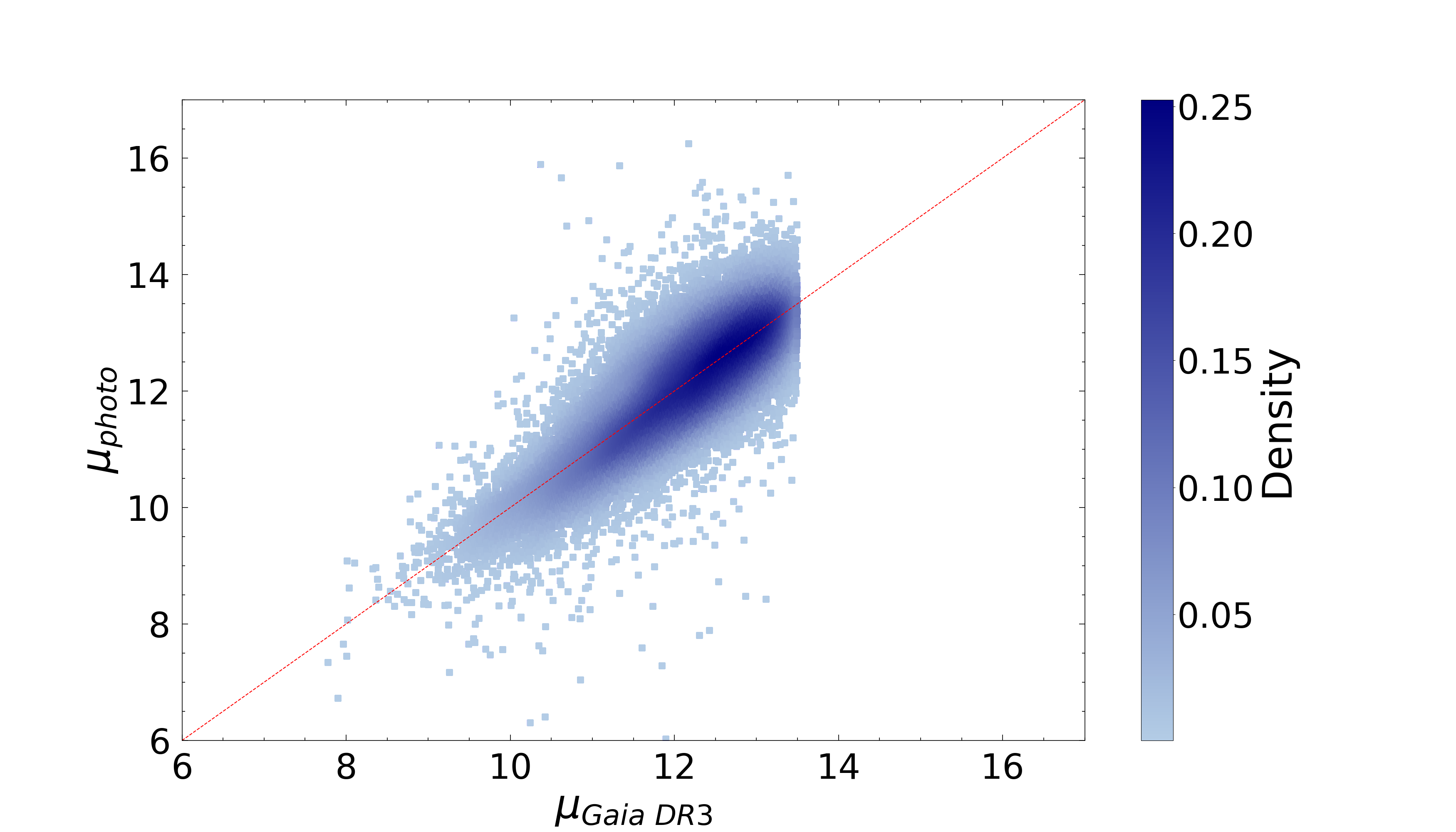}}
		\subfigure{
			\includegraphics[width=0.49\textwidth]{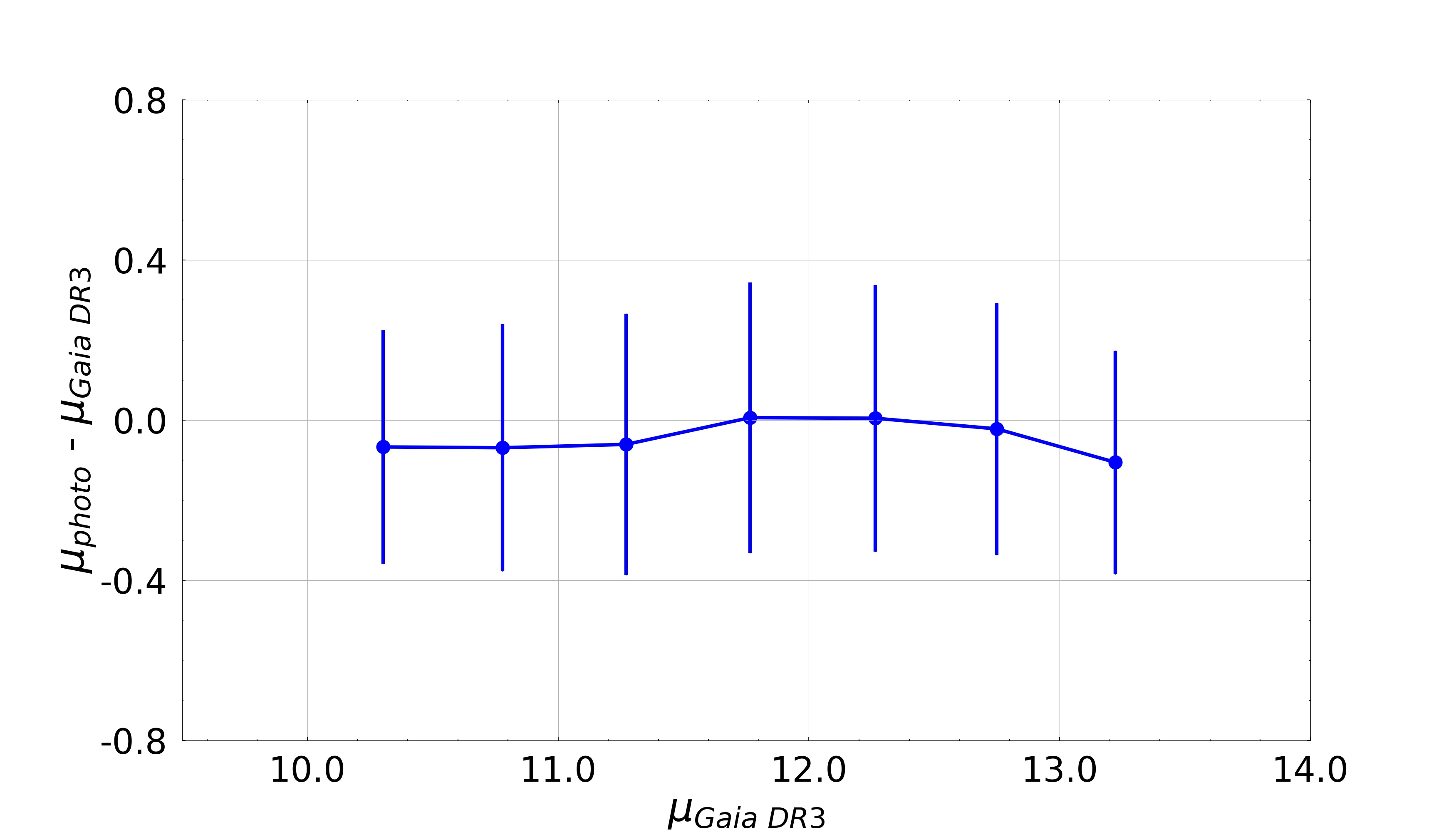}}\\
		\centering  
		\subfigure{
			\includegraphics[width=0.49\textwidth]{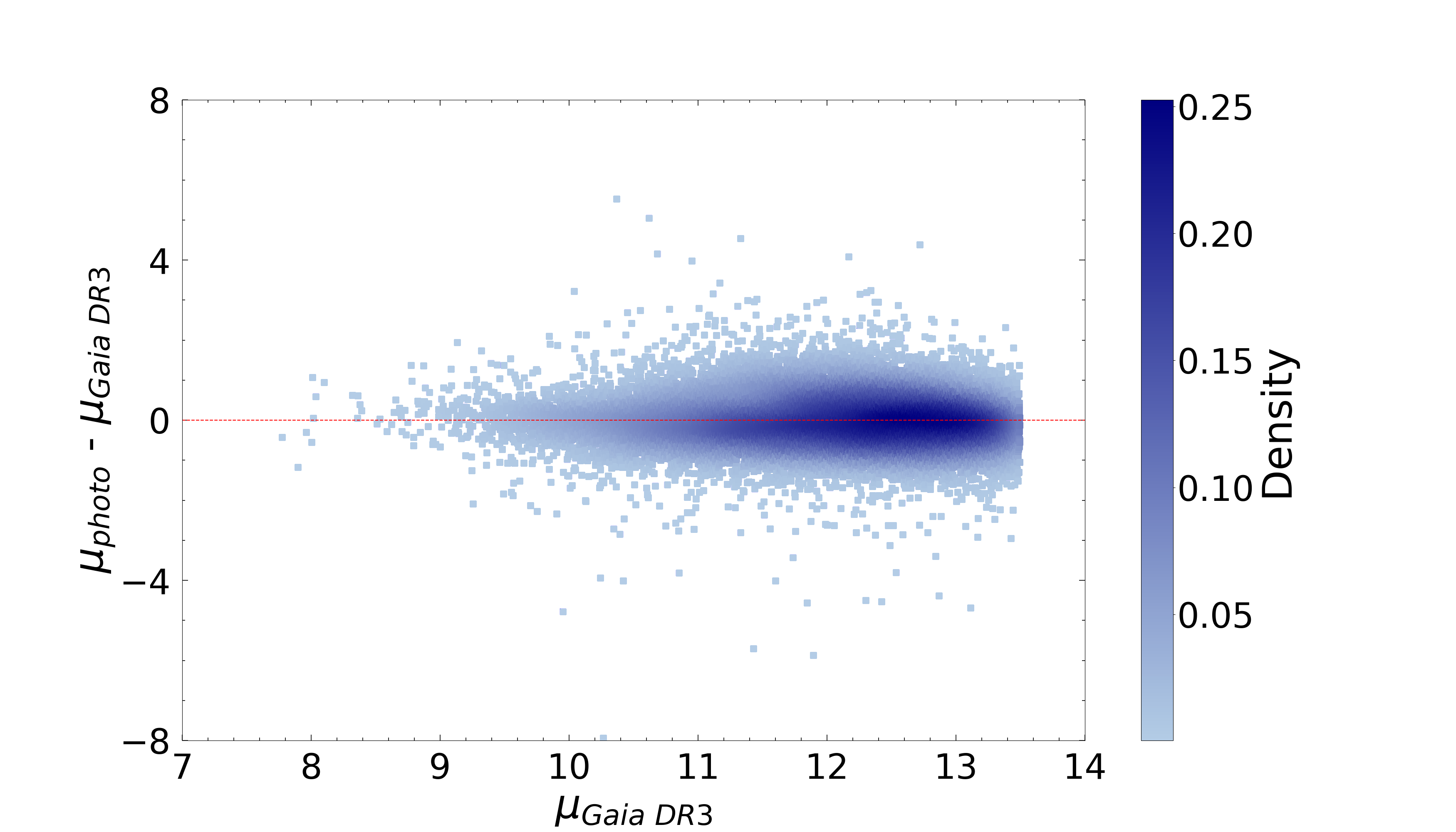}}
		\subfigure{
			\includegraphics[width=0.49\textwidth]{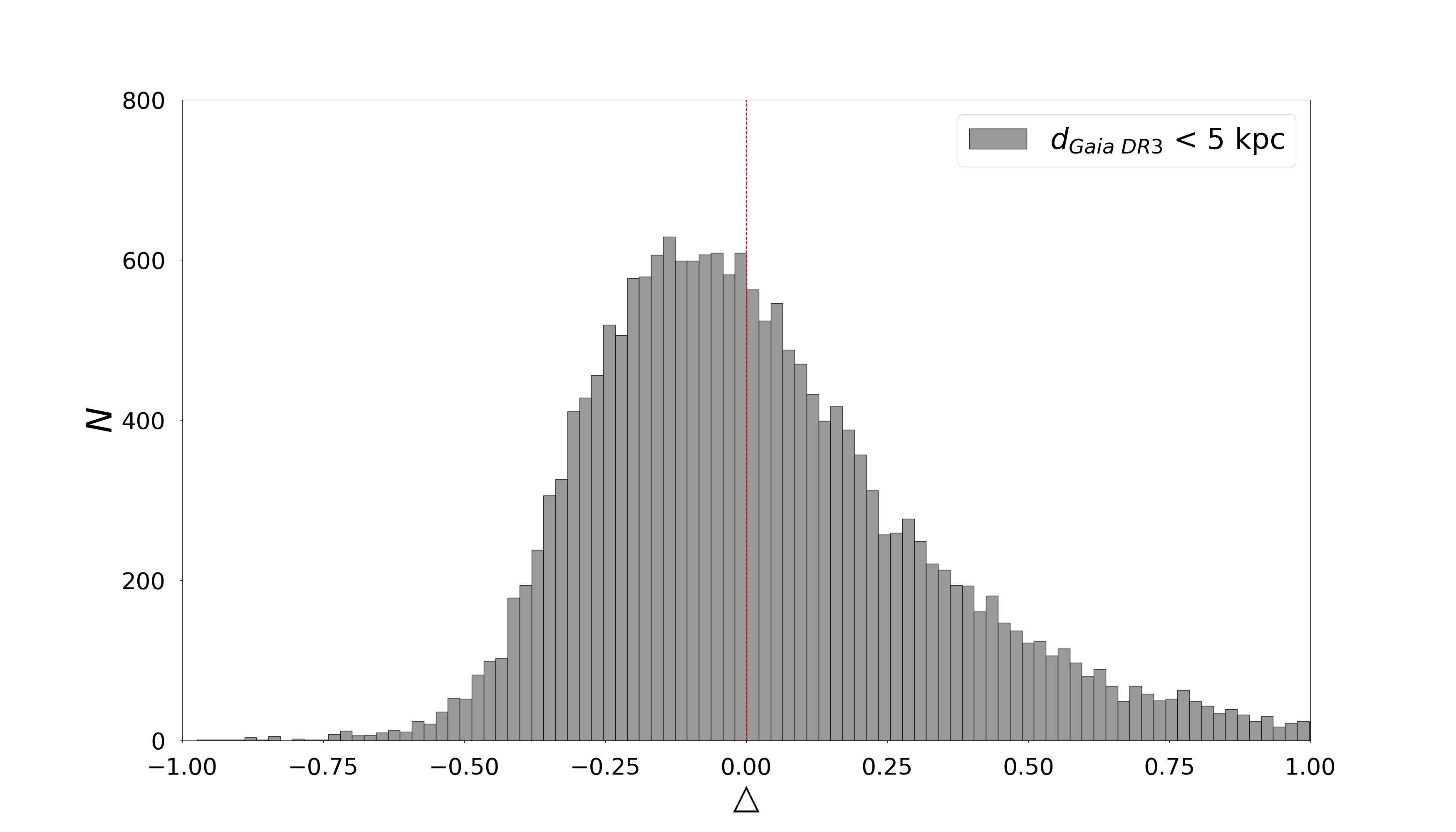}}
		\caption{Left and top-right panels: distance modulus comparison between the photometric and parallax distances, segmented based on the latter's distance modulus. The blue dots and error bars in the top-right panel represent the median and standard deviation of each bin. Bottom-right panel: the histogram distribution of the relative difference $\Delta = (d_{\text{photo}} - d_{\text{Gaia\ DR3}})/d_{\text{Gaia\ DR3}}$. The histogram is divided into 600 bins, with each bin having a binsize of approximately 0.02.}
		\label{fig7}
	\end{figure}

	To verify the accuracy of the CMR we obtained, i.e., $M_{K_s}=3.85-8.26(J-K_s)$, we compared the distance moduli derived from the CMR with those obtained from Gaia DR3 parallaxes, as shown in Figure~\ref{fig7}.~The comparison between the photometric distances of M giants from the CMR and the parallax distances from Gaia DR3 shows a median deviation of 1.5\%, with the 16th and 84th percentile values of 25.3\% and 33.6\%, respectively. This represents the high accuracy and feasibility of the CMR.
	
\begin{figure}[htp]
		\centering  
		\subfigure{
			\includegraphics[width=0.49\textwidth]{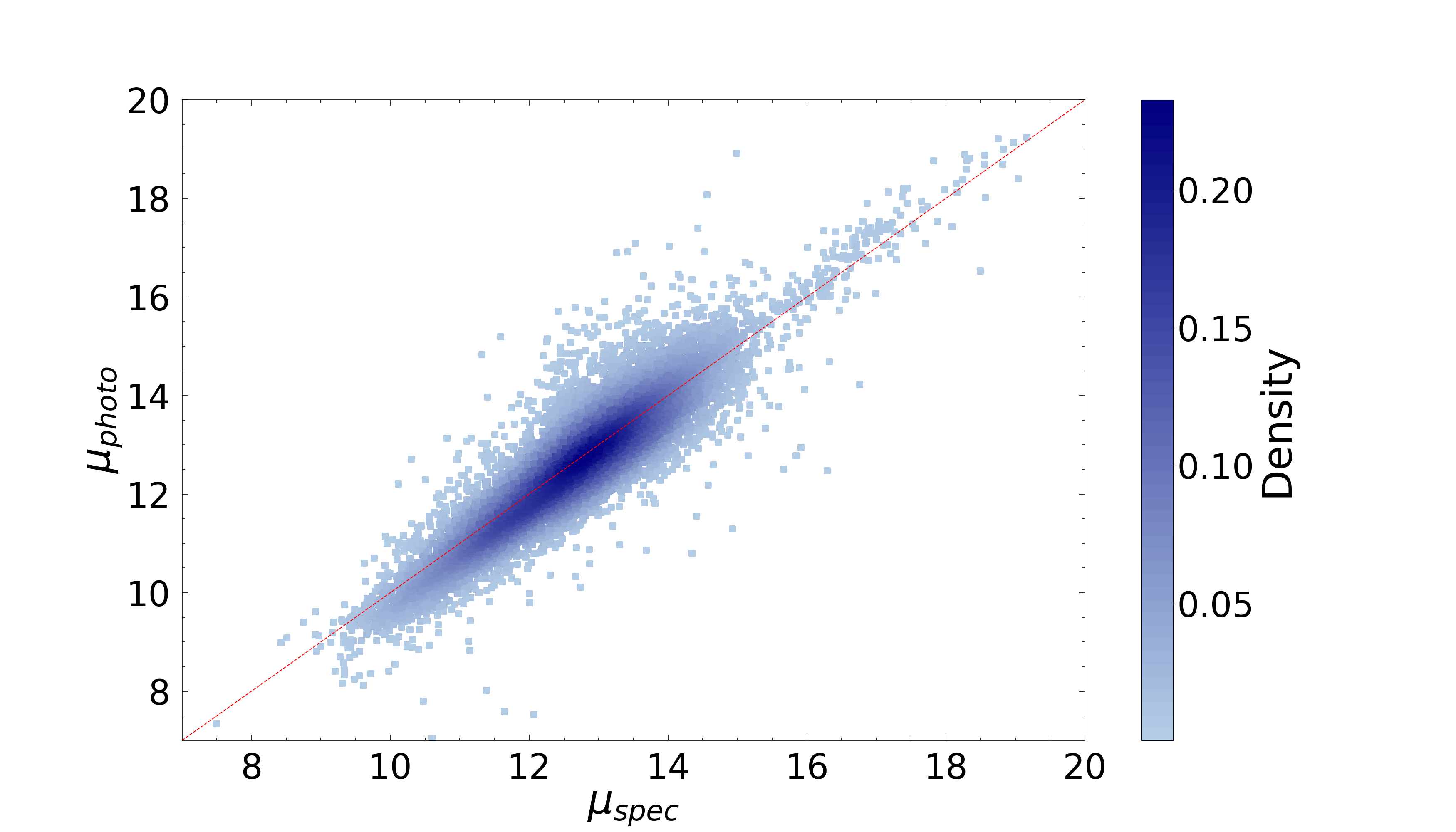}}
		\subfigure{
			\includegraphics[width=0.49\textwidth]{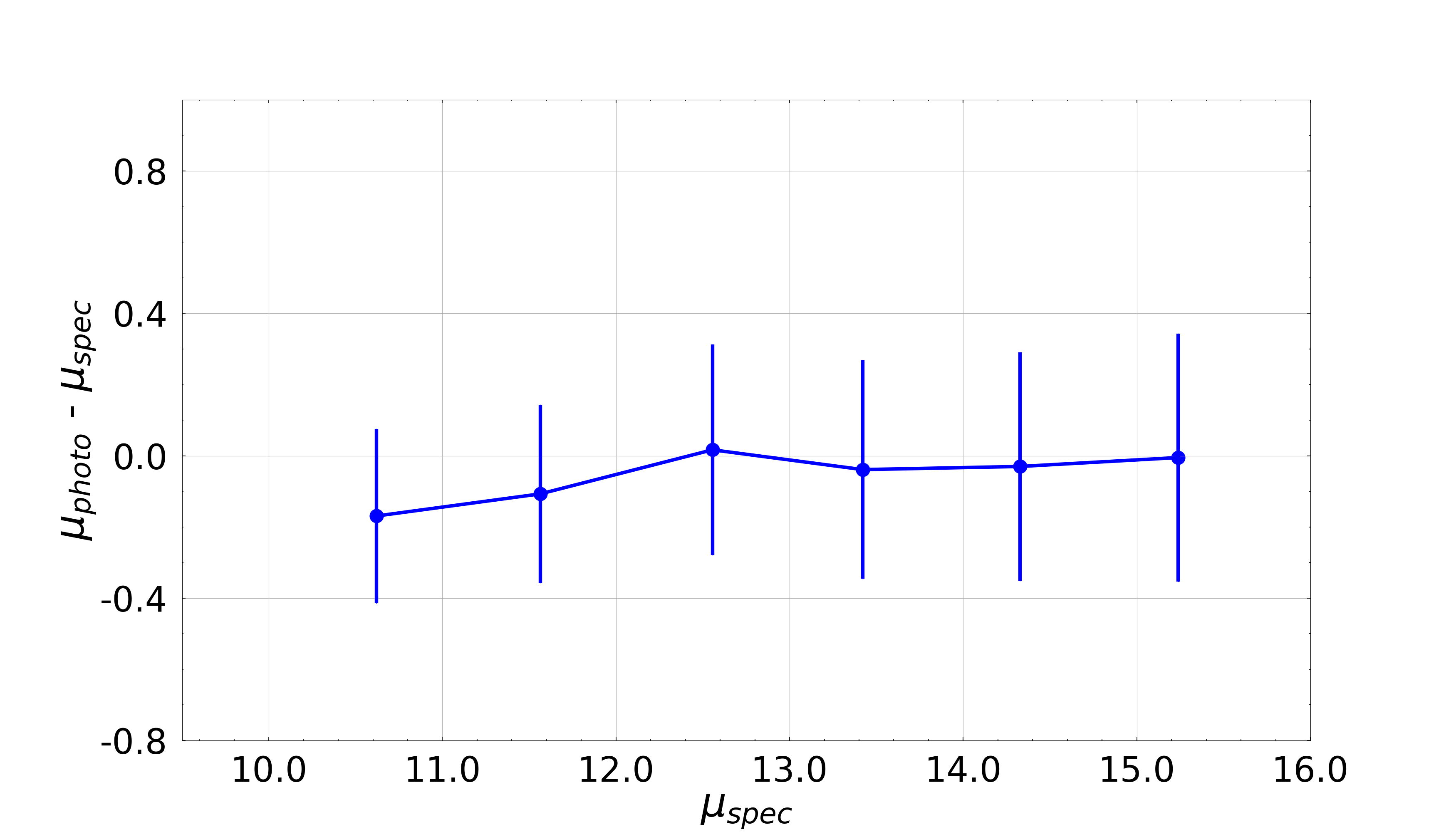}}\\
		\centering  
		\subfigure{
			\includegraphics[width=0.49\textwidth]{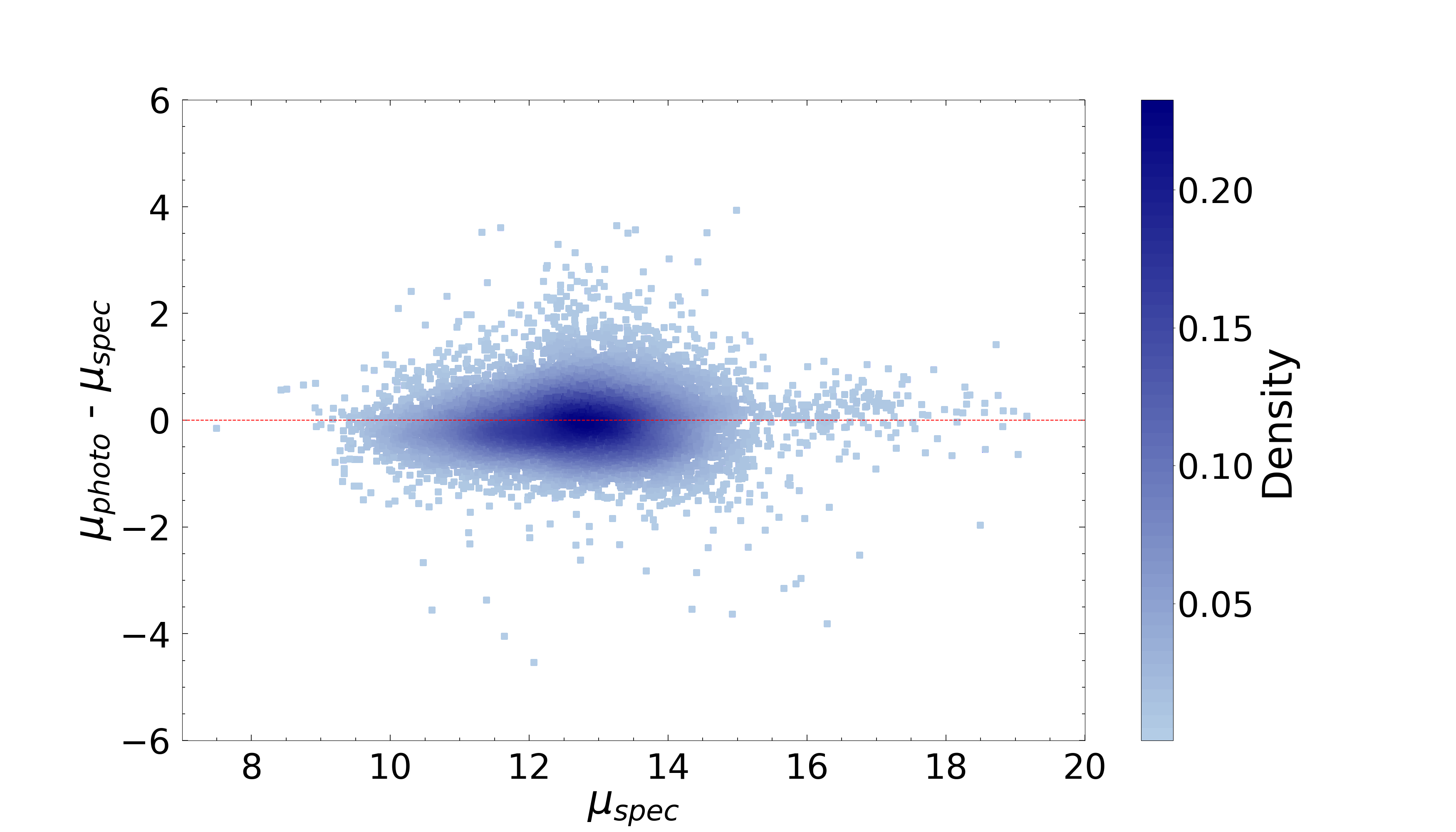}}
		\subfigure{
			\includegraphics[width=0.49\textwidth]{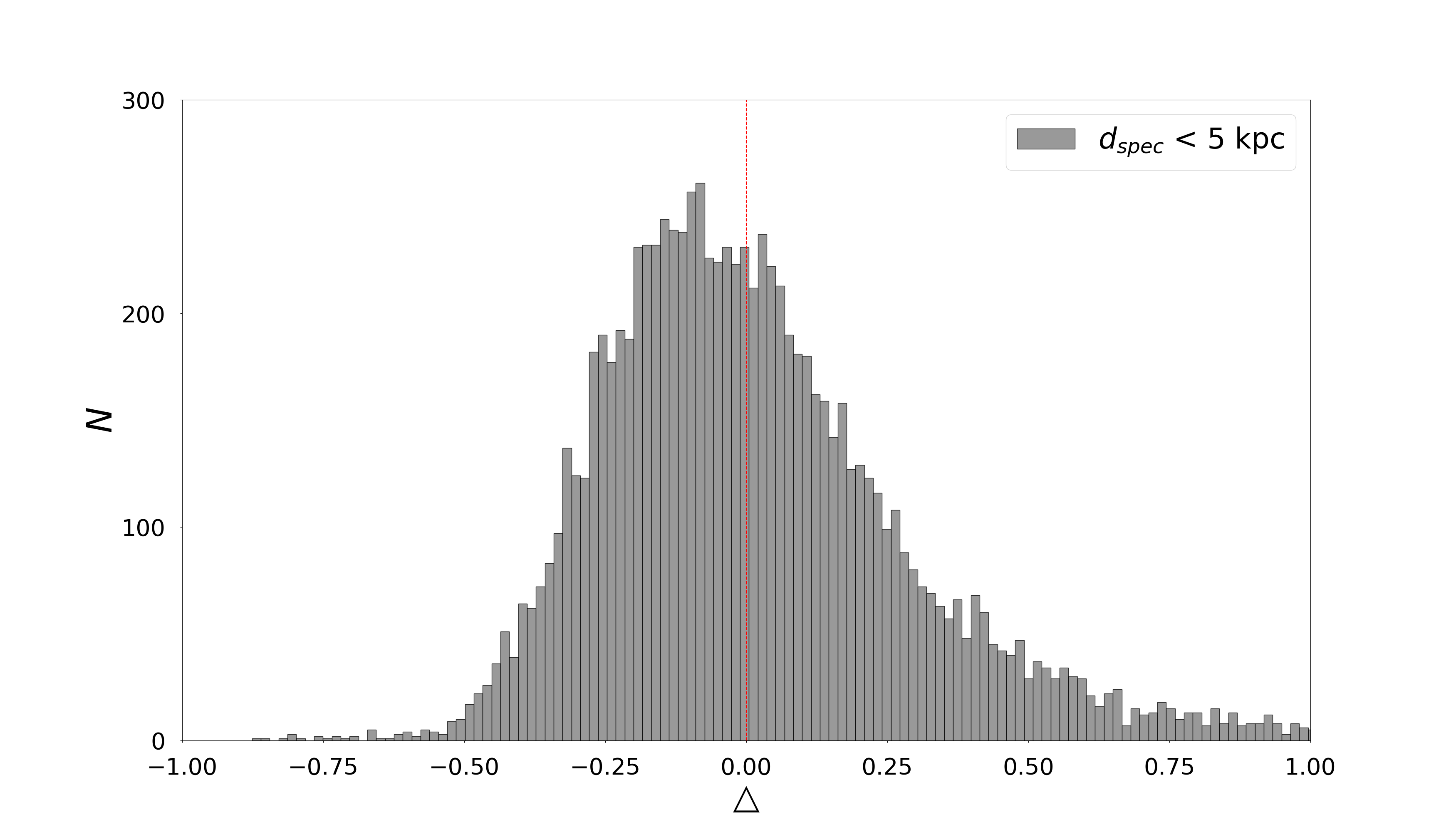}}
		\caption{Left and top-right panels:~distance modulus comparison between the photometric and spectroscopic distances, segmented based on the latter's distance modulus. The blue dots and error bars in the top-right panel represent the median and standard deviation of each bin. Bottom-right panel: the histogram distribution of the relative difference $\Delta = (d_{\text{photo}} - d_{\text{spec}})/d_{\text{spec}}$. The histogram is divided into 150 bins, with each bin having a binsize of approximately 0.02.}
		\label{fig8}
	\end{figure}

	In addition, we validated the applicability of the CMR obtained for M giants without parallax measurements. The sample consists of 9365 M giants selected by \cite{Li_2023}. We calculated their spectroscopic distances based on the relationship between the absolute magnitude~($M_J$) and the spectroscopic type ($SpTy$) following~\cite{Zhong_2015}. For these M giants, we compared the distance moduli derived from the CMR with those obtained from the spectroscopic distances, as shown in Figure~\ref{fig8}.~The comparison between the photometric distances of M giants from the CMR and the spectroscopic distances shows a median deviation of 2.3\%, with the 16th and 84th percentile values of 23.9\% and 26.7\%, respectively. This demonstrates that our derived CMR can accurately determine the distances of M giants that lack parallax measurements. Furthermore, it is expected that we can precisely calculate the distances of a larger number of M giants, better determine their parameters, and further unveil the structure of the Milky Way using this CMR.
	
	\vspace{5\baselineskip}
	
	\section{Application} \label{sec6}
	
\begin{figure}[h]
	\centering  
	\subfigure{
		\includegraphics[width=0.32\textwidth]{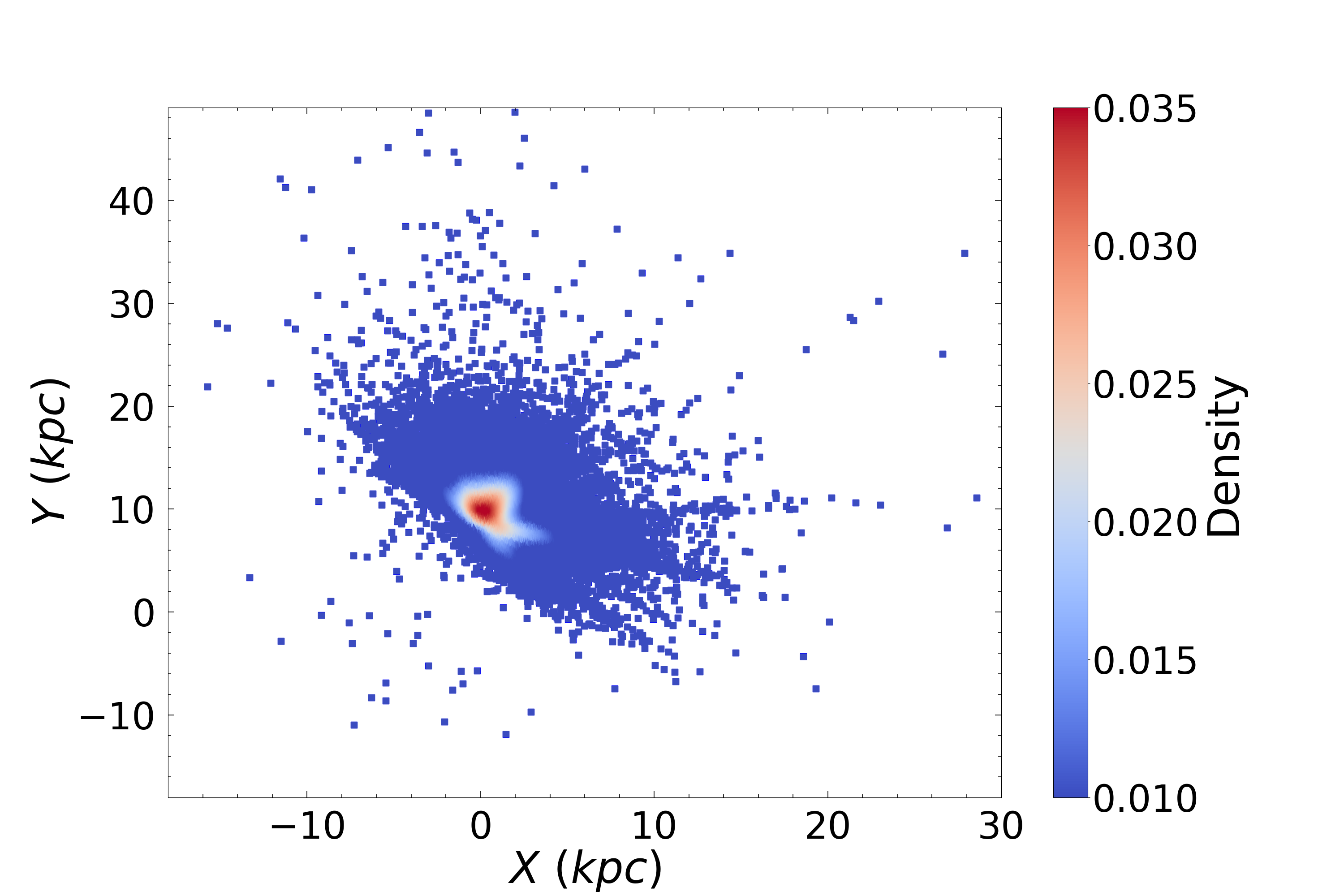}}
	\subfigure{
		\includegraphics[width=0.32\textwidth]{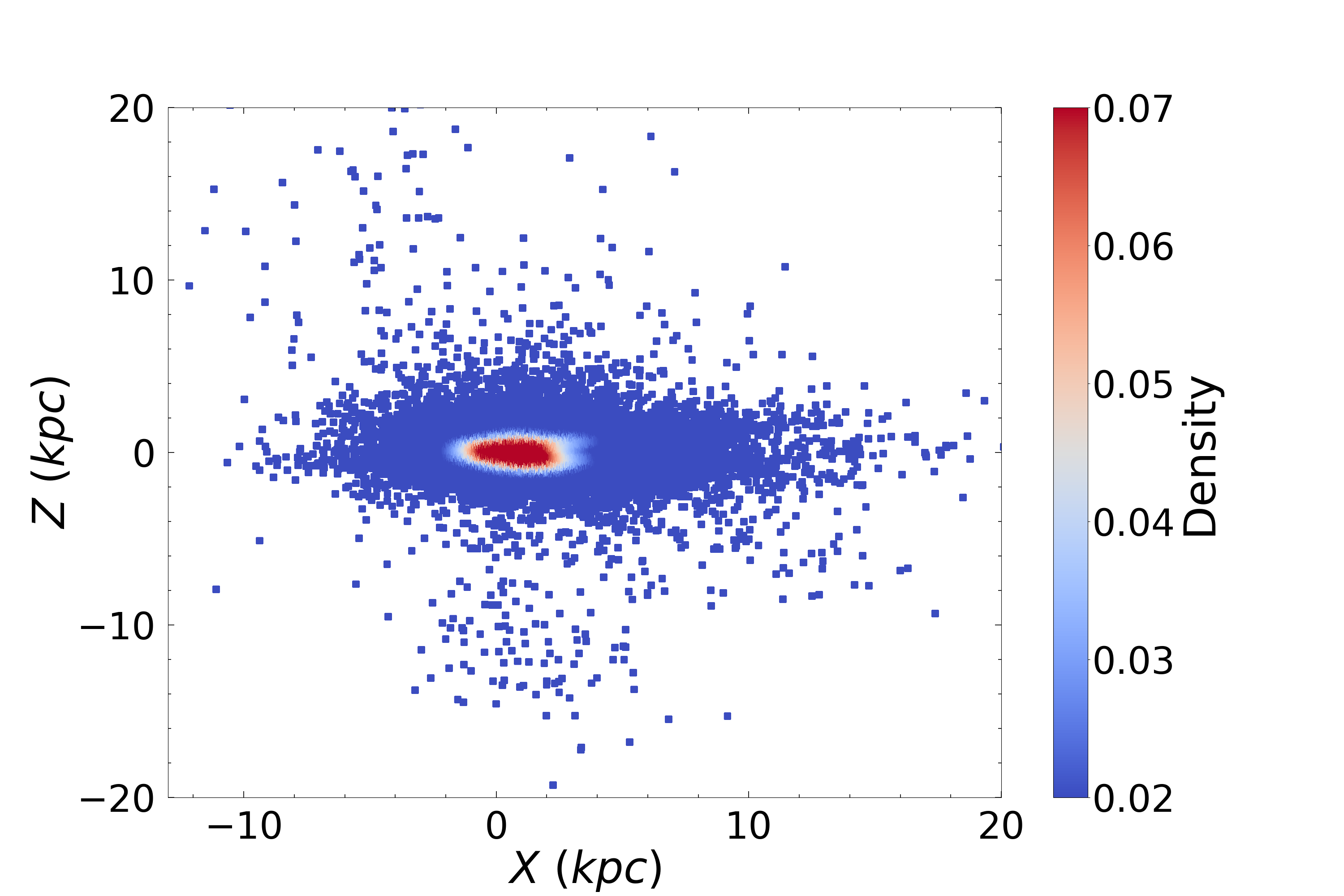}}
	\subfigure{
		\includegraphics[width=0.32\textwidth]{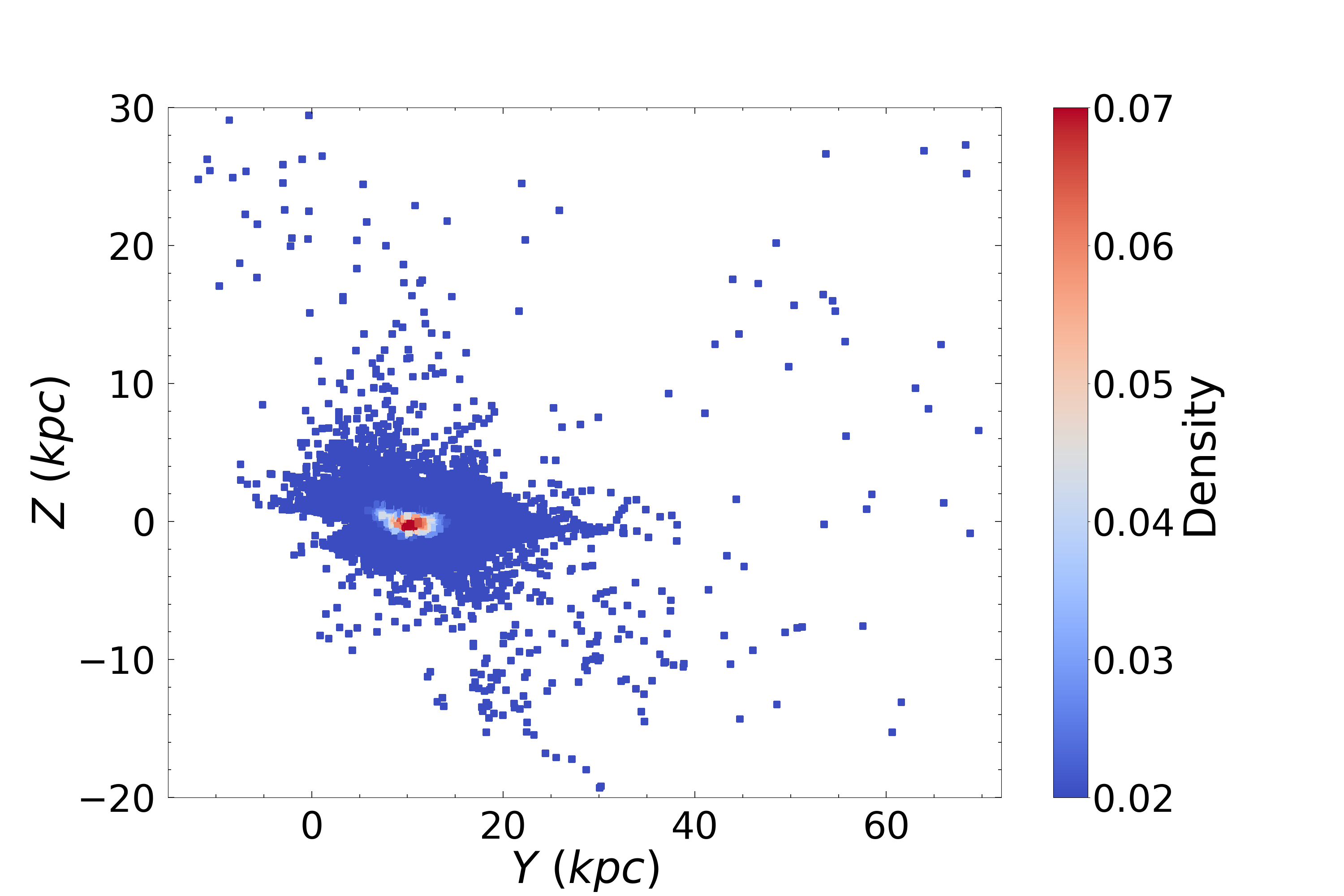}}
	\caption{The space distributions of M giants in different spaces, where the Sun is located at ($X, Y, Z$) = (0, 8.34, 0) kpc.}
	\label{fig10}
    \end{figure}

	We applied the CMR derived in this study to investigate the variations in the Galactic disk. Using the distances of 34,657 M giants obtained from the CMR, we calculated their complete six-dimensional (6D) information
	\footnote{ The Galactocentric distance of the Sun and the local circular velocity we adopted were assumed as $R_{\odot}$ = 8.34 kpc \citep{Reid_2014} and $V_{c,0} = 238~\mathrm{~km~s}^{-1}$~\citep[e.g.,][]{Reid_2004, Sch_2010, Sch_2012, Huang_2015, Bland_2016, Huang_2016, Zhou_2023}, respectively. The solar motions of $U_{\odot} = 13.00~\mathrm{~km~s}^{-1}$, $V_{\odot} = 12.24~\mathrm{~km~s}^{-1}$, and $W_{\odot} = 7.24~\mathrm{~km~s}^{-1}$ were assumed \citep{Sch_2018}, which are the velocity components toward the Galactic centre in the direction of Galactic rotation and toward the north Galactic pole, respectively.} 
	, including positions ($X, Y, Z$) and velocities ($U, V, W$). The spatial distribution of these M giants, shown in Figure~\ref{fig10}, indicates that most of them are situated within the Galactic disk.
	
	\cite{Sun_2024} analyzed the variations in the Galactic disk by examining the radial velocity ($V_R$), azimuthal velocity ($V_\phi$), and vertical velocity ($V_Z$) as functions of the projected Galactocentric distance ($R$) using red clump (RC) stars, predominantly located in the disk. Following \cite{Sun_2024}, we applied the sample selection criteria to our M-giant sample. We selected M giants with vertical velocities $|{V_Z}|\leqslant120\mathrm{~km~s}^{-1}$ and metallicity $[\mathrm{M}/\mathrm{H}]\geqslant-1.0\mathrm{~dex}$, yielding a total of 33,889 stars. We then derived their $V_R$, $V_\phi$, and $V_Z$ distributions to investigate the disk’s variations.
	
\begin{figure}[h]
    	\centering  
    	\subfigure{
    		\includegraphics[width=0.49\textwidth]{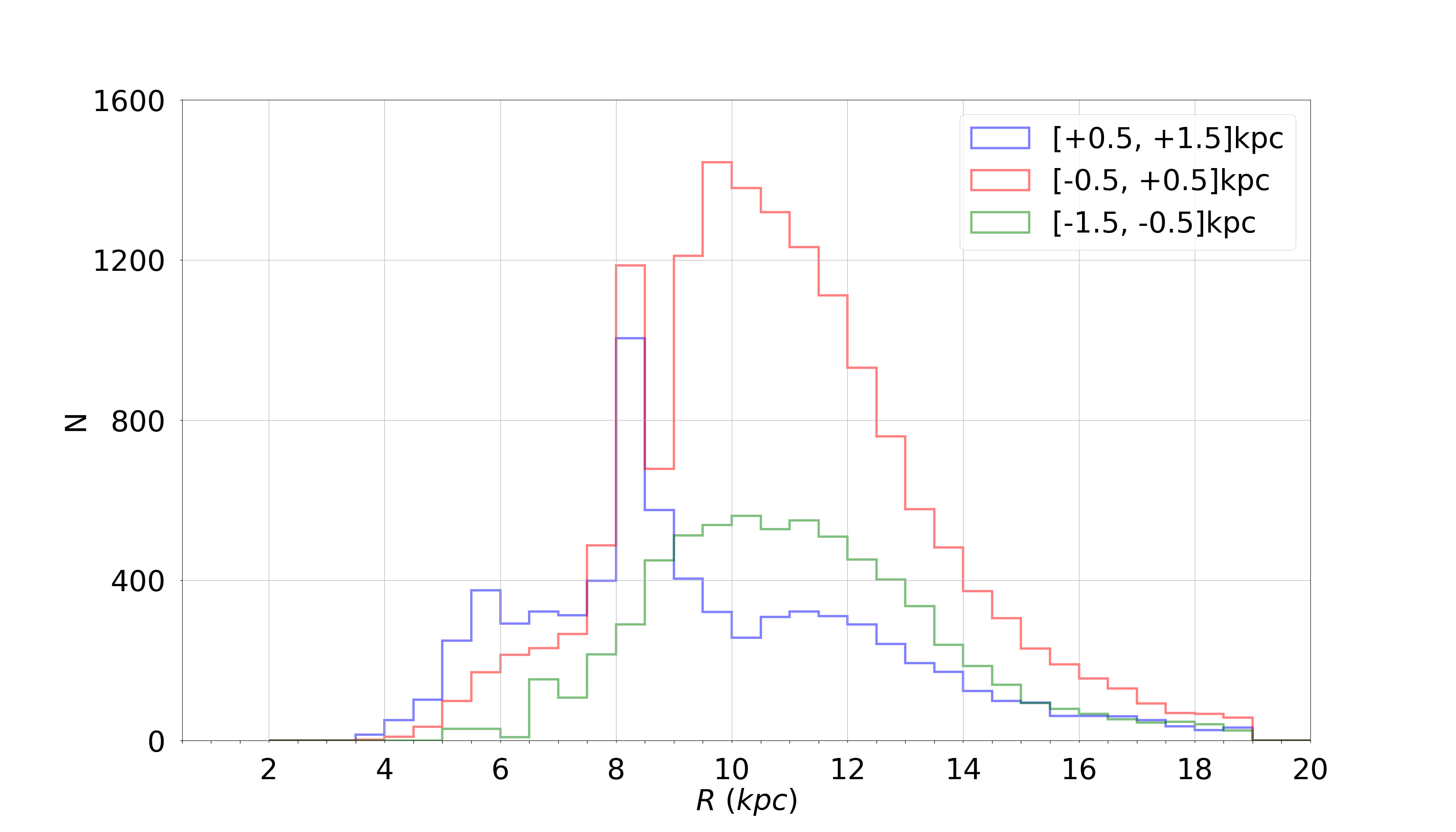}}
    	\subfigure{
    		\includegraphics[width=0.49\textwidth]{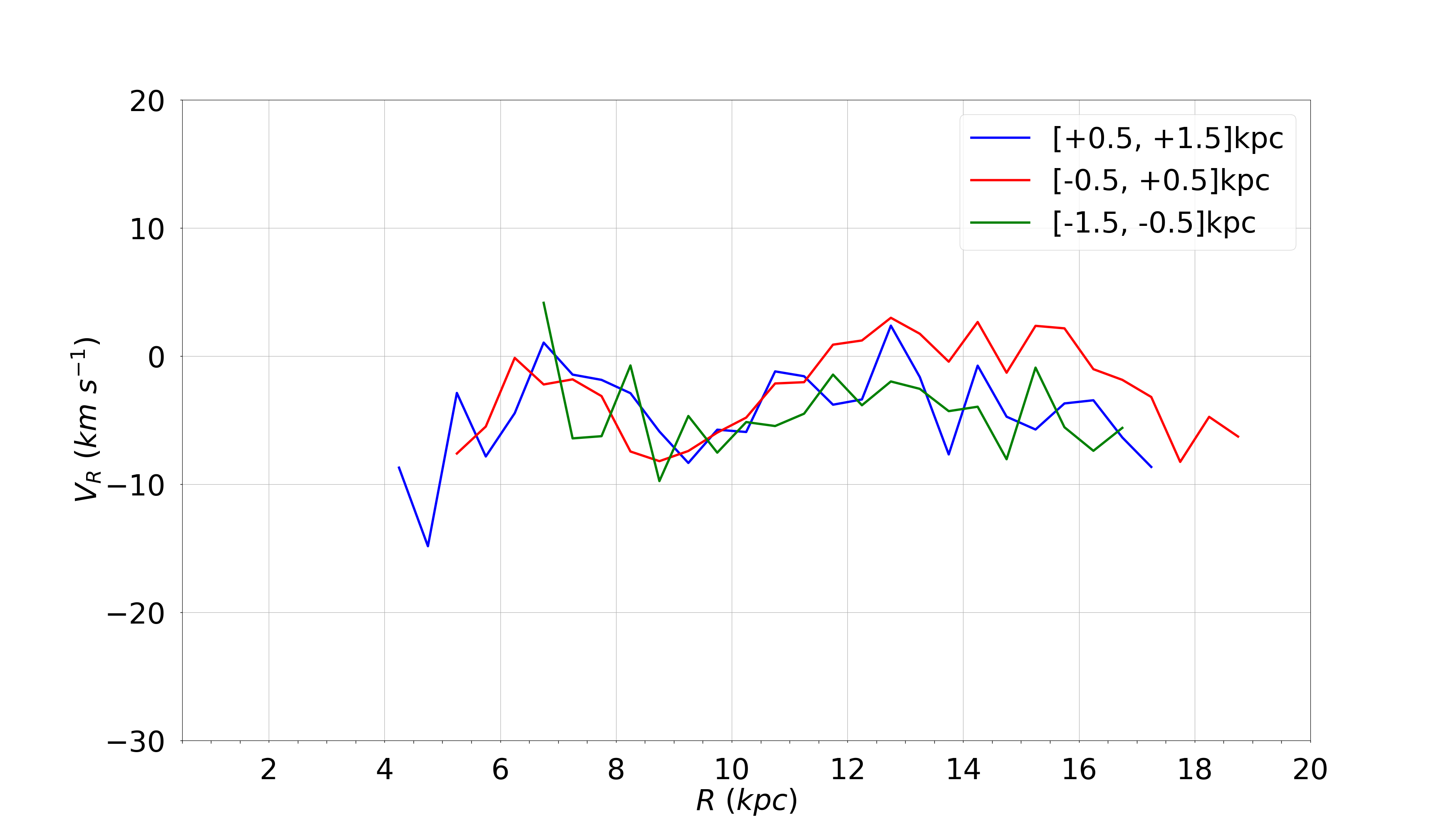}}\\
    	\centering  
    	\subfigure{
    		\includegraphics[width=0.49\textwidth]{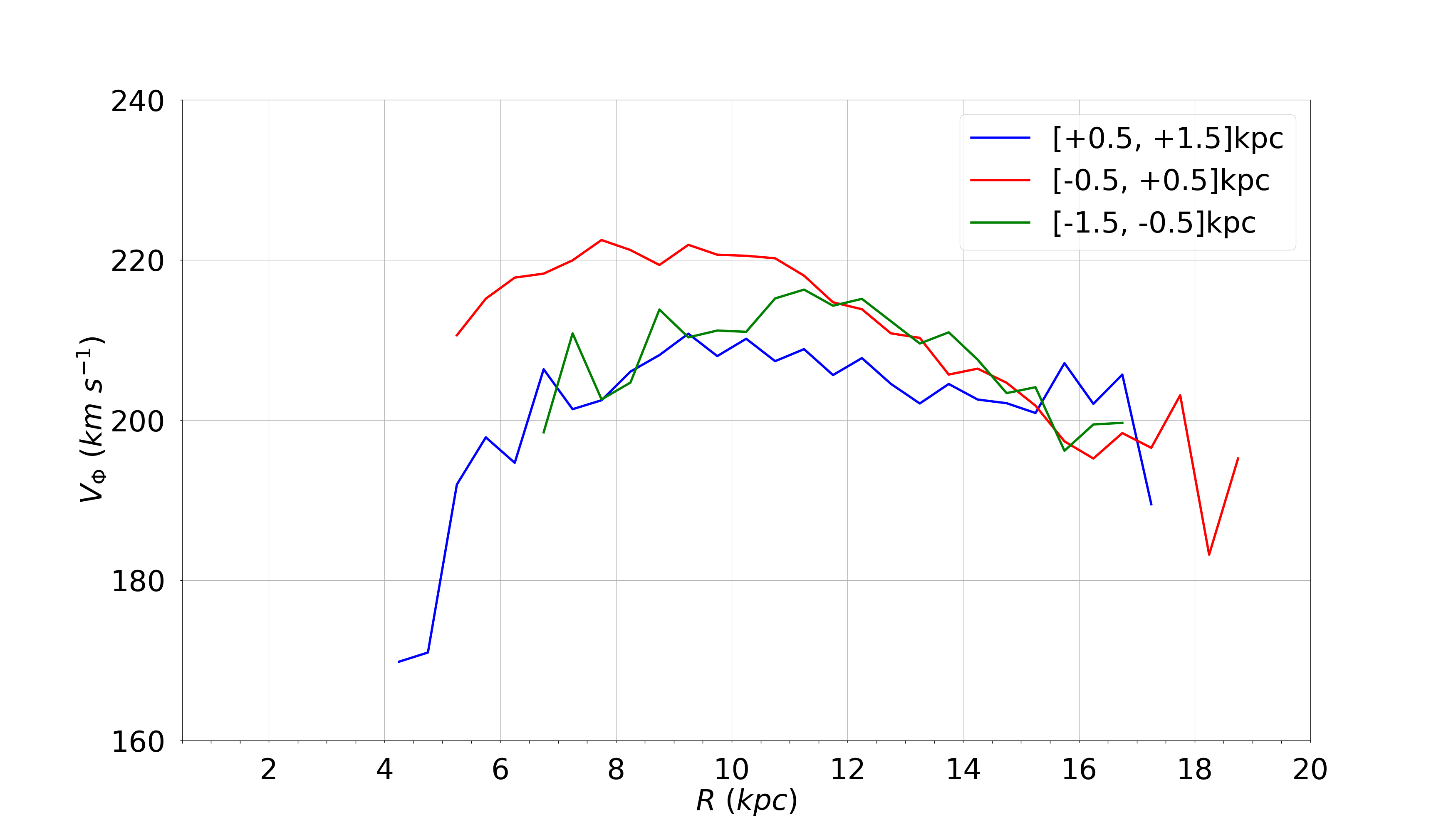}}
    	\subfigure{
    		\includegraphics[width=0.49\textwidth]{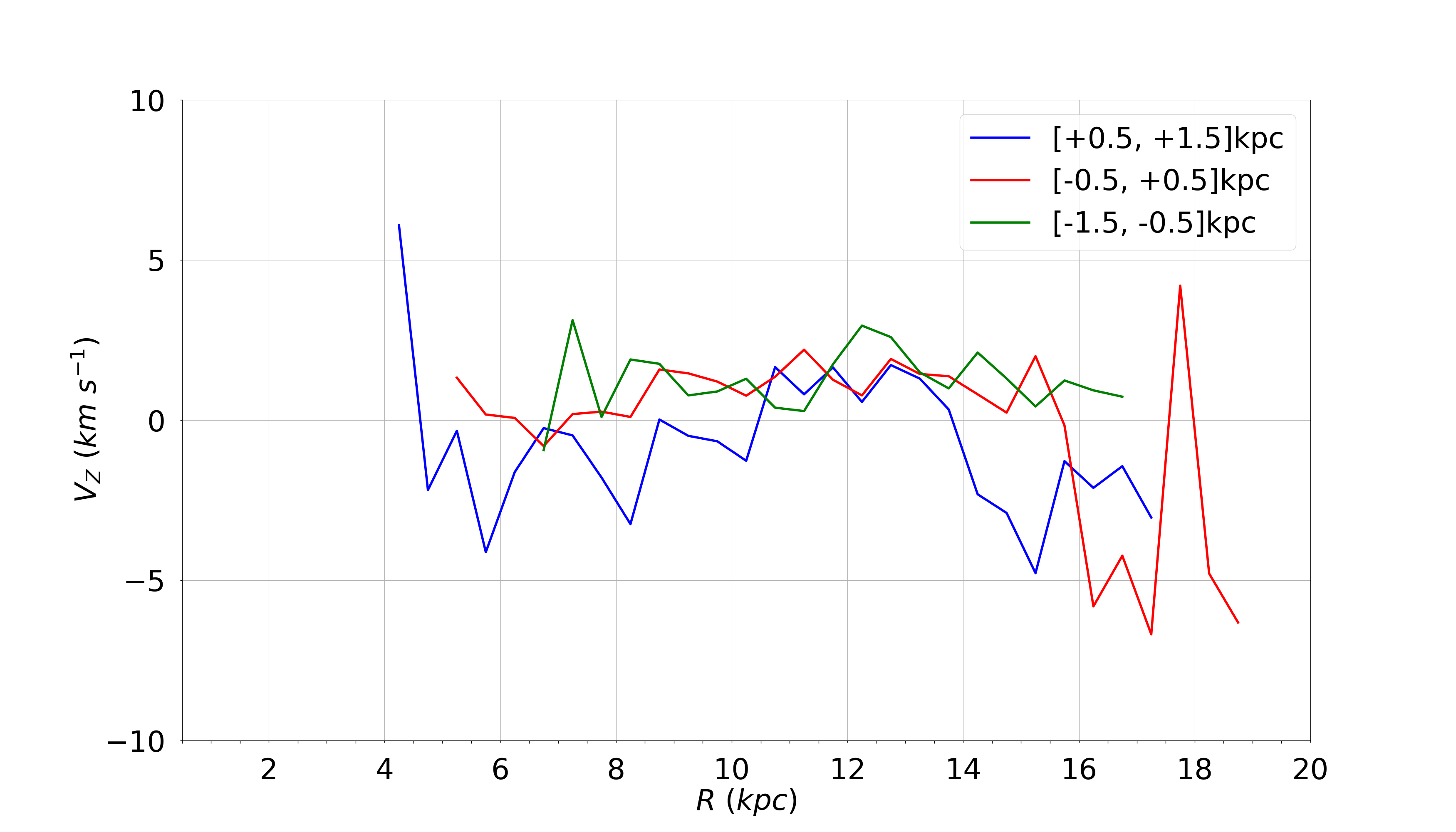}}
    	\caption{Top-left panel: the histogram distribution of the projected Galactocentric distances ($R$) of M giants with the bin width of 0.5 kpc, and different colors reprent different disk heights. Top-right and bottom panels: The variations of the radial velocity ($V_R$), azimuthal velocity ($V_\phi$), and vertical velocity ($V_z$) with $R$ at different disk heights, respectively. The bin width is also 0.5 kpc and the number of M giants in each bin is greater than 50.}
    	\label{fig9}
    \end{figure}

	The variations in $V_R$, $V_\phi$, and $V_Z$ obtained in this study are presented in Figure~\ref{fig9}. The top-left panel illustrates the distribution of M giants as a function of $R$, revealing two prominent peaks at approximately 8 and 10 kpc. In the top-right panel, the radial velocity ($V_R$) exhibits distinct trends: it is negative within the range $R \sim$7.0--9.0$\mathrm{~kpc}$, positive between $R\sim$9.0--13.0$\mathrm{~kpc}$, and reverts to negative values beyond $R\sim$13.0$\mathrm{~kpc}$. The bottom-left panel shows the azimuthal velocity ($V_\phi$) as a function of $R$ across different vertical height intervals. The results indicate a relatively smooth inverted U-shaped trend, with $V_\phi$ reaching its maximum in the $[-0.5, 0.5]\mathrm{~kpc}$ height range and decreasing with increasing $|Z|$ for $R<11\mathrm{~kpc}$. Additionally, $V_\phi$ values for M giants in the southern Galactic plane exceed those in the northern plane for $9.5 < R < 15\mathrm{~kpc}$. The bottom-right panel reveals that the vertical velocity ($V_Z$) exhibits a subtle increasing trend with $R$ for $R < 13\mathrm{~kpc}$, transitioning from the inner to the outer Galactic disk.
	
    The velocity variations of $V_R$, $V_\phi$, and $V_Z$ with $R$ at different heights obtained in this study exhibit trends comparable to those reported by \cite{Sun_2024}, who analyzed the RC star sample. Specifically, \cite{Sun_2024} observed a negative radial gradient in $V_R$ for $R\sim$5.5--9.0$\mathrm{~kpc}$, positive values for $R\sim$9.0--12.0$\mathrm{~kpc}$, and a return to negative values at greater distances. Similarly, their analysis showed that $V_\phi$ for RC stars in the southern Galactic plane exceeds that of stars in the northern plane, while $V_Z$ demonstrates an increasing trend from the inner to the outer disk. As highlighted by \cite{Sun_2024}, these velocity patterns reflect the oscillation, north–south rotational asymmetry, and warp of the Galactic disk. The consistent trends observed in our results further confirm these three key characteristics of the Galactic disk, providing additional validation of the calibrated CMR derived in this study. Moreover, the data employed in our analysis extend to even more distant regions of the Galactic disk, reaching up to $R$ = 18 kpc.%\vspace{1\baselineskip}

	\section{Summary} \label{sec5}
	Based on Gaia DR3 data, we performed a cencus of OC-M giant pairs using a sample of 34,657 M giants and 6879 OCs. In total, we identified 58 M giants associated with 43 OCs. These M giants are either member stars of OCs or have 5D astrometric parameters consistent within 3 times the corresponding standard deviations of OCs. 
	
	Employing both the full sample of 58 OC-M giant pairs and a subsample of 17 pairs with the ages of OCs all over 1 Gyr, we calibrated the CMR for M giants. Our analysis revealed that the subsample provide a more accurate calibration, and the linear relation is superior to the empirical distance relation in characterizing the CMR of M giants within 5 kpc from the Sun. Incorporating various levels of error into the reddening values of M giants, we refitted the CMR accordingly. The best-fitting CMR derived in this study is $M_{K_s}=3.85-8.26(J-K_s)$. The calibration was performed using OC-M giant pairs from the subsample, with a 5\% error applied to the reddening values of the M giants. The median deviations between the photometric distances of M giants derived from our CMR and parallax distances from Gaia DR3, alongside known spectroscopic distances, are 1.5\% and 2.2\%, respectively. This indicates that the distances of M giants inferred from the CMR are accurate, affirming the robust validity and feasibility of the CMR we obtained. Additionally, we calculated the radial velocity ($V_R$), azimuthal velocity ($V_\phi$), and vertical velocity ($V_Z$) for the M giants using distances derived from the calibrated CMR. The resulting velocity variations are consistent with previous studies, revealing key features of the Galactic disk, including oscillation, north–south rotational asymmetry, and warp. These findings further validate the reliability of the CMR developed in this work.
	
	With the continuous release of Gaia high-precision data and LAMOST spectral data, an increasing number of OC-M giant pairs will be discovered, and the CMR of M giants will be better constrained, allowing for more accurate distance determinations. Meanwhile, it is expected that in the future, the radial velocities and metallicities of OC member stars in the identified OC-M giant pairs can be used as a reference to better study the intrinsic properties of M giants.
	
	\section*{Acknowledgements}
	We thank the anonymous referee for the instructive comments and suggestions that greatly helped us to improve the paper. This study is supported by the National Natural Science Foundation of China under grant Nos. 12273027, the science research grants from the China Manned Space Project with NO. CMS-CSST-2021-B03 and the Innovation Team Funds of China West Normal University grant No. KCXTD2022-6. Y.X. is supported by the NSFC grant 11933011, National SKA Program of China (grant No. 2022SKA0120103), and the Key Laboratory for Radio Astronomy. C.J.H. acknowledges support from the National Postdoctoral Program for Innovative Talents of the Oﬀice of China Postdoc Council (grant No. BX20240414) and the NSFC grant No. 12403041. J.Z. would like to acknowledge the NSFC under grants 12073060, and the Youth Innovation Promotion Association CAS. This work has made use of data from the Guoshoujing Telescope (the Large Sky Area Multi-Object Fiber Spectroscopic Telescope, LAMOST, \url{https://www.lamost.org/lmusers/}), the Sloan Digital Sky Survey(SDSS, \url{https://www.sdss.org/}) and the European Space Agency (ESA) mission Gaia (\url{https://www.cosmos.esa.int/gaia}).
	\clearpage
	\appendix
	%\section{Analysis of the pairs where the M giants are simultaneously harbored in 2 different OCs }
	\section{OCs containing M giants}\label{appendixA}
	
	Initially, 46 OCs containing M giants were identified across four catalogs: \cite{Cantat-Gaudin_2020}, \cite{Castro-Ginard_2022}, \cite{Hao_2022_1}, and \cite{Hunt_2024}. Among these, 12 clusters were reported in both \cite{Hunt_2024} and at least one of the other three catalogs. For these overlapping clusters, we adopted the parameters from \cite{Hunt_2024}, as they are derived from the most recent Gaia DR3 data.
	
	Additionally, by cross-referencing the “Name” and “AllNames” fields in \cite{Hunt_2024}, we identified that LP 1375 and UBC 434, Waterloo 1 and UBC 609, as well as UBC 1246 and OC 0268, are each the same clusters reported under different names. To maintain consistency, we retained only LP 1375, Waterloo 1, and UBC 1246, as reported in \cite{Hunt_2024}.
	
	After these refinements, a total of 43 OCs harboring M giants were ultimately included in our analysis.
    %\vspace{1\baselineskip}
	
	\renewcommand\thefigure{\Alph{section}\arabic{figure}}  
	\setcounter{figure}{0}

	\clearpage
	\renewcommand\thefigure{\Alph{section}\arabic{figure}}    
	%\newpage
	
	\section{Distributions of identified OCs and M giants}
	\label{appendixB}
	\setcounter{figure}{0}
	Open clusters (blue dots) harboring M giants (red stars) are
	as shown in Figures \ref{figB1}–\ref{figB5}.The columns of each panel
	represent the distributions of the member stars of OCs and M
	giants for positions in positions in $RA\left(\alpha\right)$ and $Dec\left(\delta\right)$, proper motions in $\mathsf{\mu}_{\alpha^*}$ and $\mathsf{\mu}_\delta$, and the CMD in $G~\text{vs.}~G_{\text{BP}}-G_{\text{RP}}$.
	
	\begin{figure}[h]
		\centering  
		\subfigure{
			\includegraphics[width=0.32\textwidth]{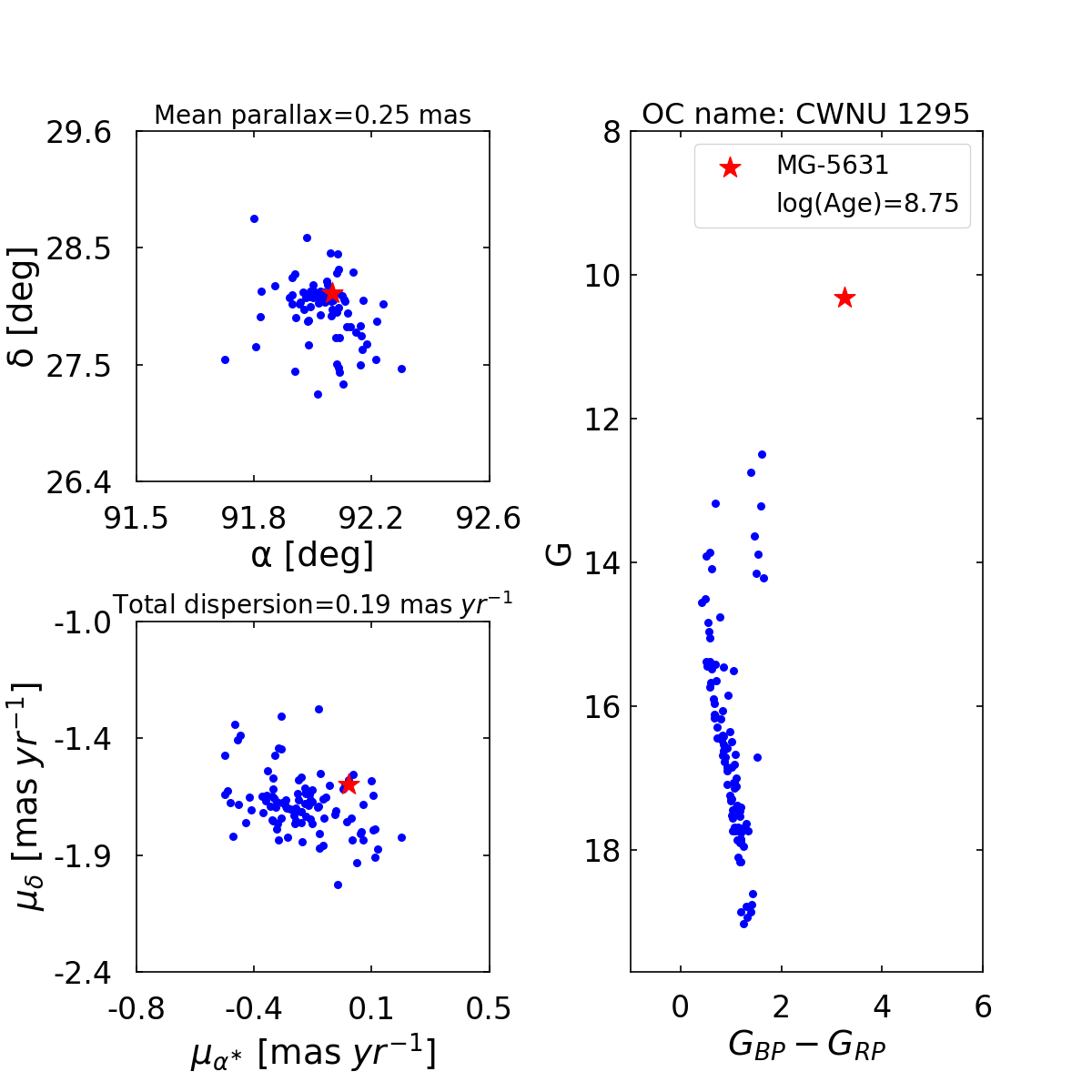}}
		\subfigure{
			\includegraphics[width=0.32\textwidth]{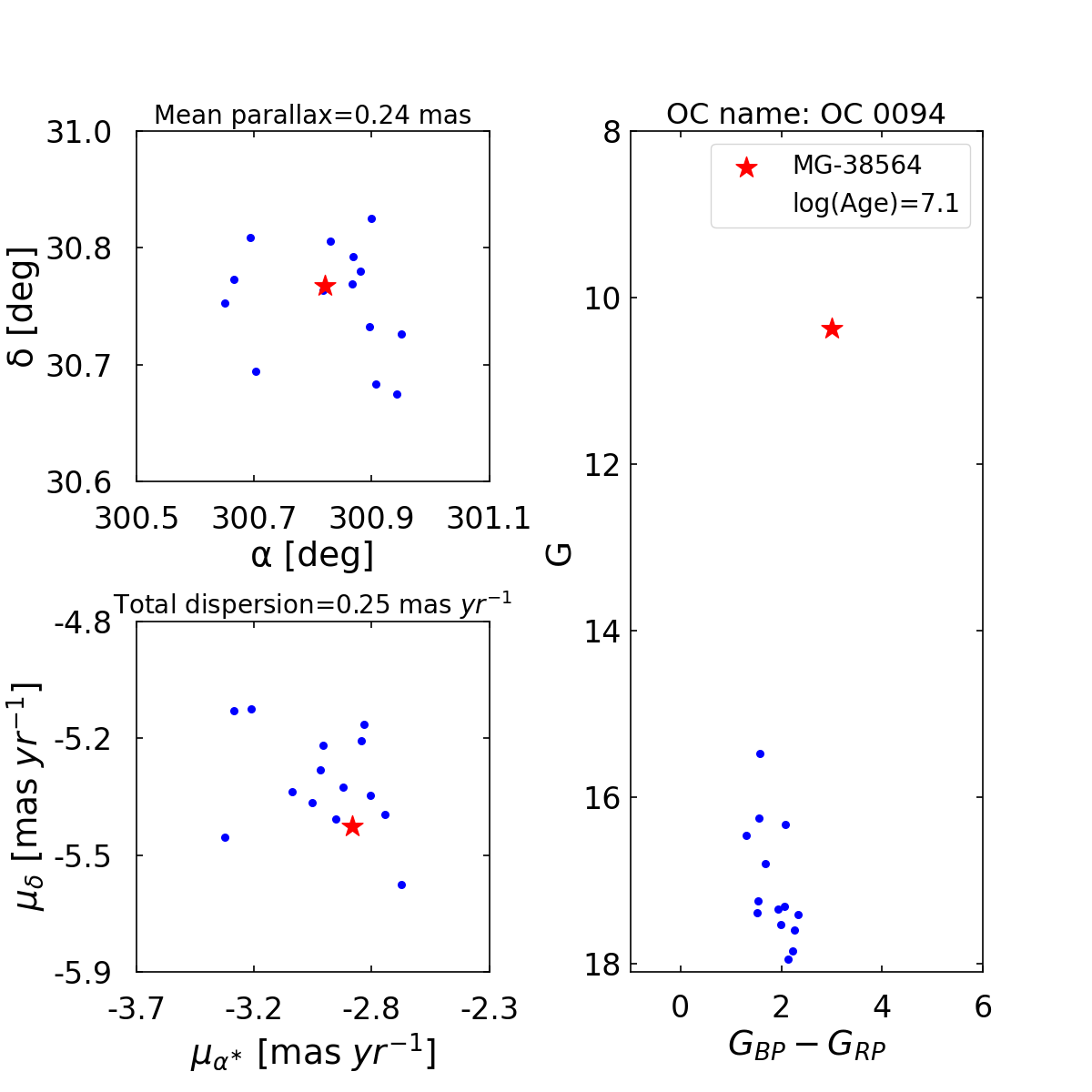}}
		\subfigure{
			\includegraphics[width=0.32\textwidth]{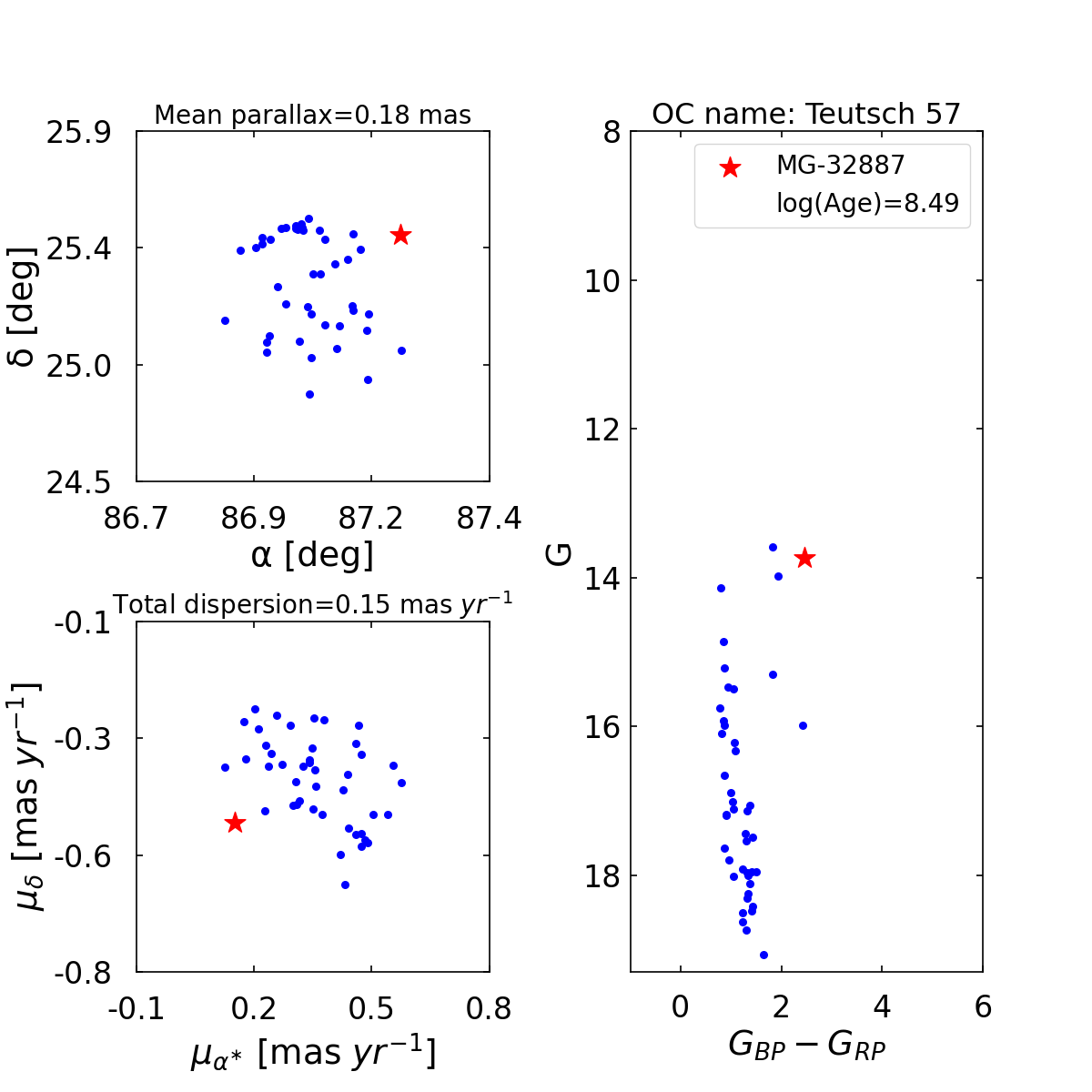}}
		\\
		\centering  
		\subfigure{
			\includegraphics[width=0.32\textwidth]{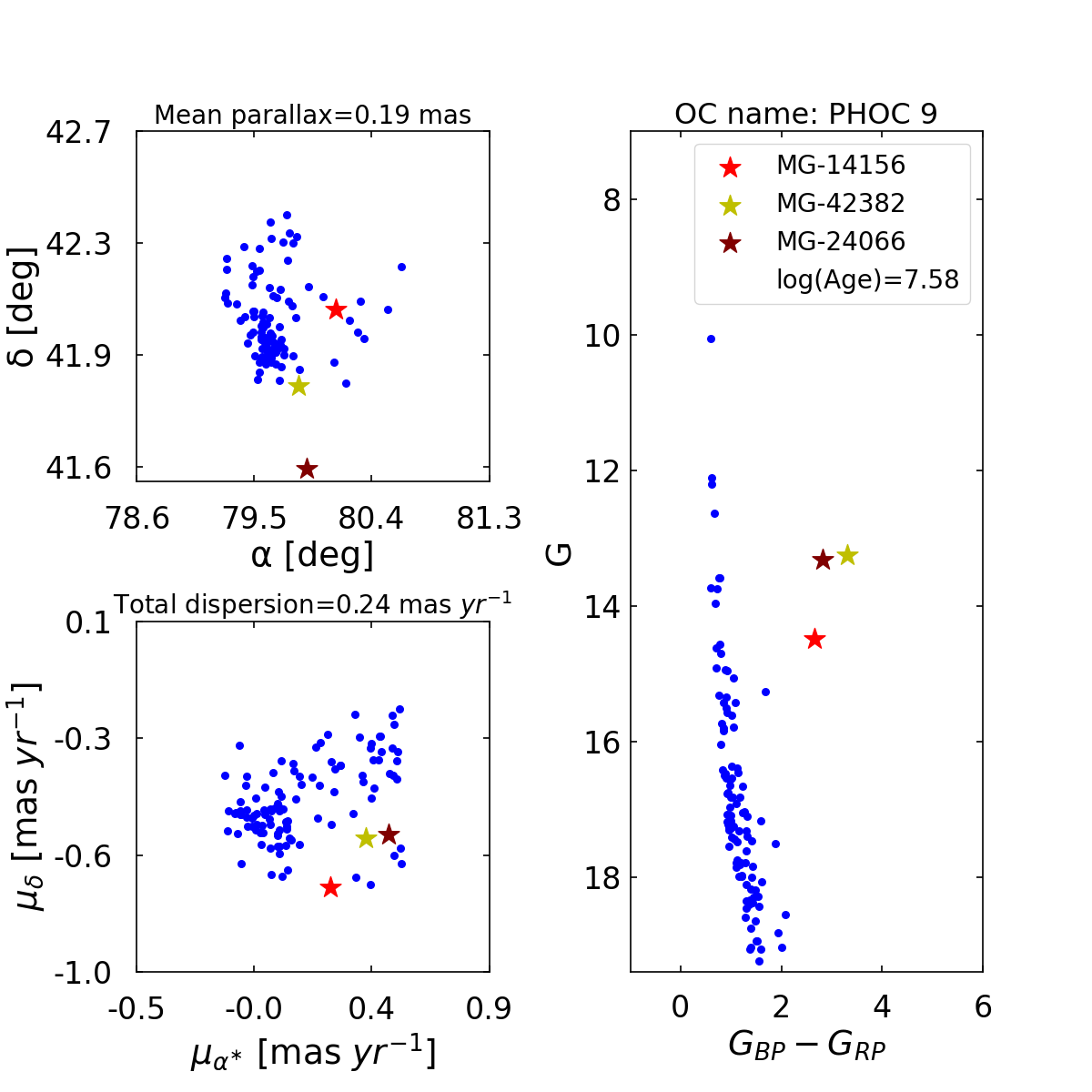}}
		\subfigure{
			\includegraphics[width=0.32\textwidth]{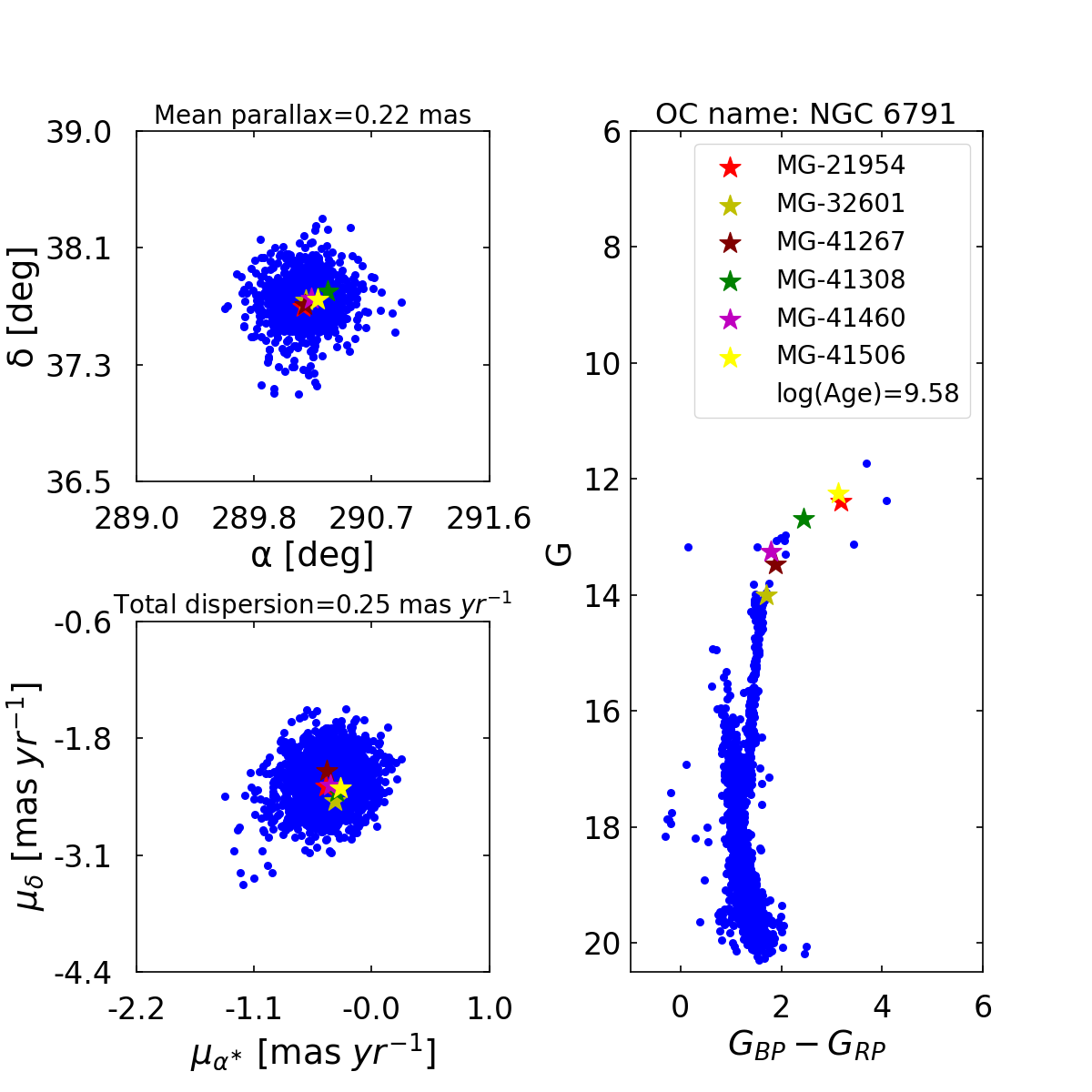}}
		\subfigure{
			\includegraphics[width=0.32\textwidth]{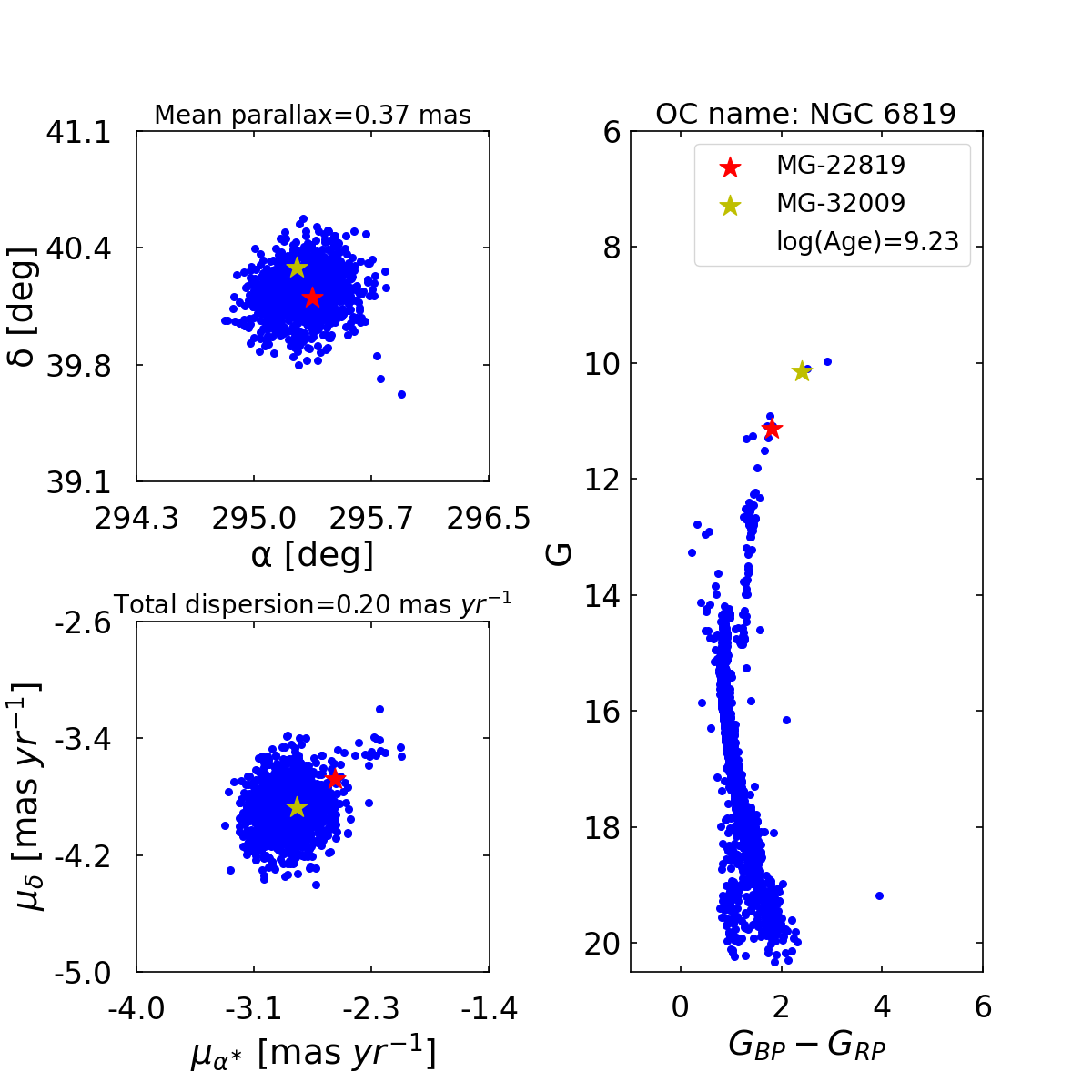}}
		\\
		\centering  
		\subfigure{
			\includegraphics[width=0.32\textwidth]{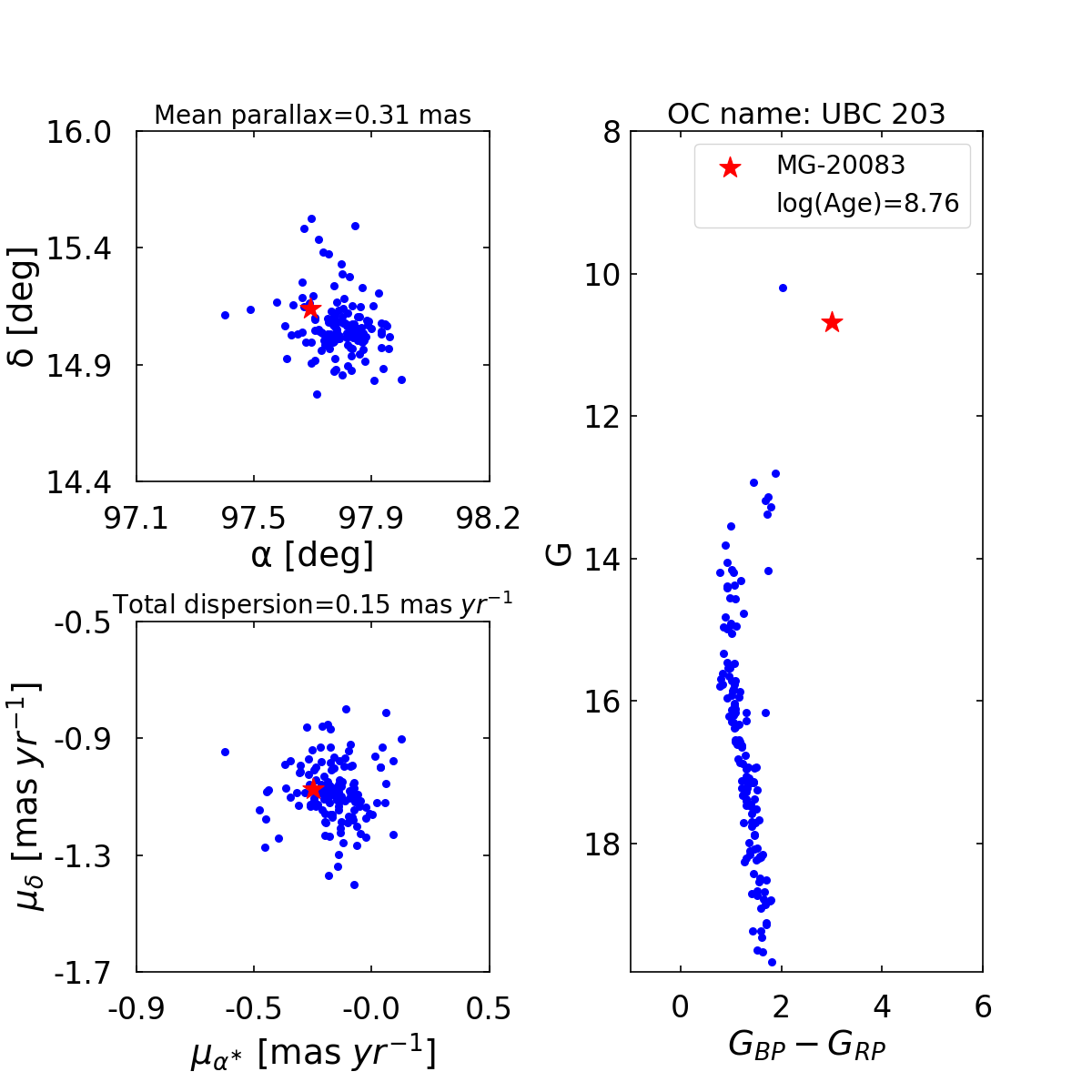}}
		\subfigure{
			\includegraphics[width=0.32\textwidth]{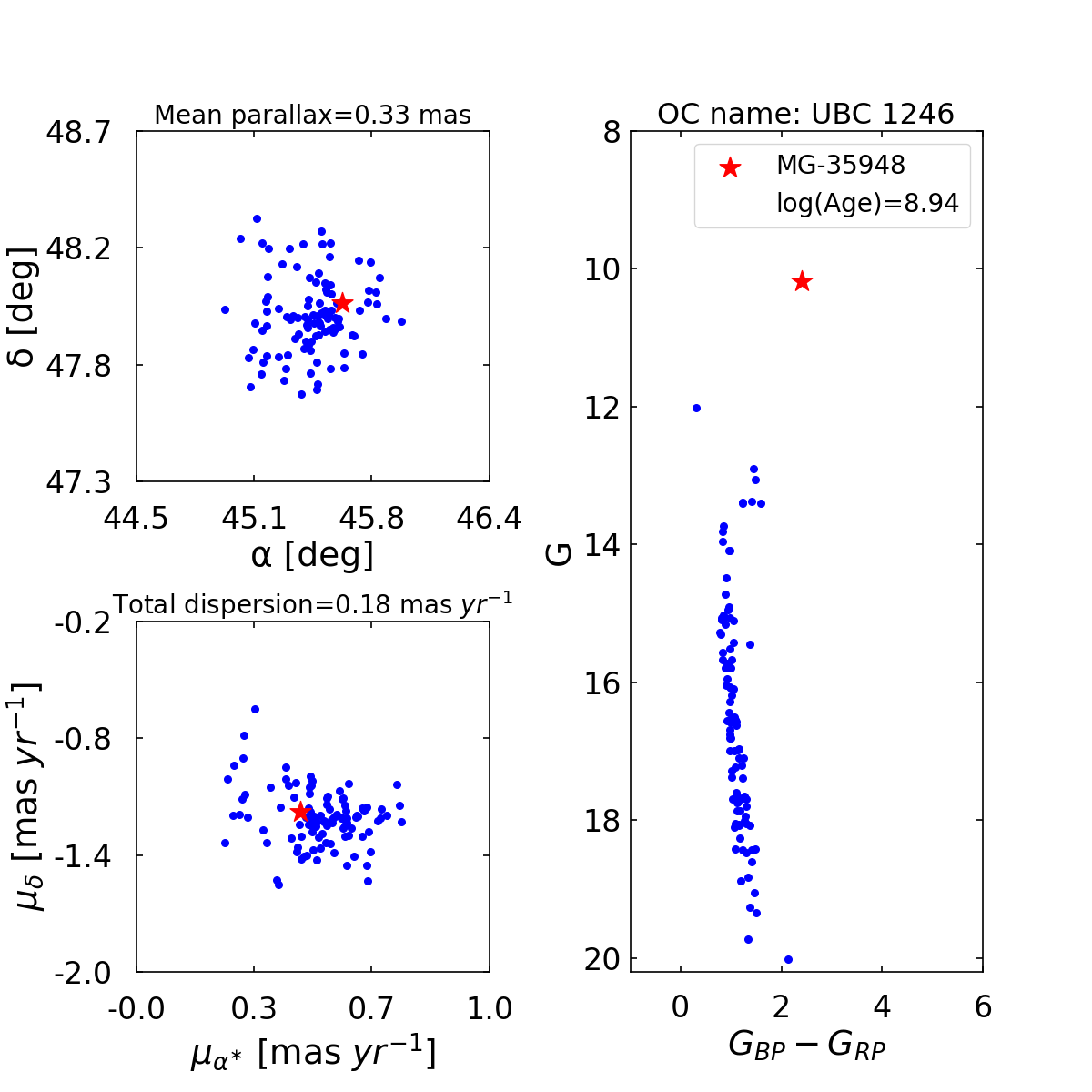}}
		\subfigure{
			\includegraphics[width=0.32\textwidth]{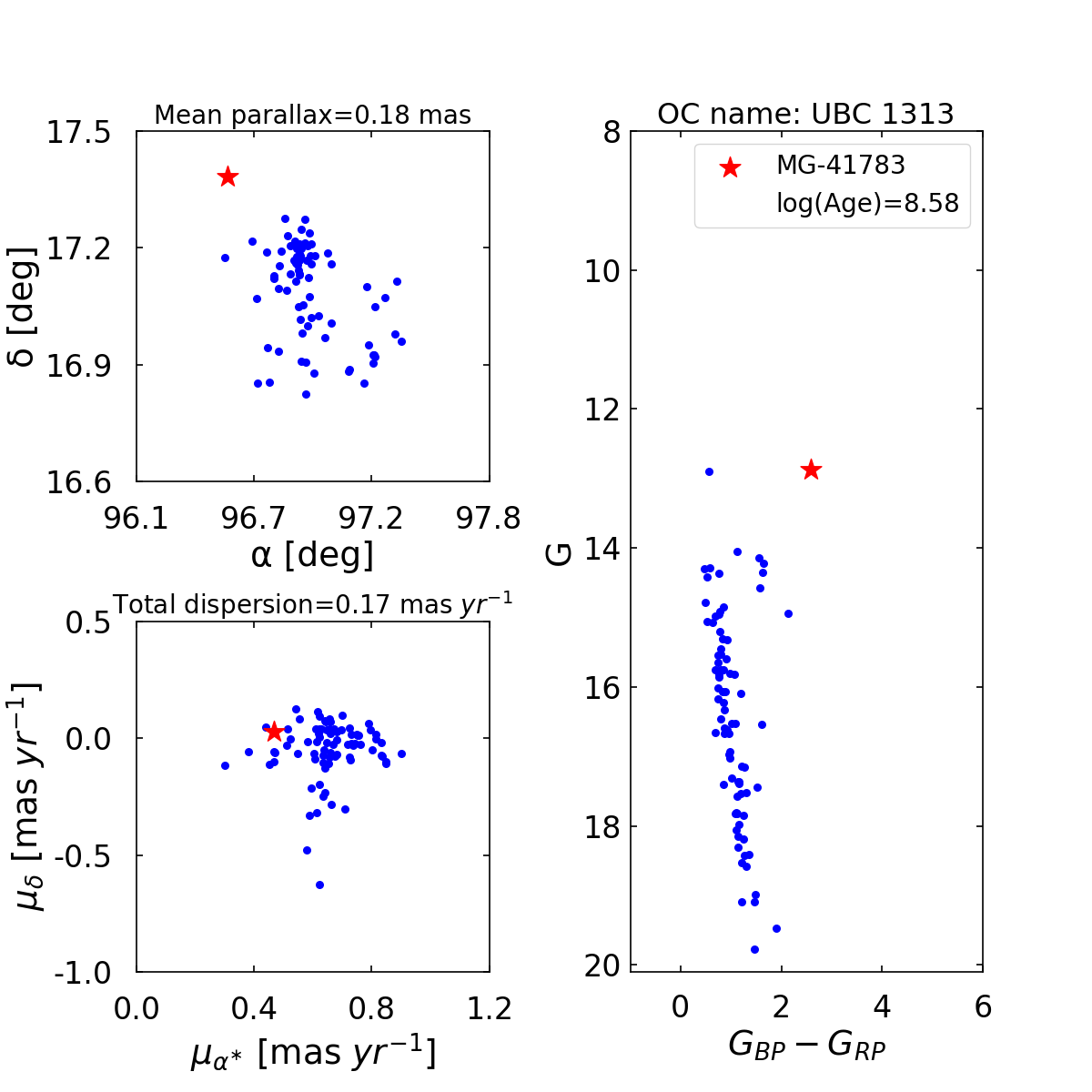}}
		\caption{The columns of each panel represent the distributions of the
			member stars of OCs(blue dots) and M giants(red prntagram) for , as well as the mean parallax and total proper-motion dispersion of OCs. Here, the listed OCs are CWNU 1295, OC 0094, Teutsch 57, PHOC 9, NGC 6791, NGC 6819, UBC 203, UBC 1246,~and~UBC 1313. The identification numbers of M giants and the ages of OCs are displayed in the upper right of the CMD.}
		\label{figB1}
	\end{figure}
	
	\begin{figure}[h]
		\centering  
		\subfigure{
			\includegraphics[width=0.32\textwidth]{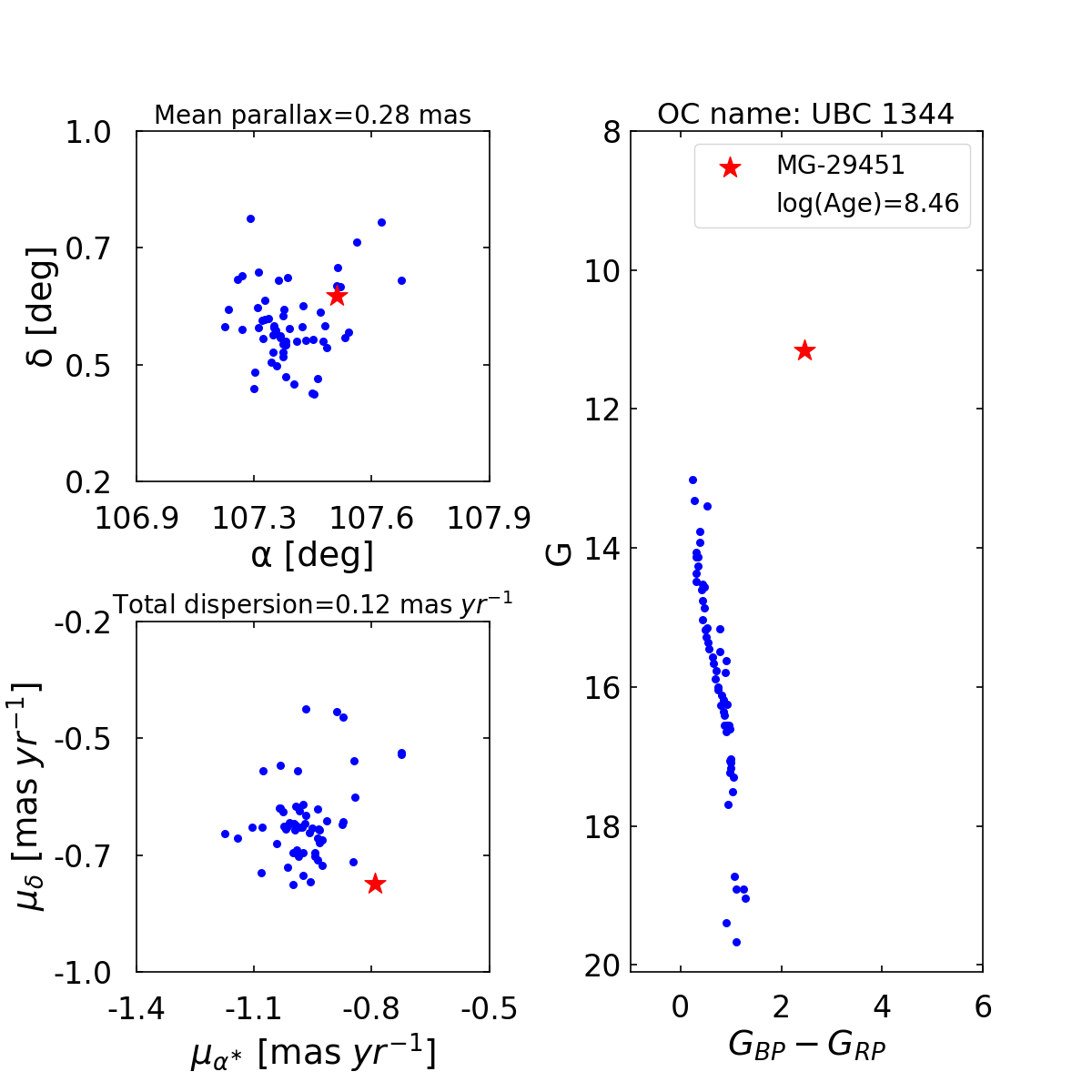}}
		\subfigure{
			\includegraphics[width=0.32\textwidth]{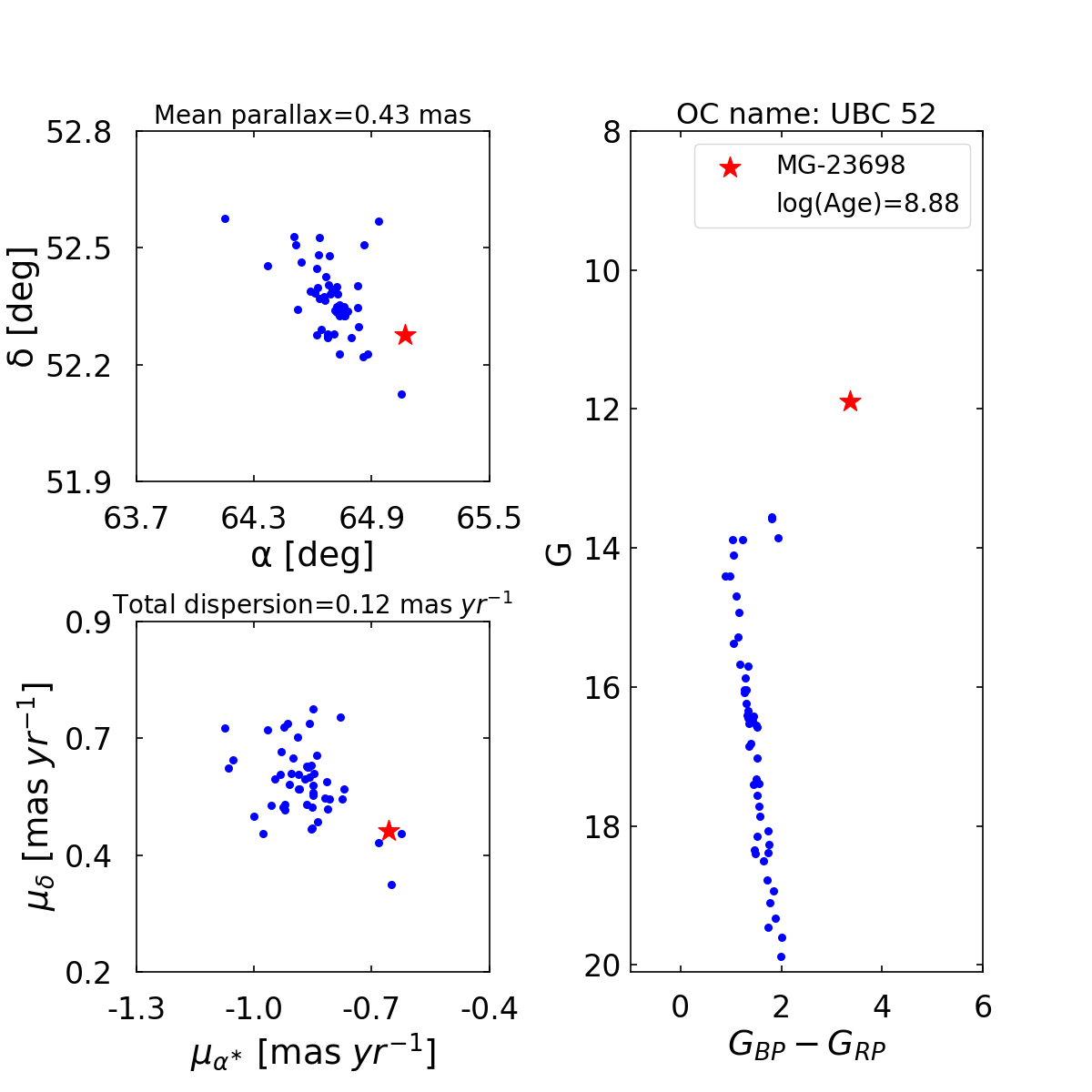}}
		\subfigure{
			\includegraphics[width=0.32\textwidth]{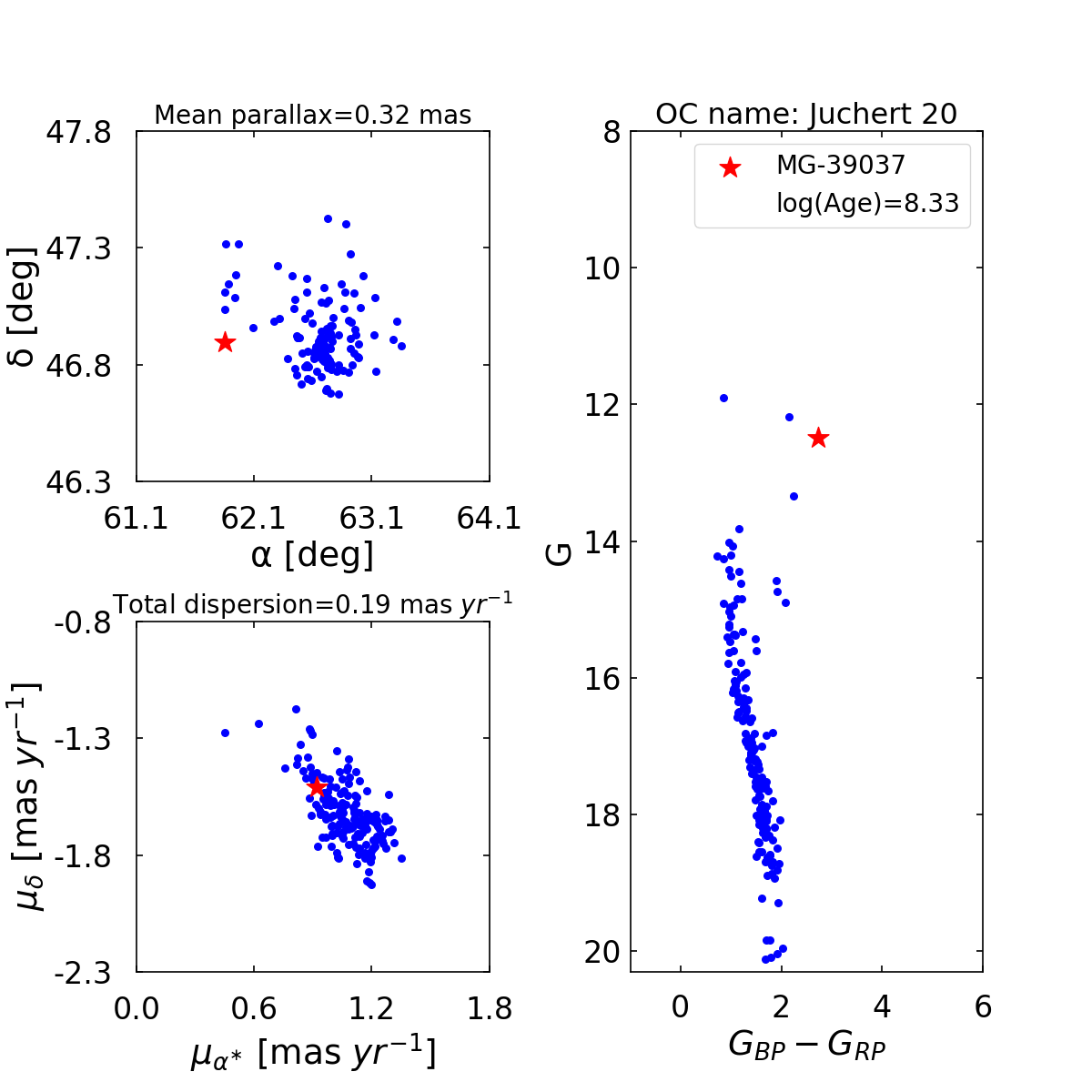}}
		\\
		\centering  
		\subfigure{
			\includegraphics[width=0.32\textwidth]{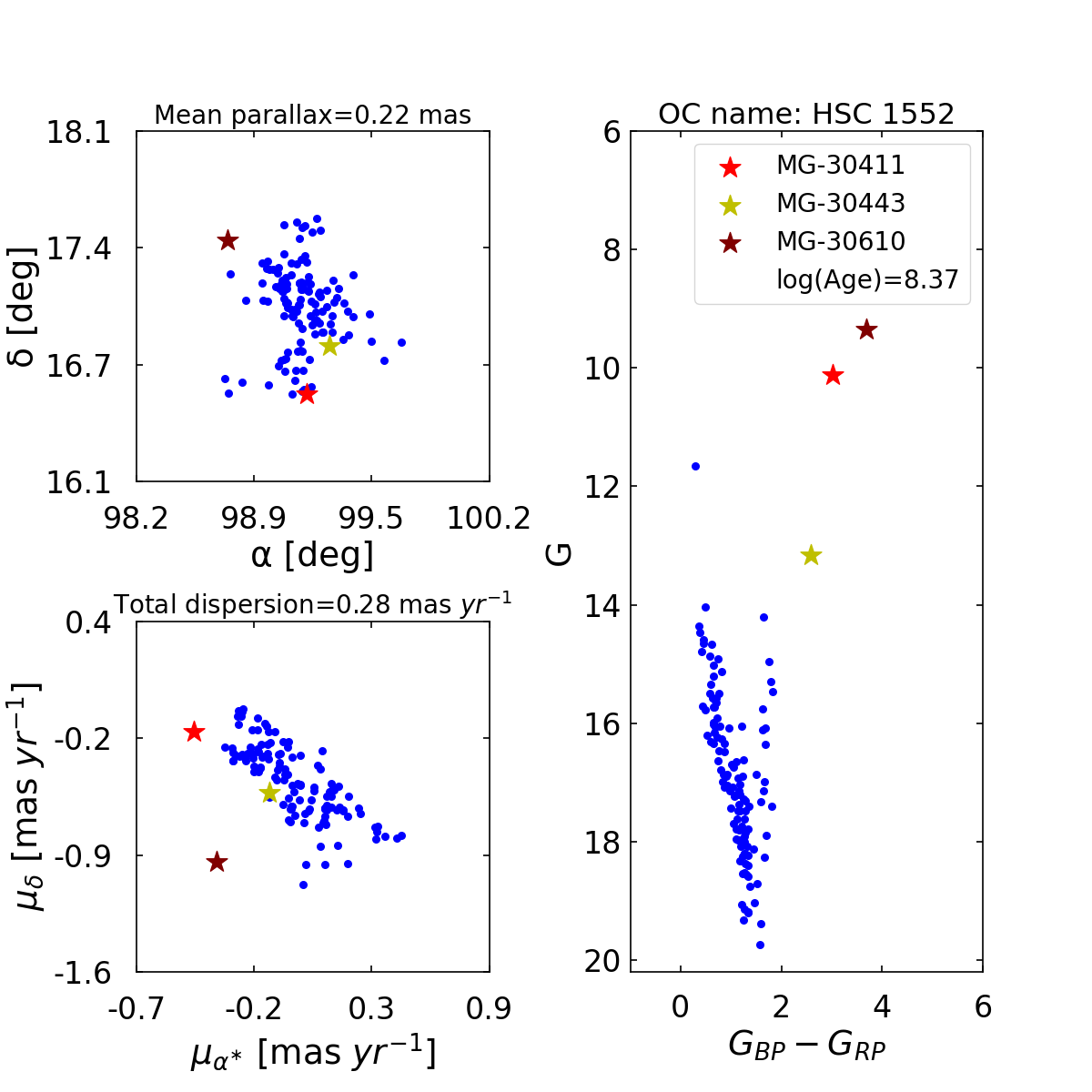}}
		\subfigure{
			\includegraphics[width=0.32\textwidth]{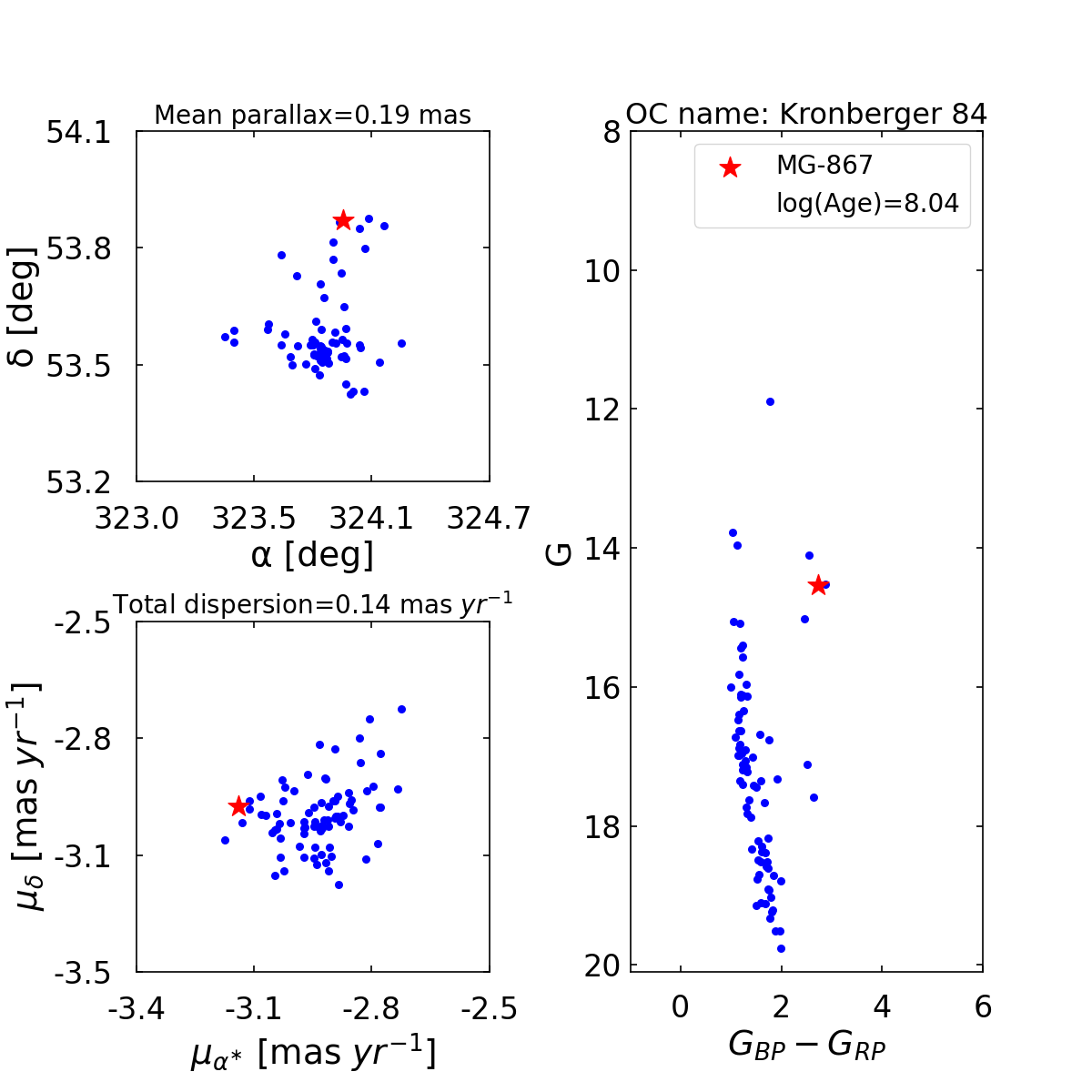}}
		\subfigure{
			\includegraphics[width=0.32\textwidth]{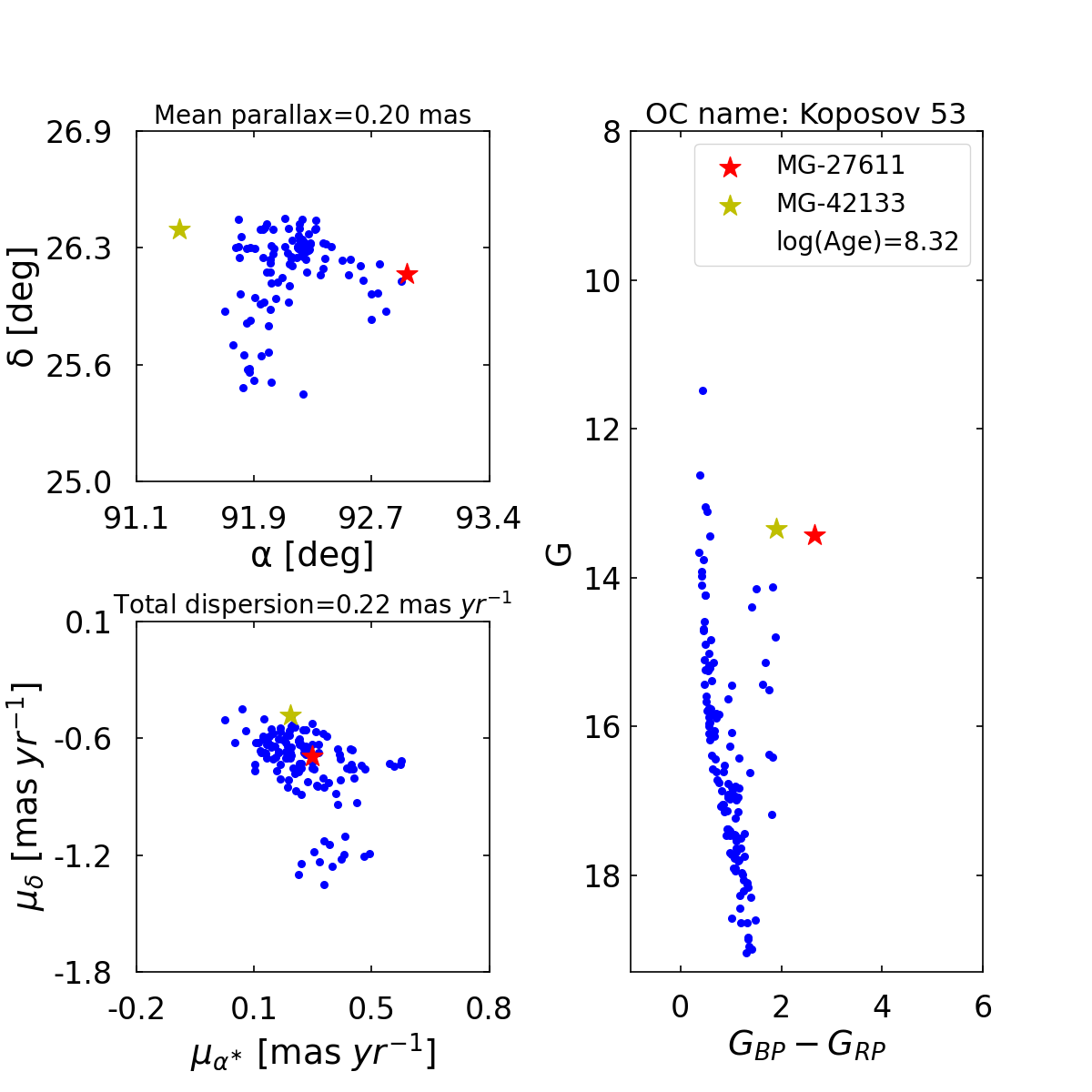}}
		\\
		\centering  
		\subfigure{
			\includegraphics[width=0.32\textwidth]{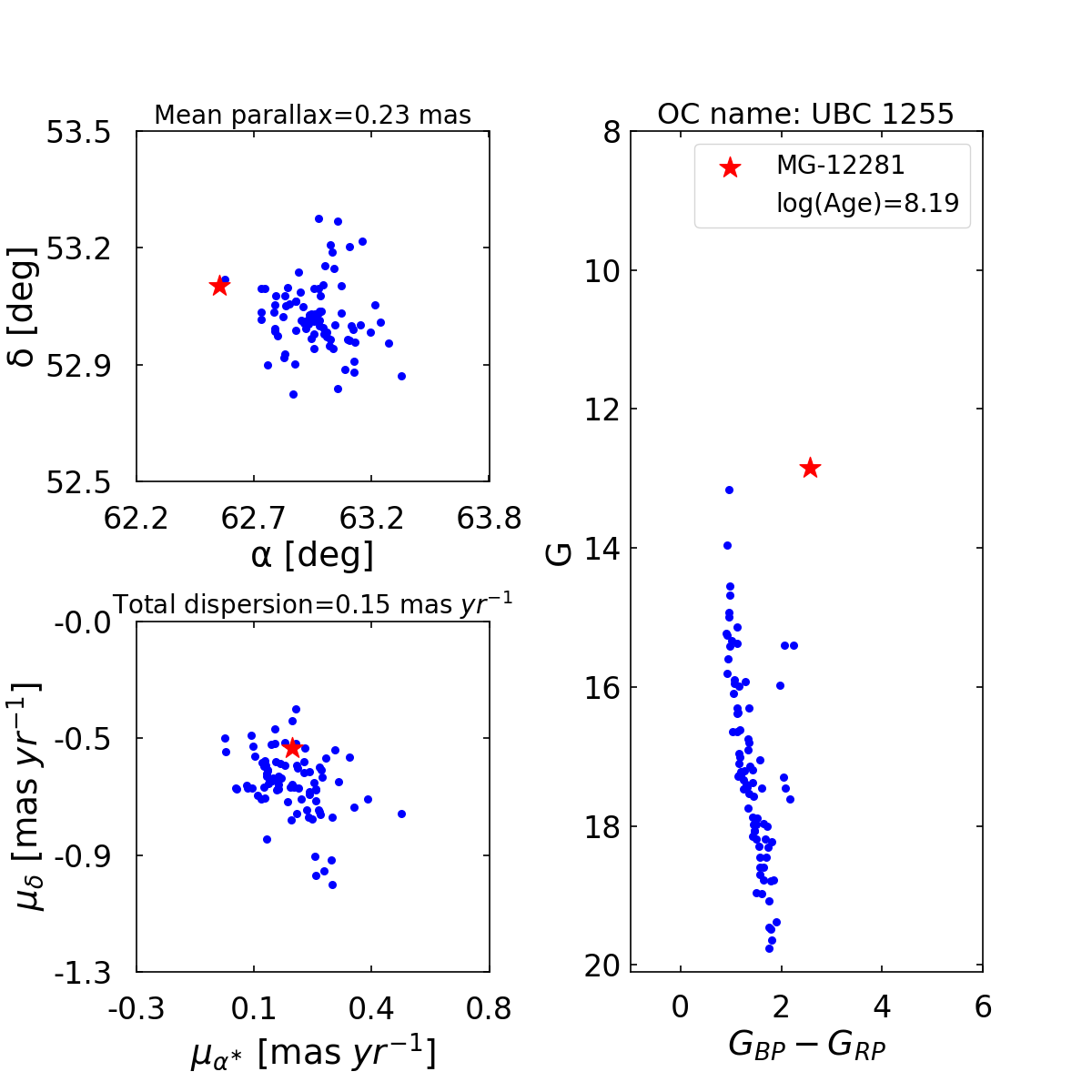}}
		\subfigure{
			\includegraphics[width=0.32\textwidth]{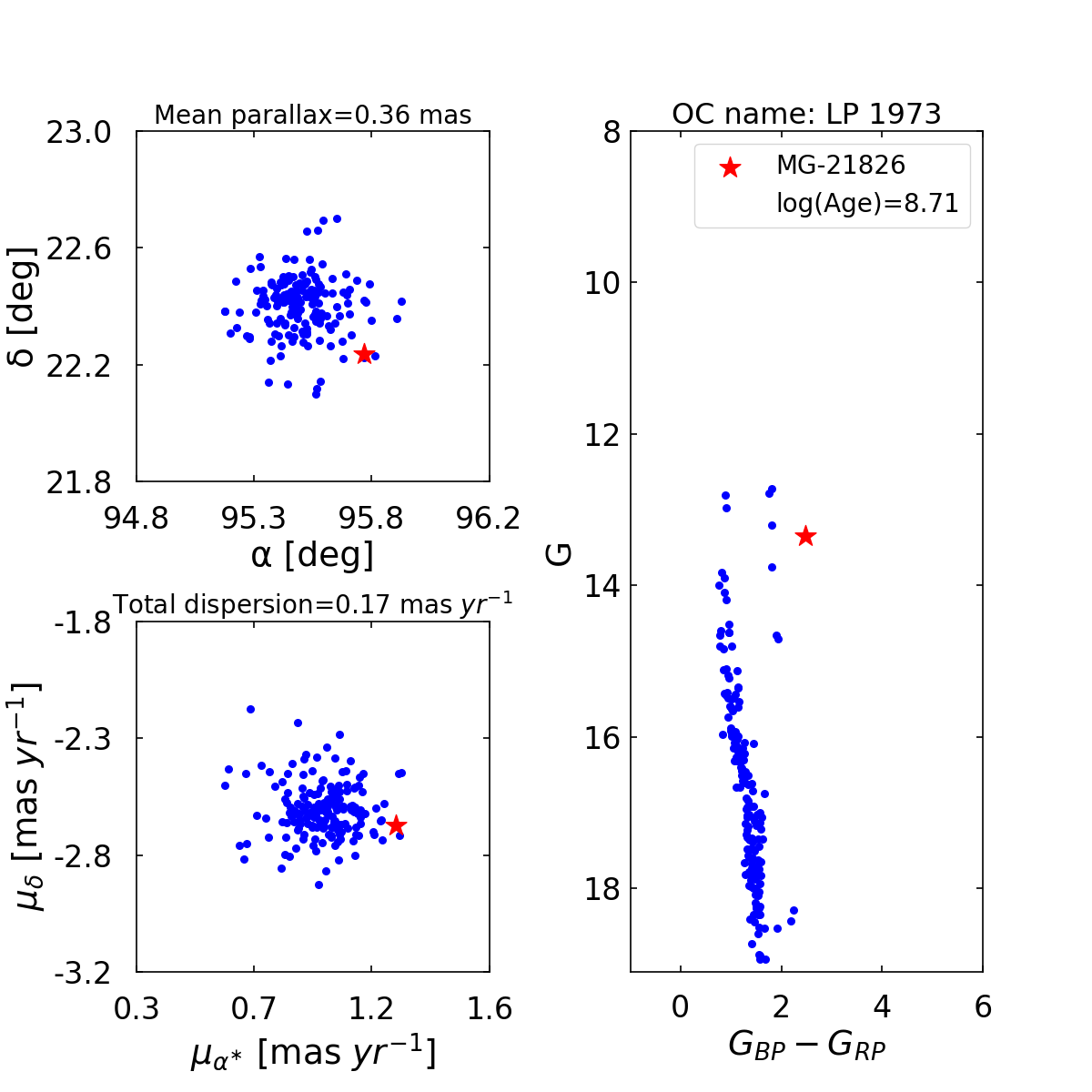}}
		\subfigure{
			\includegraphics[width=0.32\textwidth]{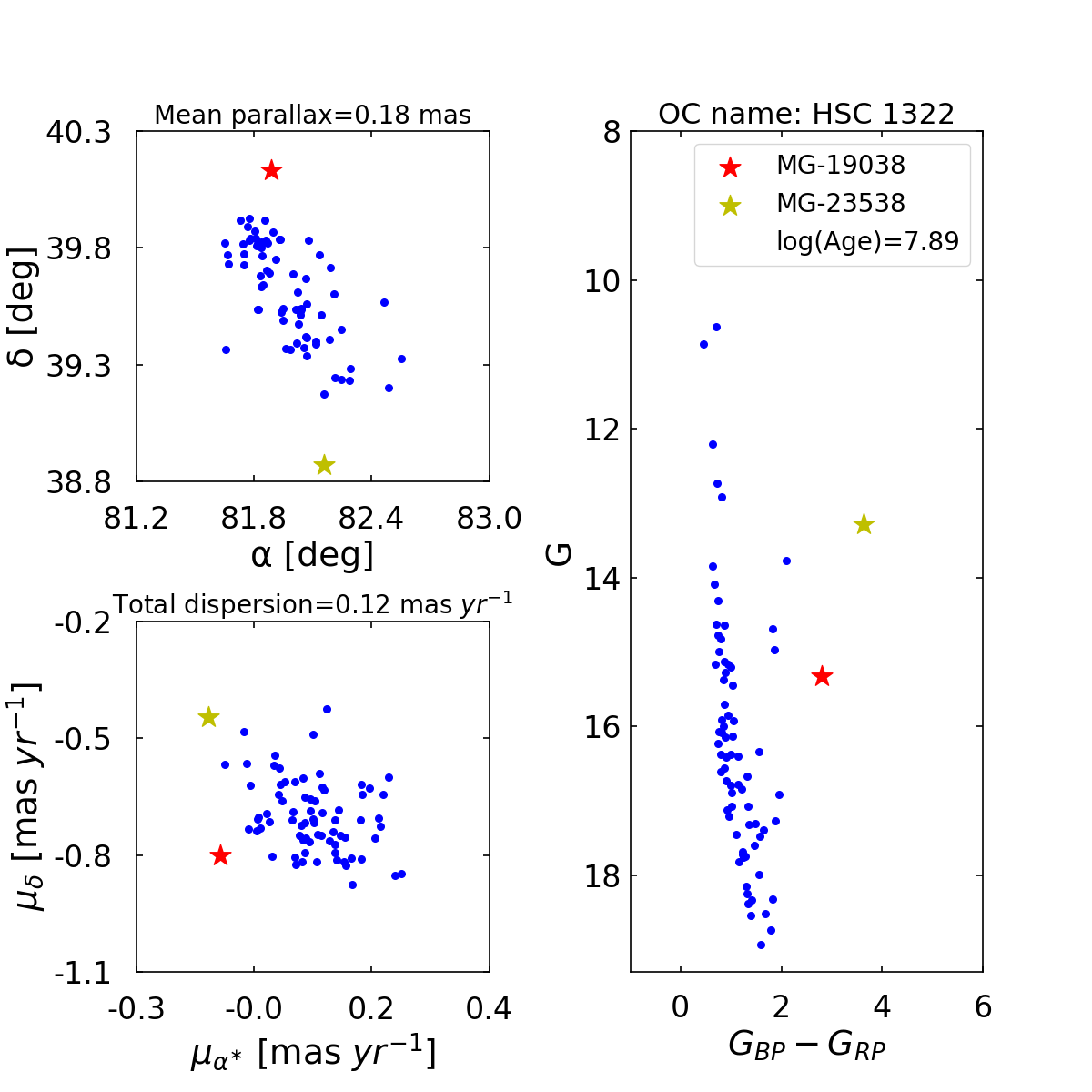}}
		\caption{Here, the listed OCs are UBC 1344, UBC 52, Juchert 20, HSC 1552, Kronberger 84, Koposov 53,~UBC 1255, LP 1973, and HSC 1322.}
	\end{figure}
	
	\begin{figure}[h]
		\centering  
		\subfigure{
			\includegraphics[width=0.32\textwidth]{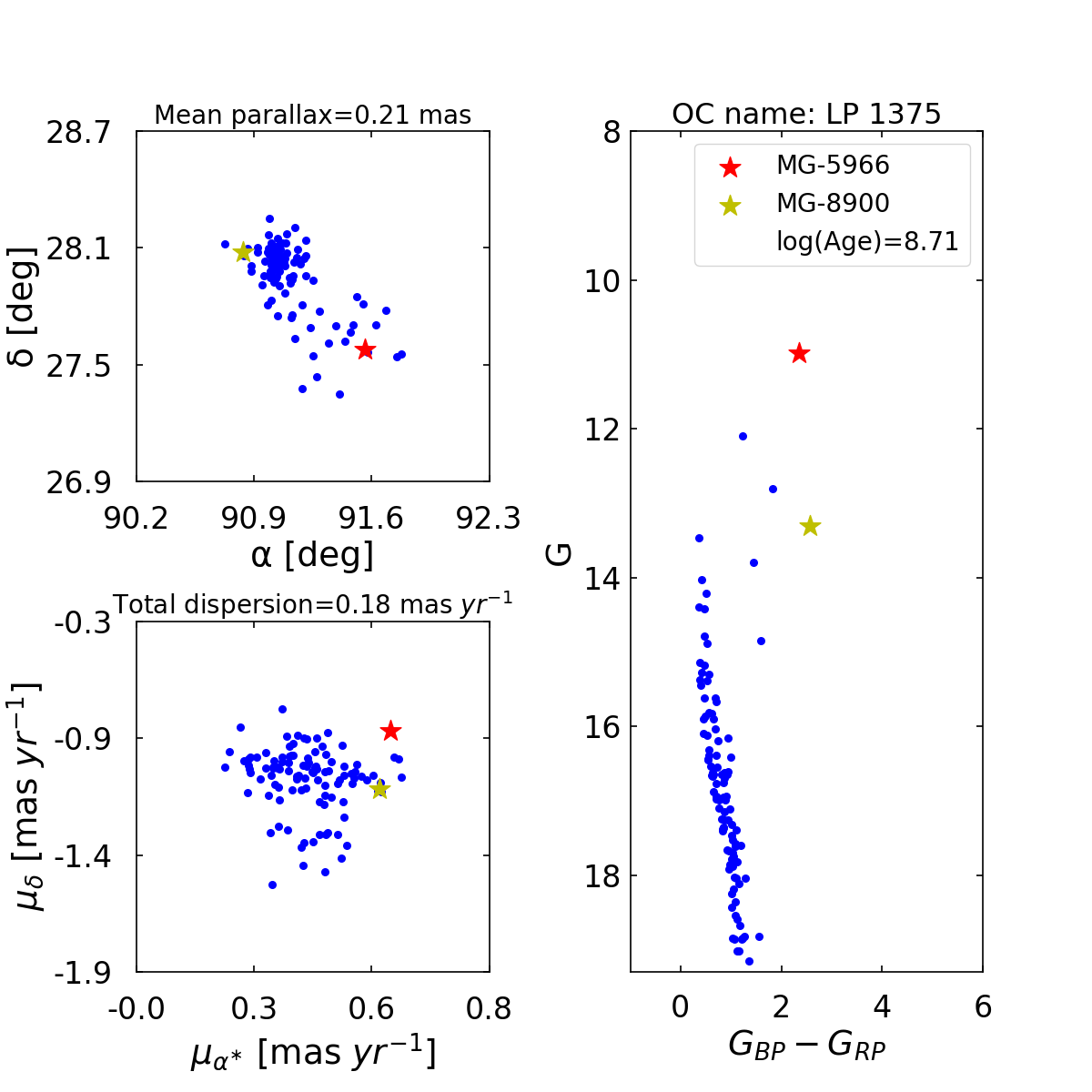}}
		\subfigure{
			\includegraphics[width=0.32\textwidth]{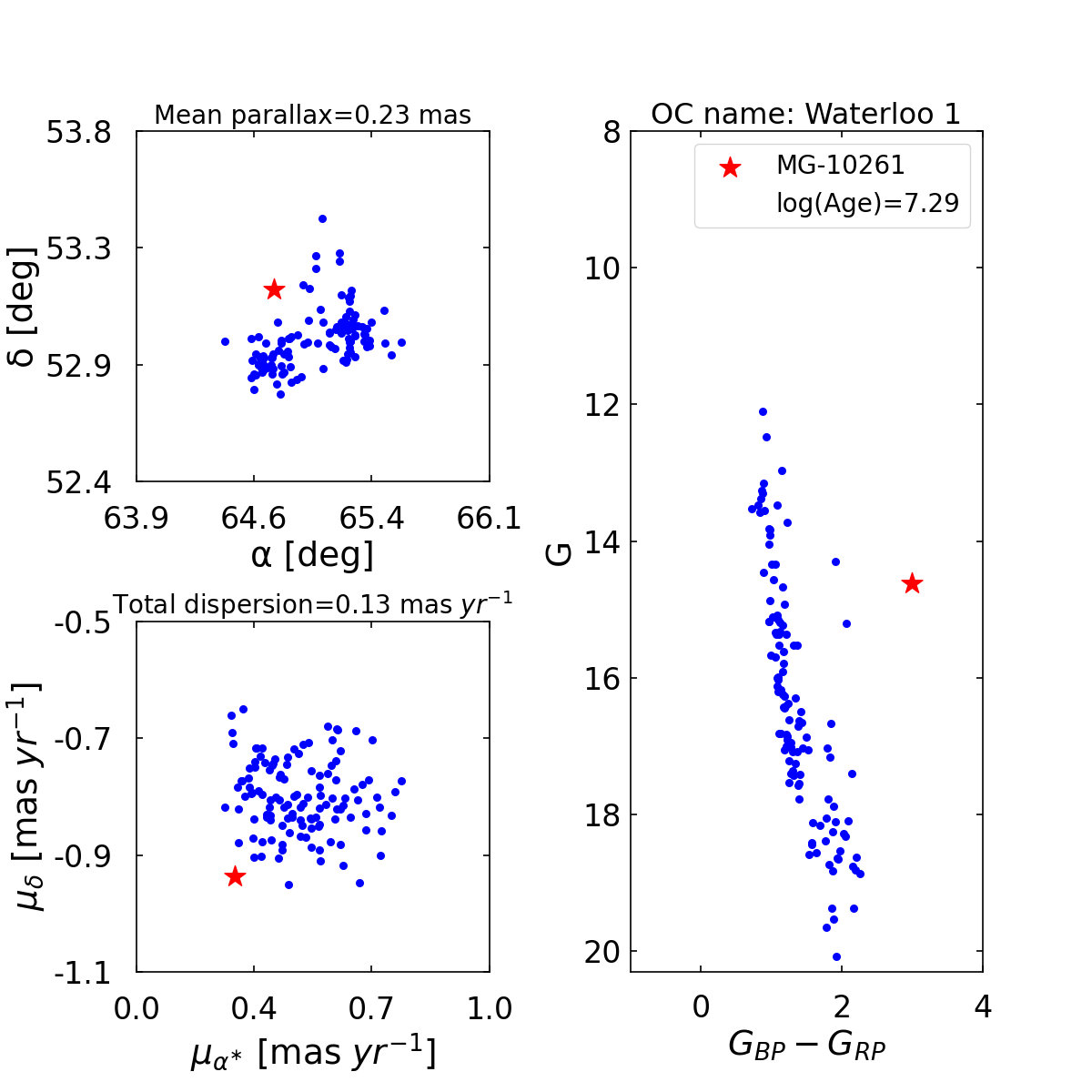}}
		\subfigure{
			\includegraphics[width=0.32\textwidth]{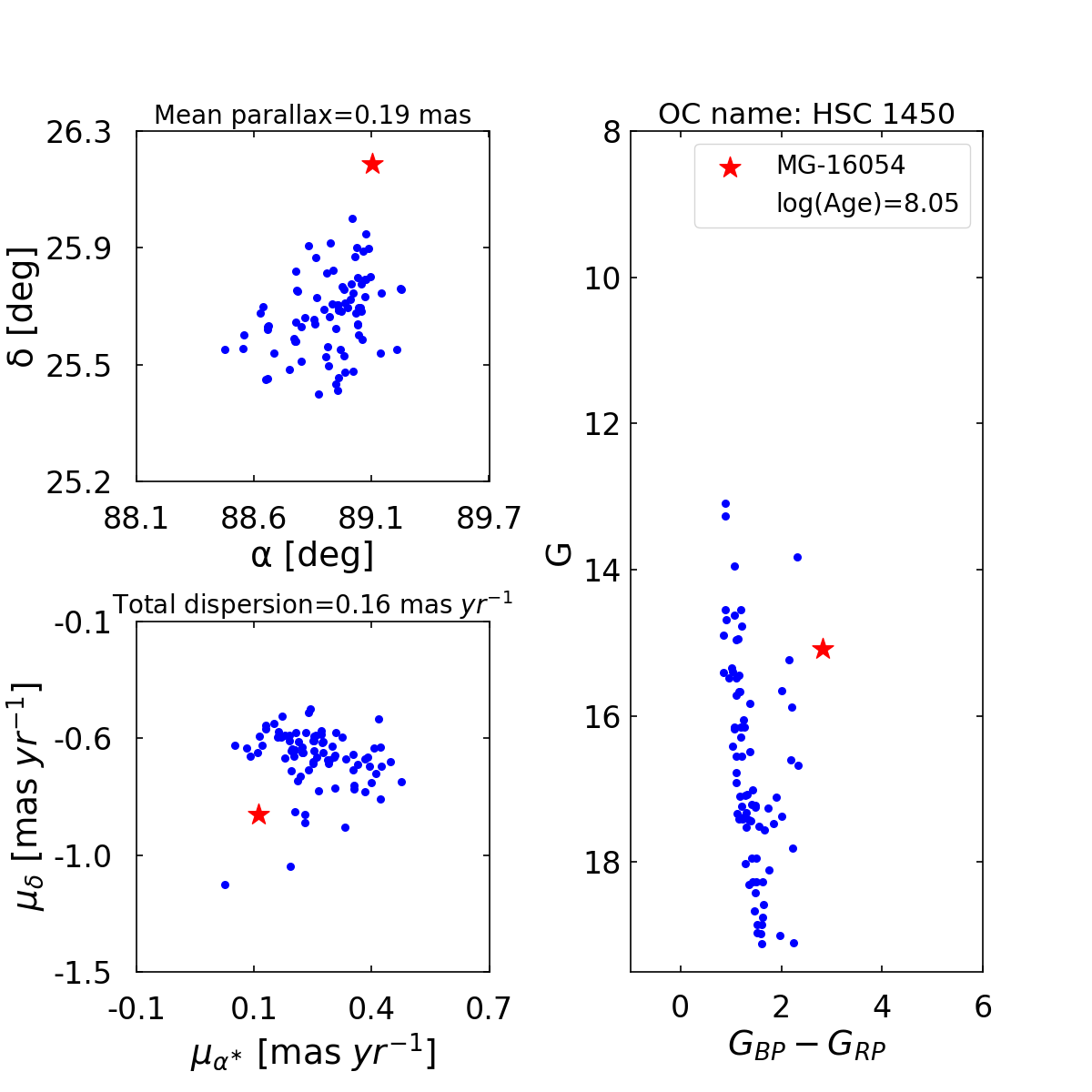}}
		\\
		\centering  
		\subfigure{
			\includegraphics[width=0.32\textwidth]{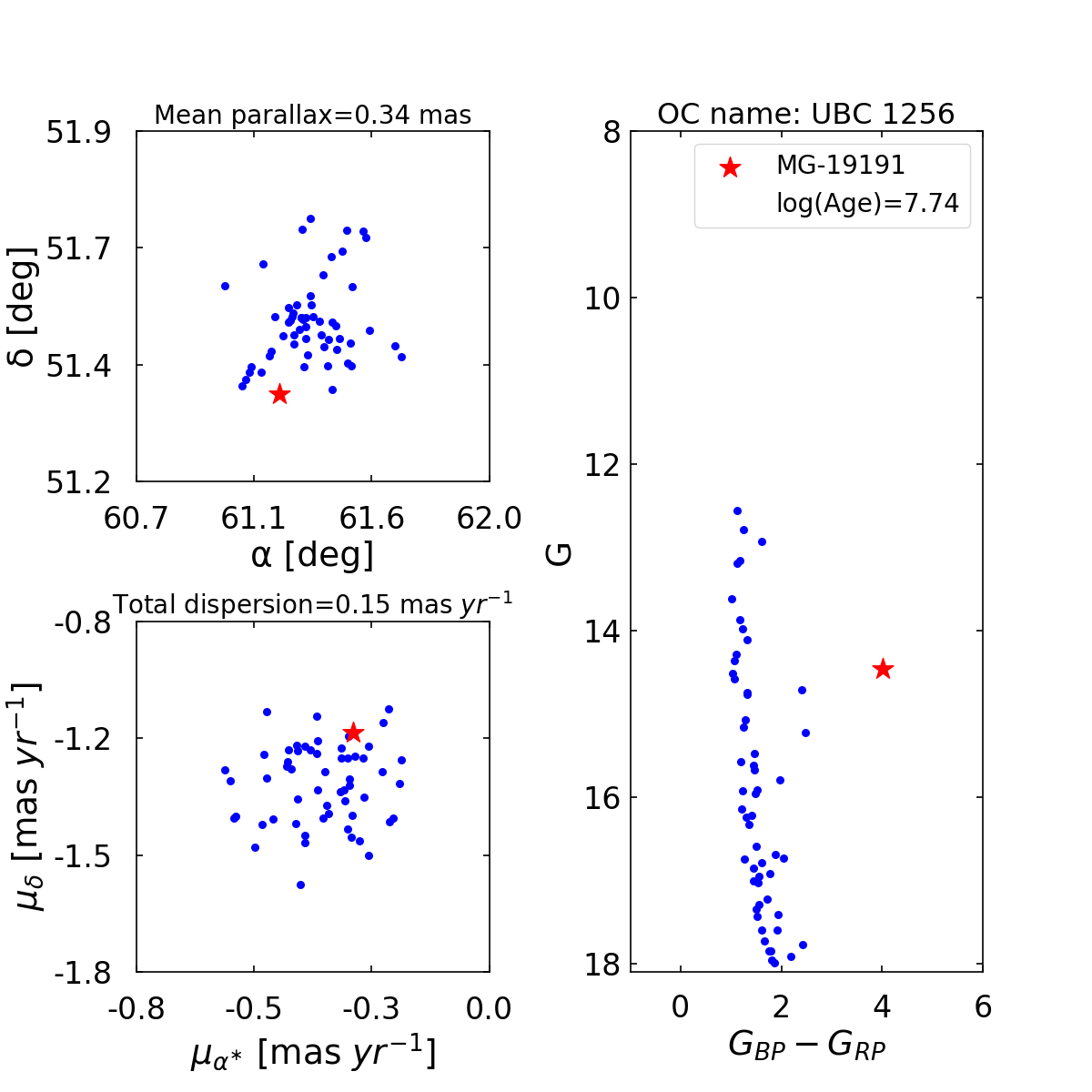}}
		\subfigure{
			\includegraphics[width=0.32\textwidth]{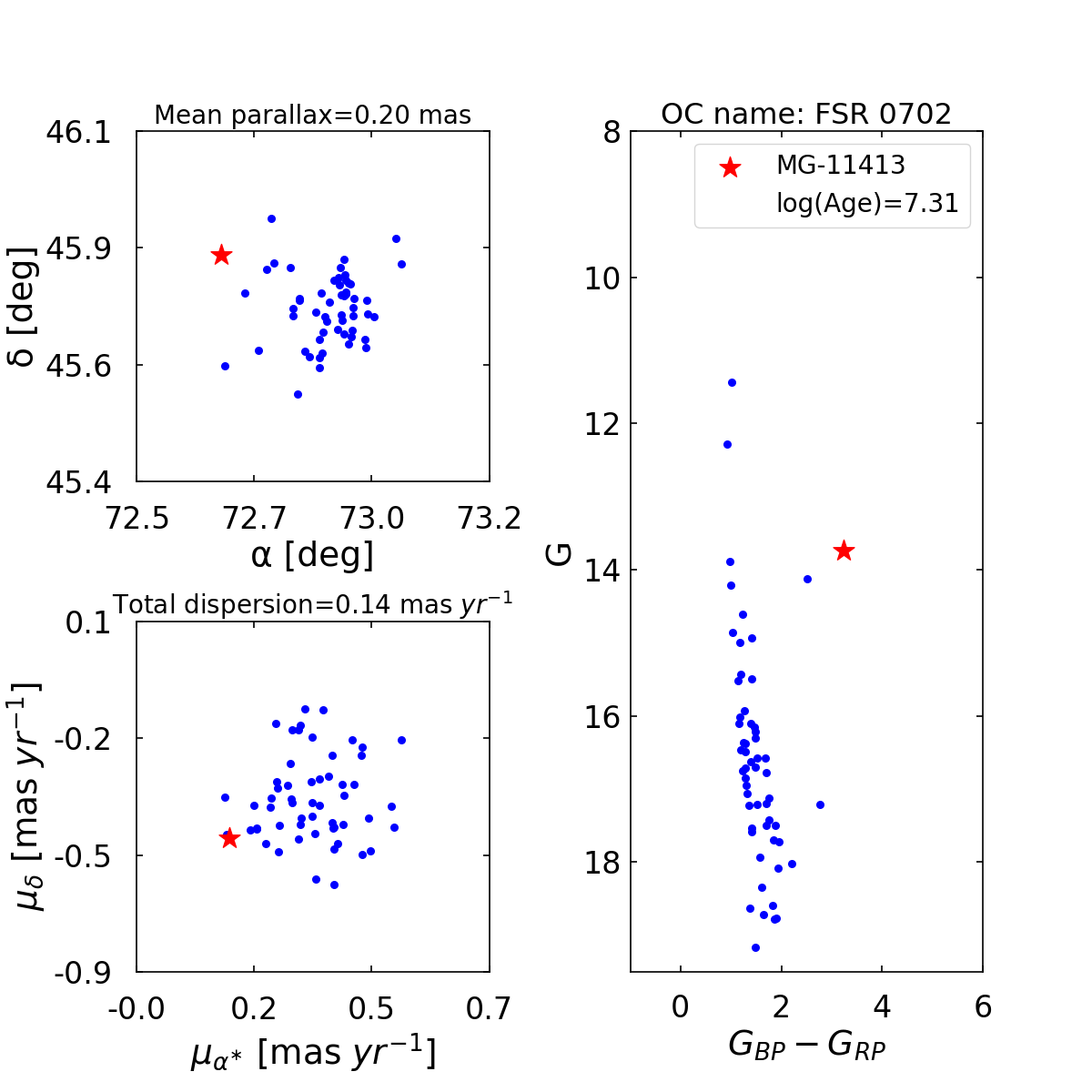}}
		\subfigure{
			\includegraphics[width=0.32\textwidth]{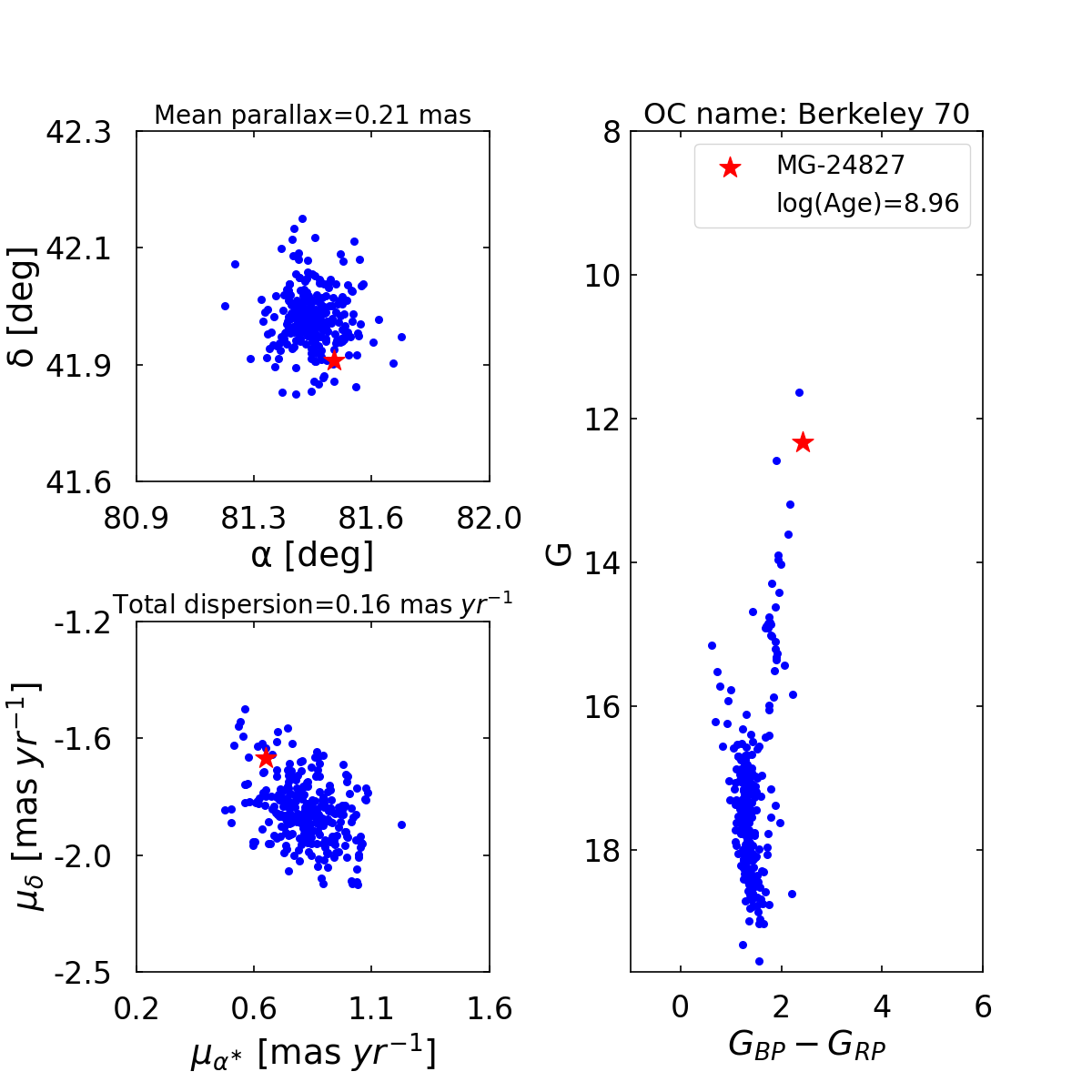}}
		\\
		\centering  
		\subfigure{
			\includegraphics[width=0.32\textwidth]{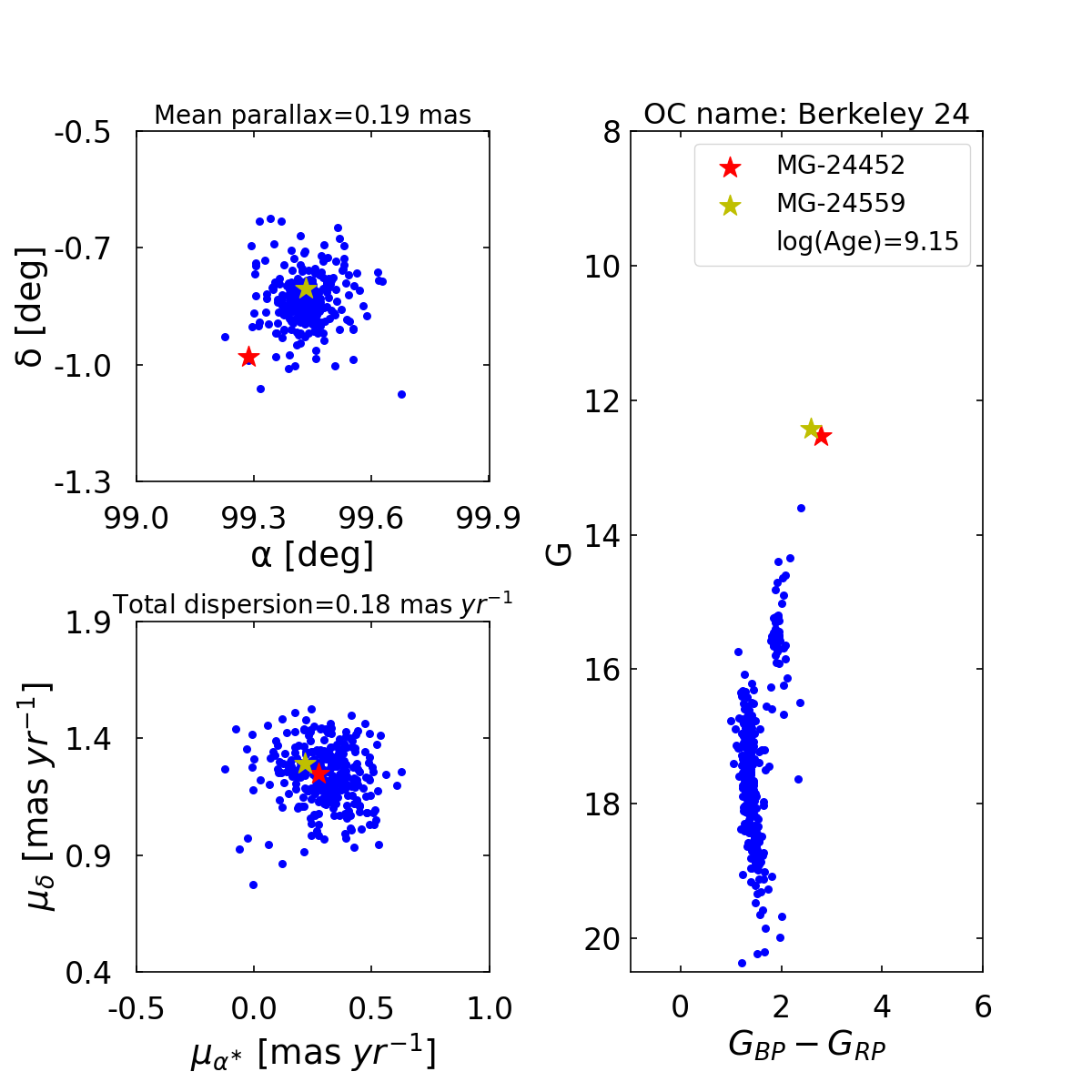}}
		\subfigure{
			\includegraphics[width=0.32\textwidth]{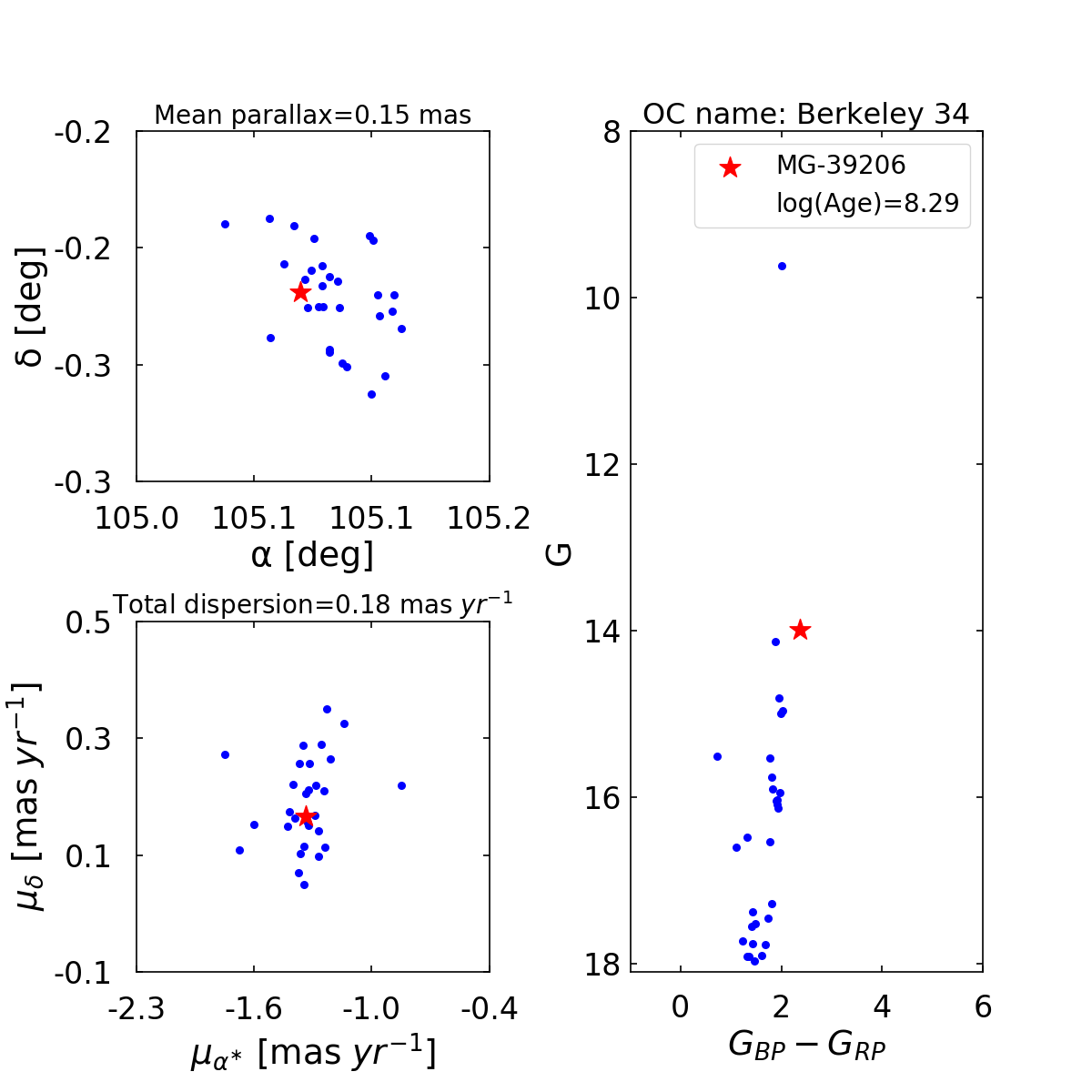}}
		\subfigure{
			\includegraphics[width=0.32\textwidth]{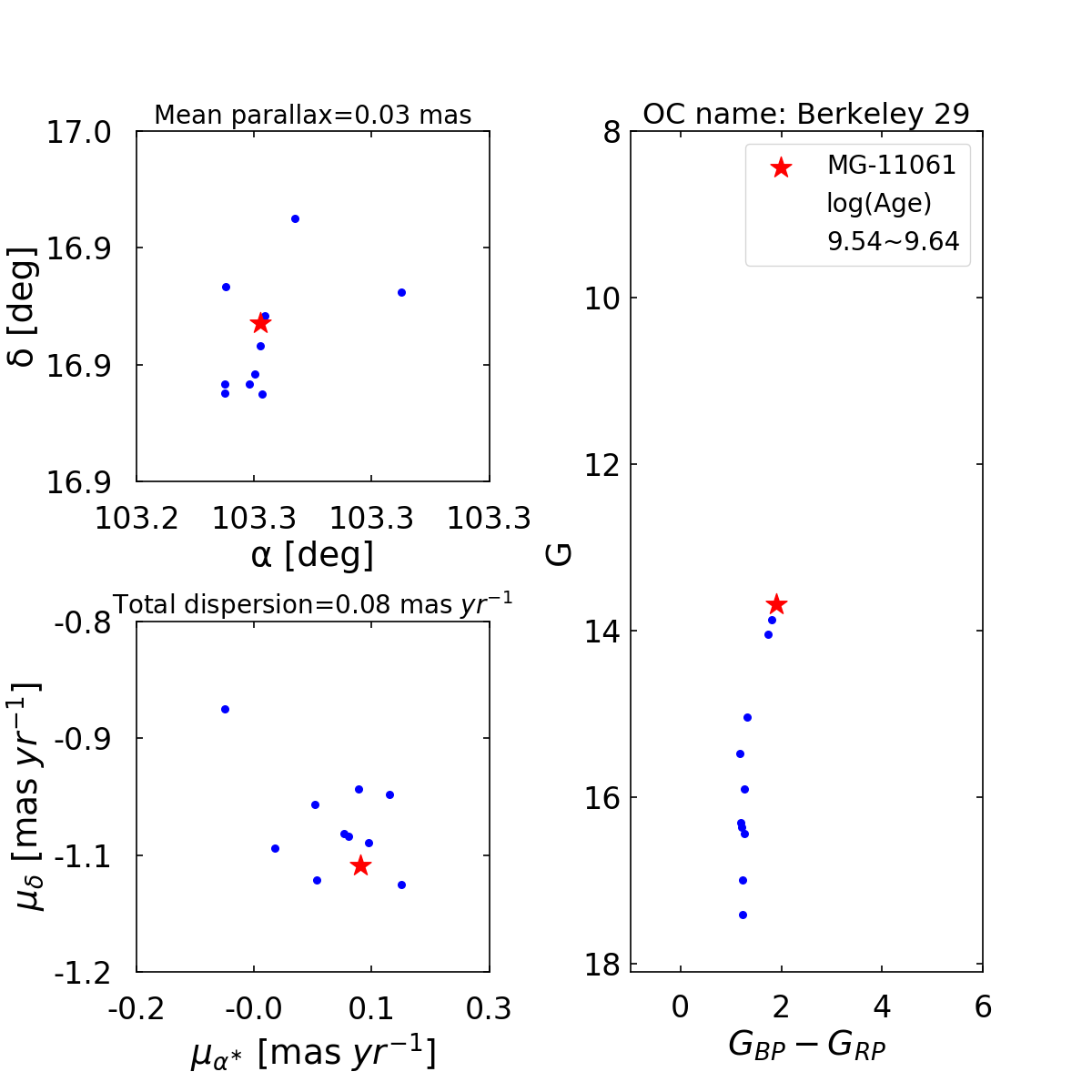}}
		\caption{Here, the listed OCs are LP 1375, Waterloo 1, HSC 1450, UBC 1256, FSR 0702, Berkeley 70, Berkeley 24, Berkeley 34, and Berkeley 29.}
	\end{figure}
	
	\begin{figure}[h]
		\centering  
		\subfigure{
			\includegraphics[width=0.32\textwidth]{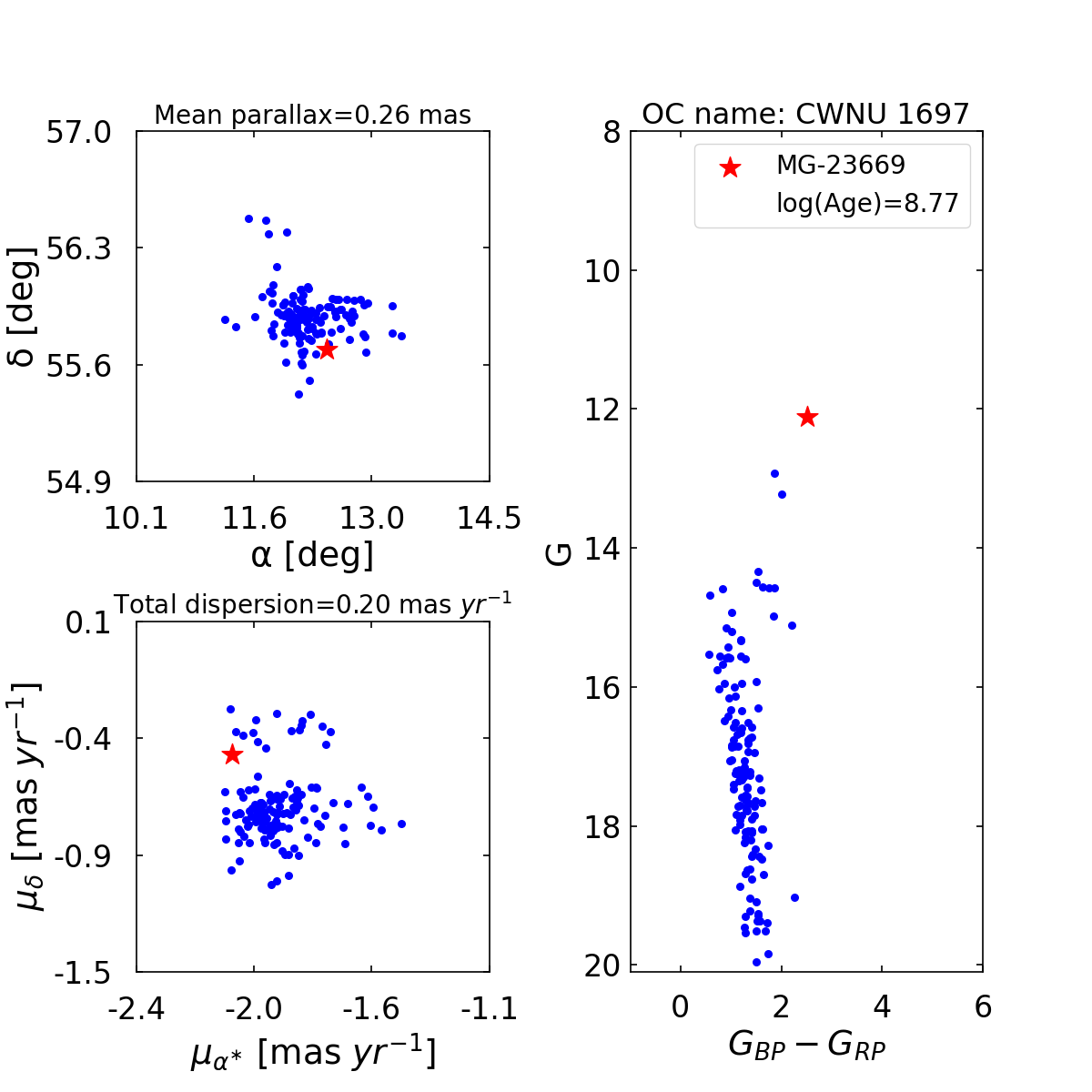}}
		\subfigure{
			\includegraphics[width=0.32\textwidth]{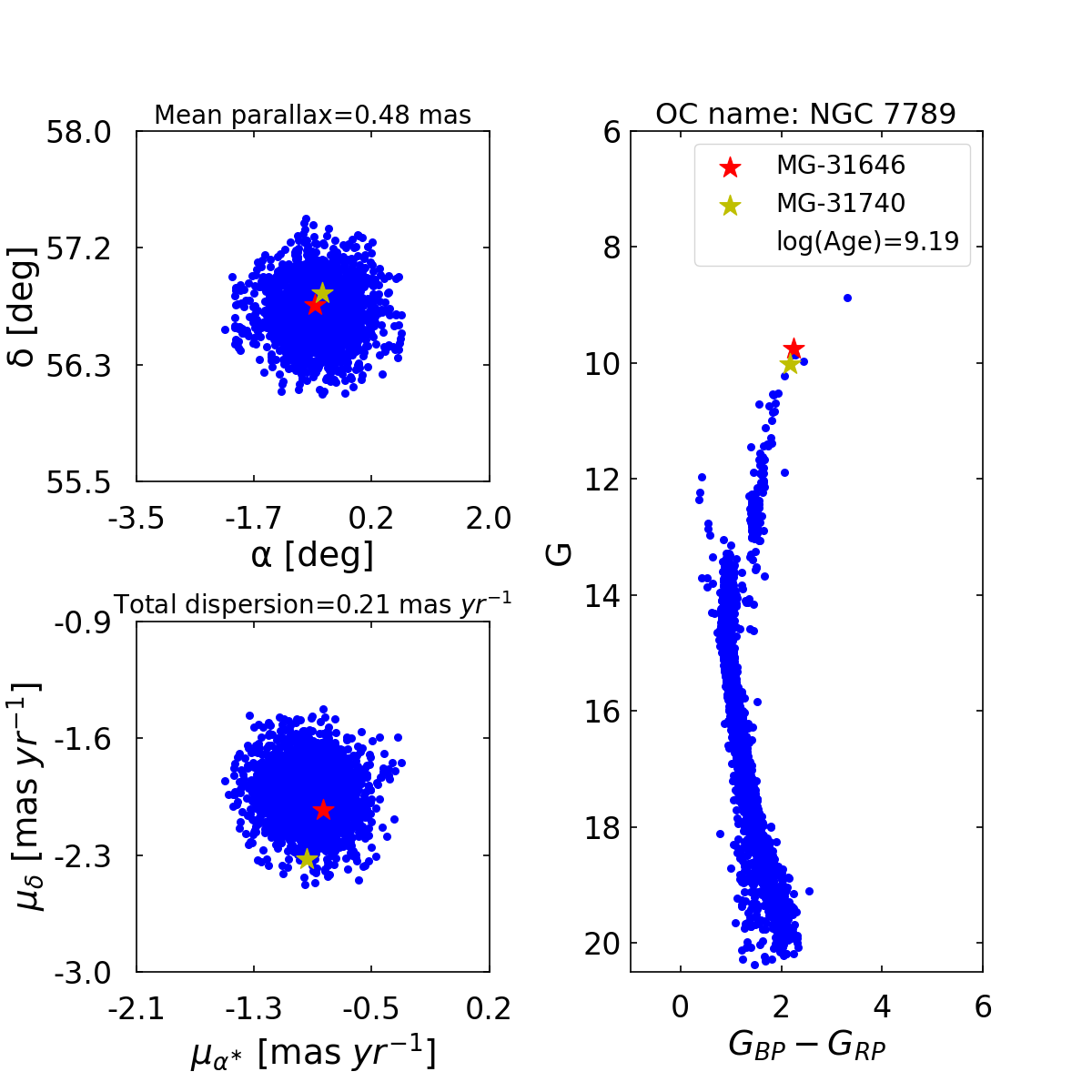}}
		\subfigure{
			\includegraphics[width=0.32\textwidth]{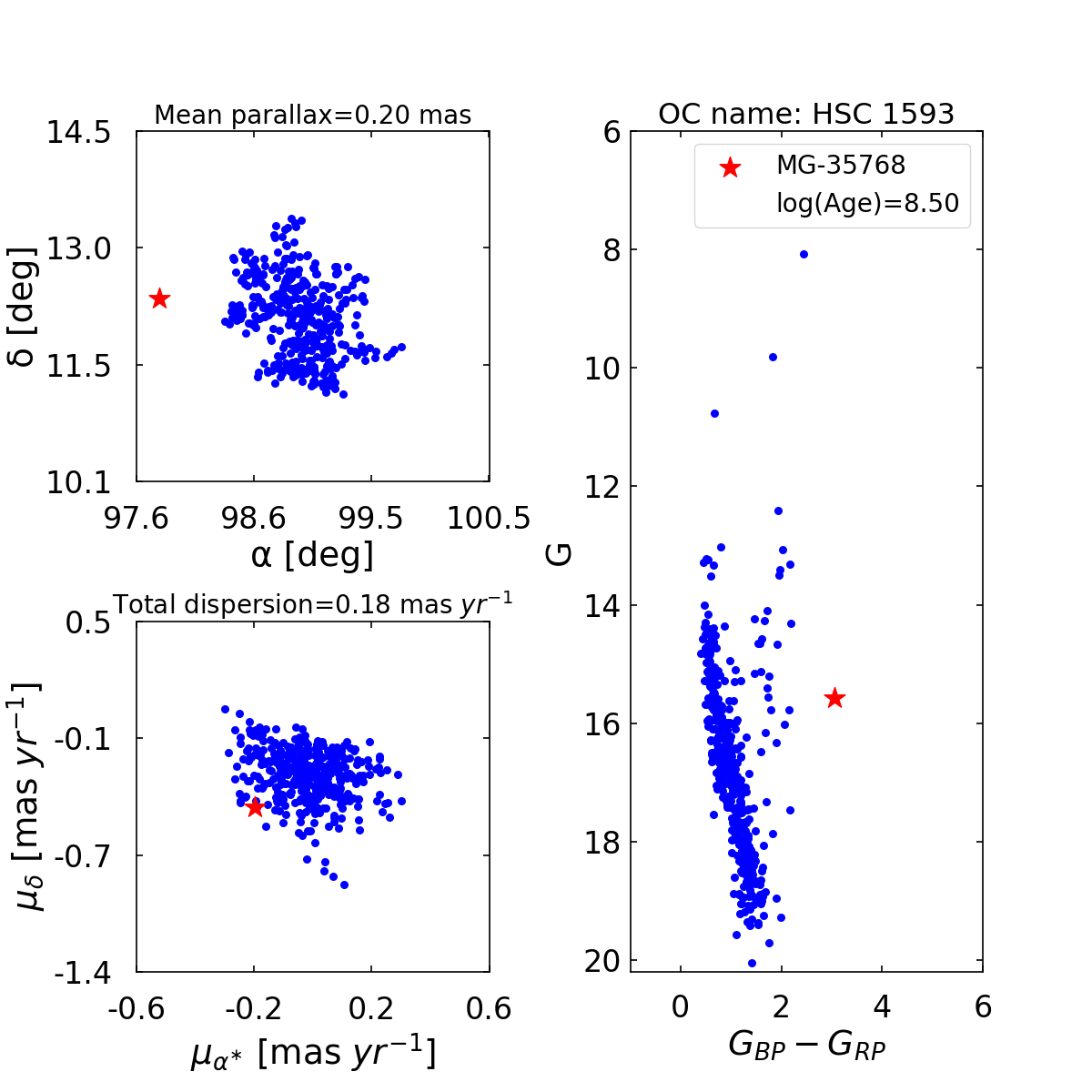}}
		\\
		\centering  
		\subfigure{
			\includegraphics[width=0.32\textwidth]{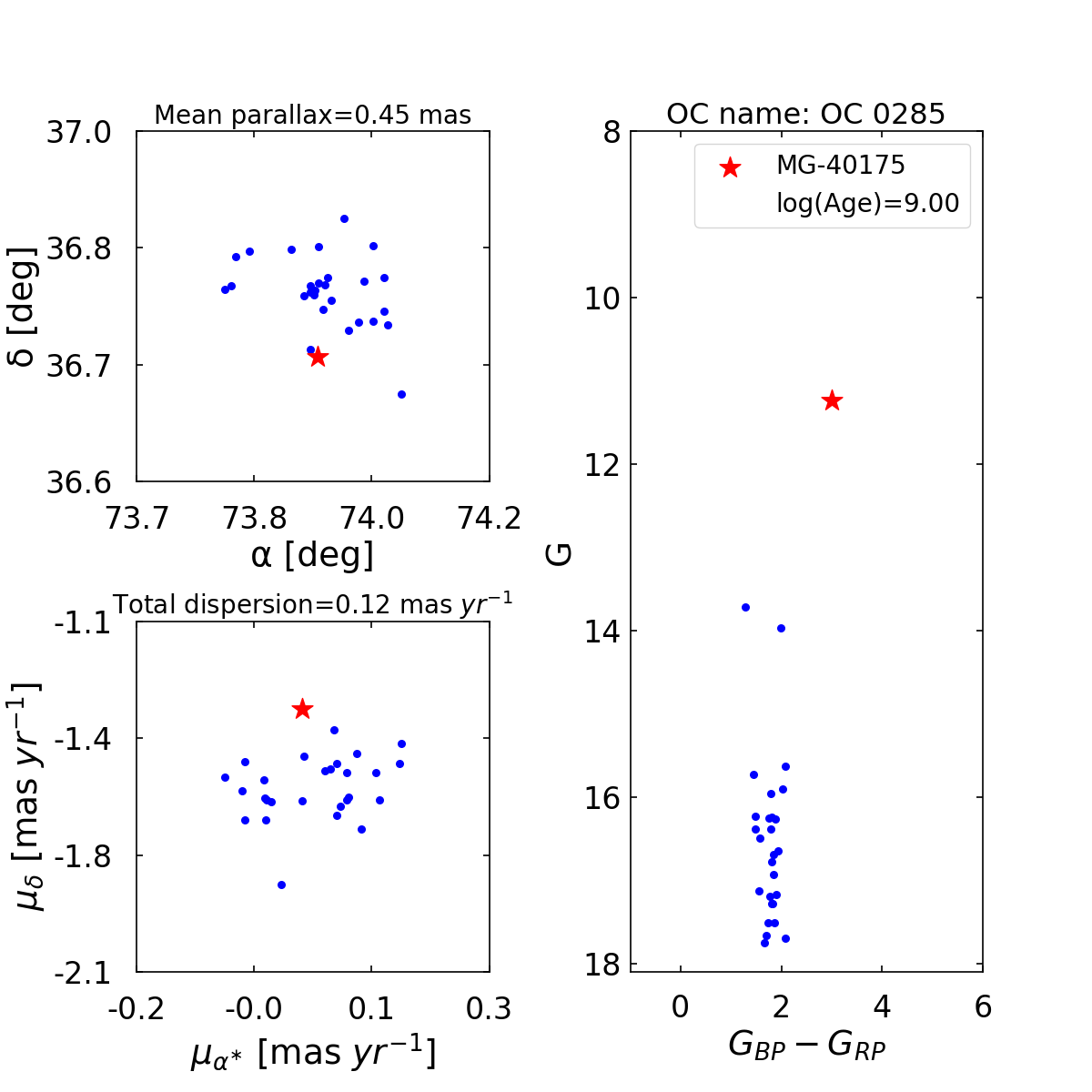}}
		\subfigure{
			\includegraphics[width=0.32\textwidth]{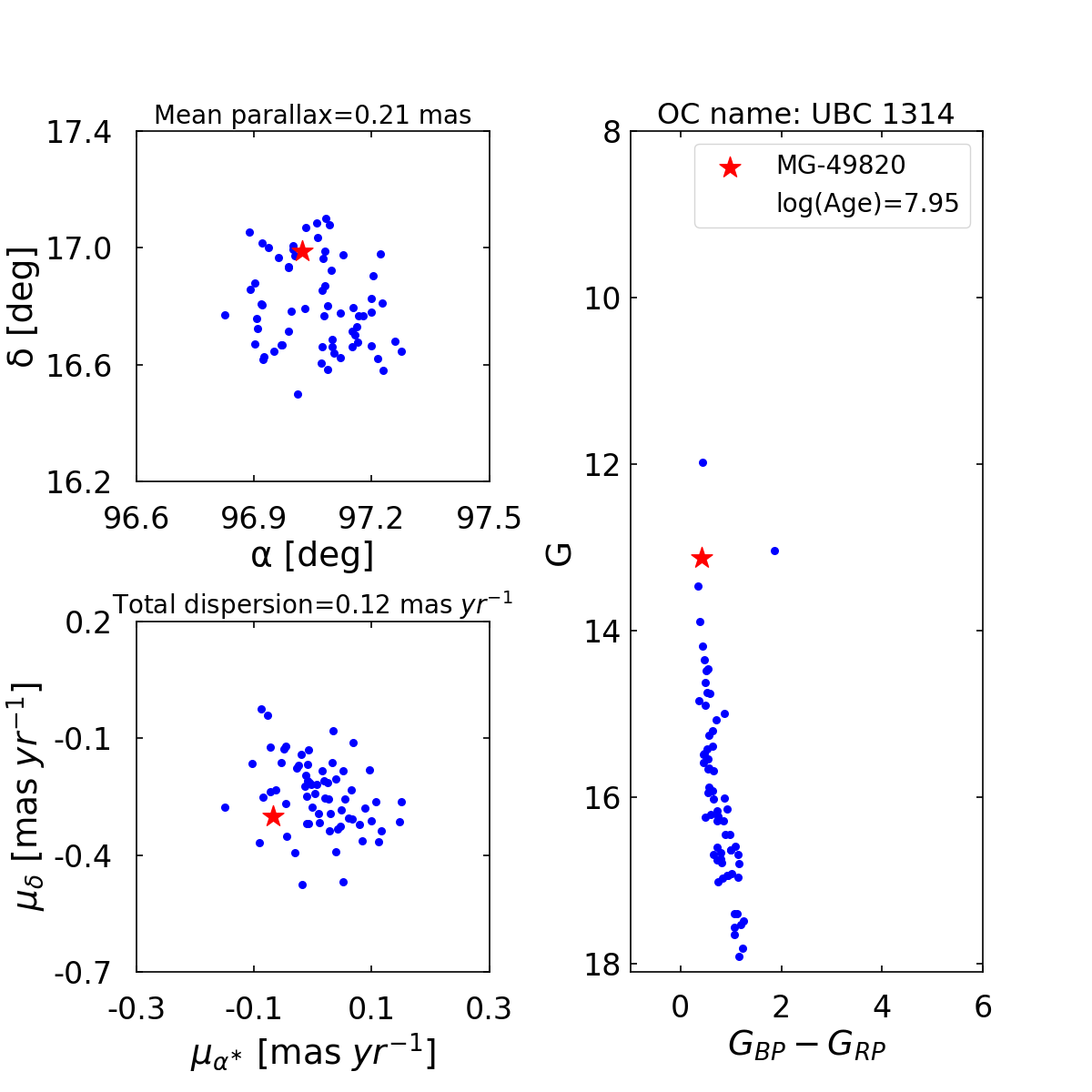}}
		\subfigure{
			\includegraphics[width=0.32\textwidth]{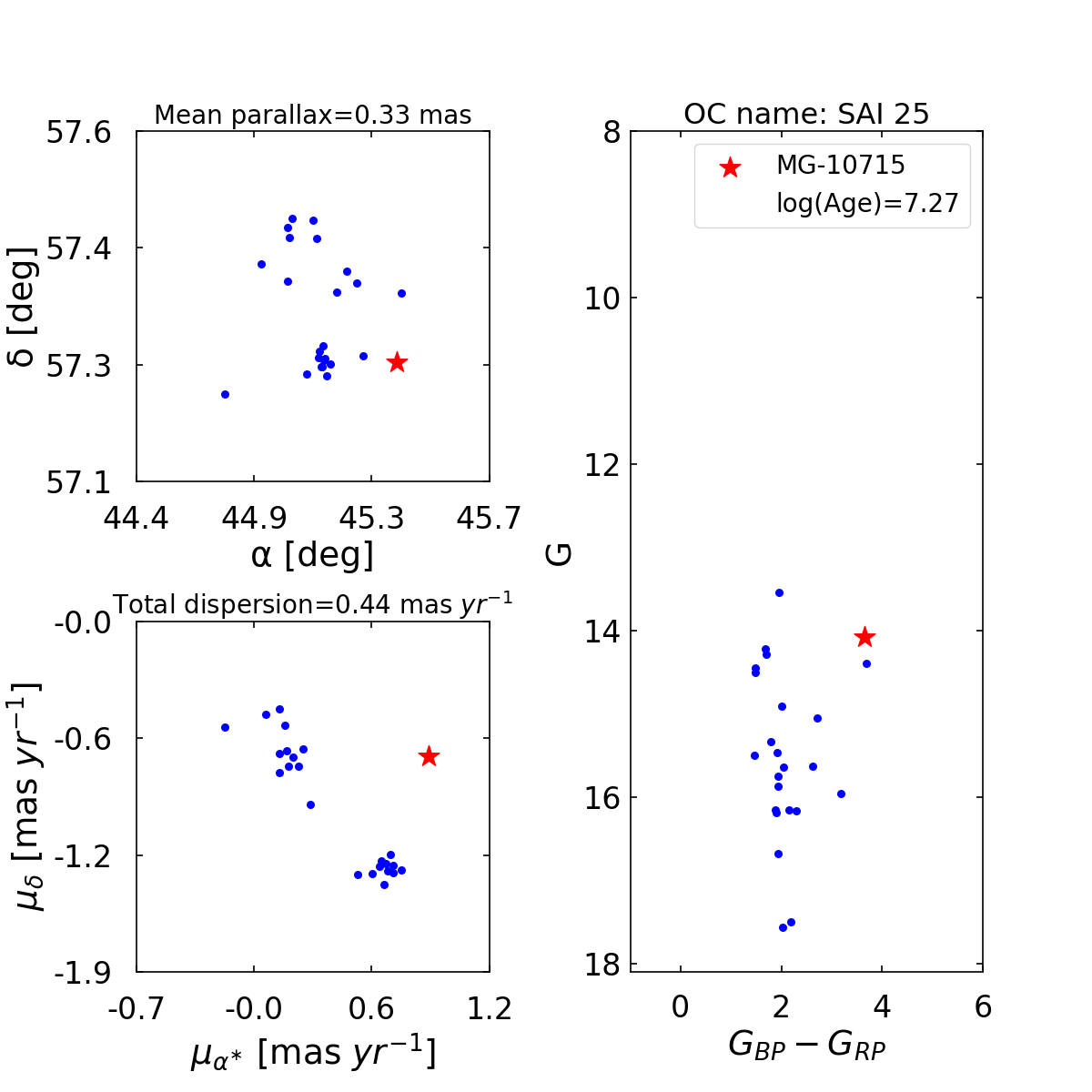}}
		\\
		\centering  
		\subfigure{
			\includegraphics[width=0.32\textwidth]{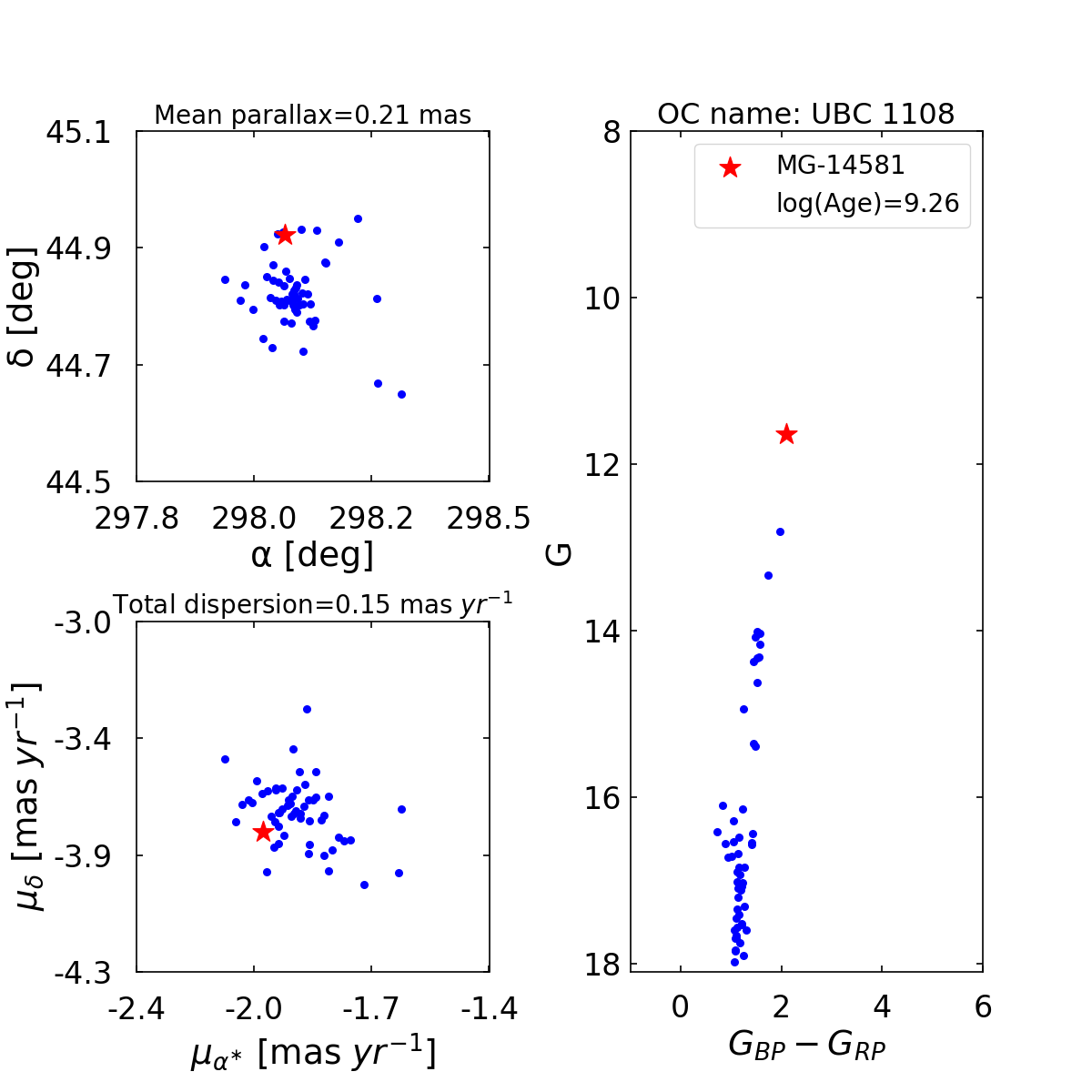}}
		\subfigure{
			\includegraphics[width=0.32\textwidth]{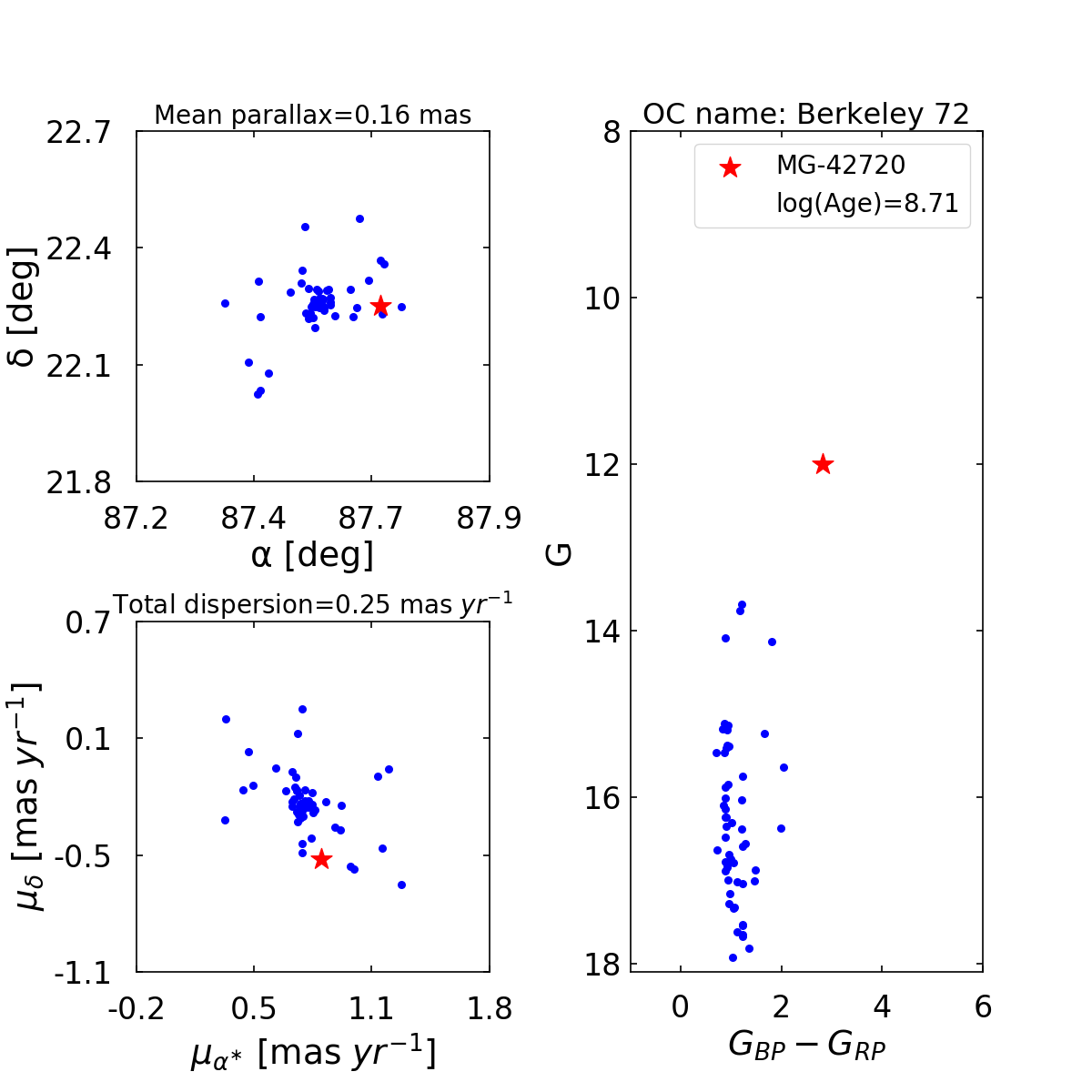}}
		\subfigure{
			\includegraphics[width=0.32\textwidth]{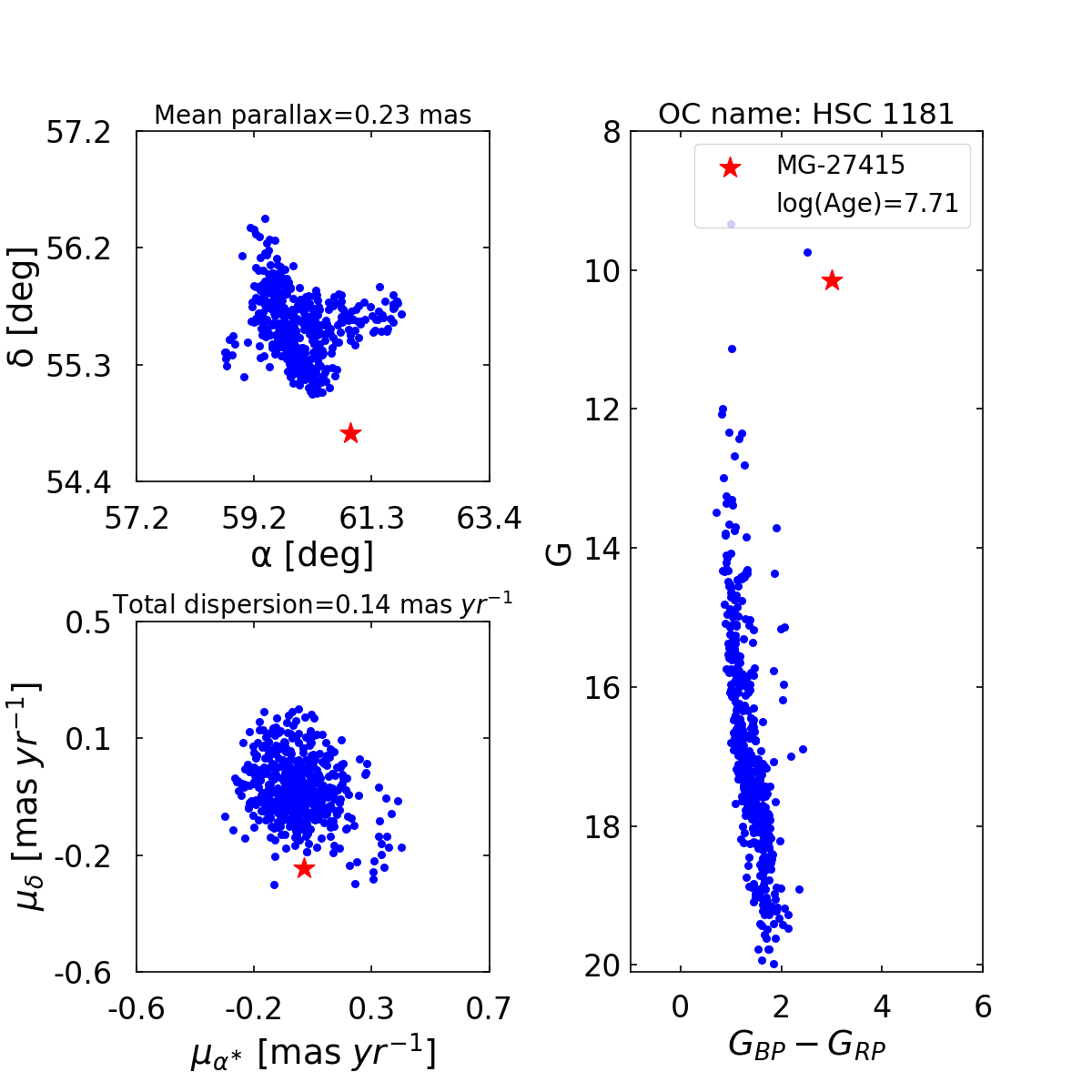}}
		\caption{Here, the listed OCs are CWNU 1697, NGC 7789, HSC 1593, OC 0285, UBC 1314, SAI 25, UBC 1108, Berkeley 72, and HSC 1181.}
	\end{figure}
	
	\begin{figure}[h] 
		\subfigure{
			\includegraphics[width=0.32\textwidth]{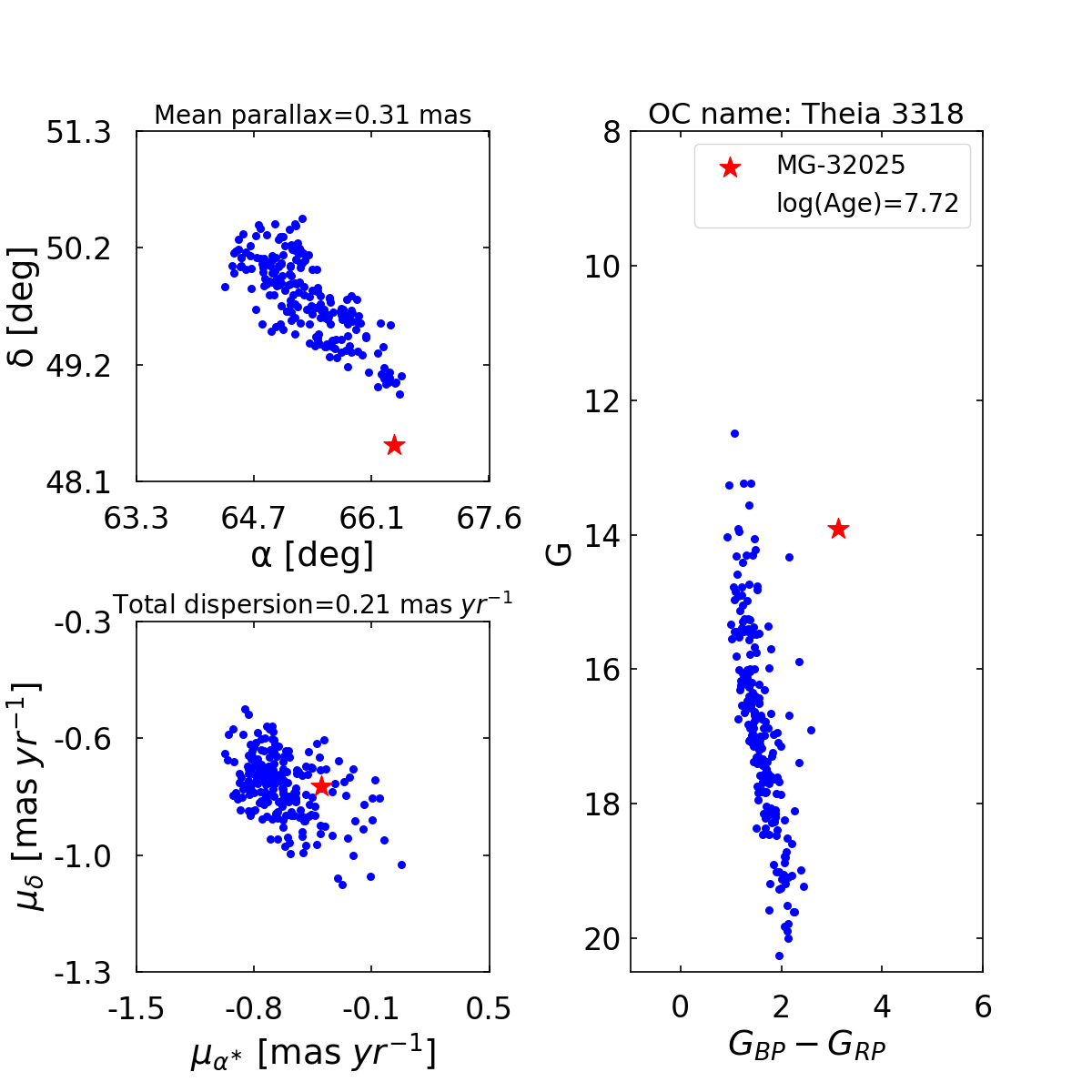}}
		\caption{Here, the listed OC is Theia 3318.}
		\label{figB5}
	\end{figure}
	
	\clearpage
	\section*{ORCID iDs}
	\noindent 
	J. Li~\orcidlink{0000-0002-4953-1545}~\url{https://orcid.org/0000-0002-4953-1545}~~~~~~~~~~
	Z. Z. Yan~\orcidlink{0000-0003-3571-6060}~\url{https://orcid.org/0000-0003-3571-6060}\\
	Y. Xu~\orcidlink{0000-0001-5602-3306}~\url{https://orcid.org/0000-0001-5602-3306}~~~~~~~~~
	J. Zhong~\orcidlink{0000-0001-5245-0335}~\url{https://orcid.org/0000-0001-5245-0335}\\
	Y. J. Li~\orcidlink{0000-0001-7526-0120}~\url{https://orcid.org/0000-0001-7526-0120}
	
	\bibliography{PASPsample631}{}
	\bibliographystyle{aasjournal}

\end{document}